\documentclass{aa}

\usepackage{natbib}
\usepackage{txfonts}
\usepackage[dvips]{graphicx}
\usepackage{subfigure}
\usepackage{color}
\usepackage{multirow}

\bibpunct{(}{)}{;}{a}{}{,}

\begin{document}

\def\mc{\multicolumn}
\def\md{\multicolumn{2}{c}}
\def\mt{\multicolumn{3}{c}}
\def\loggf{$\log{gf}$}
\def\kms{km\,s$^{-1}$}
\def\vsini{v$\sin{i}$}
\def\Vbroad{$V_{\rm broad}$}
\def\teff{$T_{\rm eff}$}
\def\tmean{$T^{\rm mean}_{\rm eff}$}
\def\tphot{$T_{\rm eff}^{\rm phot}$}
\def\texc{$T_{\rm eff}^{\rm exc}$}
\def\tHa{$T_{\rm eff}^{{\rm H}\alpha}$}
\def\logg{$\log{g}$}
\def\loggi{$\log{g_{\rm ion}}$}
\def\logge{$\log{g_{\rm evol}}$}
\def\loggf{$\log{gf}$}
\def\C2{${\rm C}_2$}
\def\bv{$B-V$}
\def\by{$b-y$}
\def\btvt{$B_{\rm T}-V_{\rm T}$}
\def\msun{${\rm M}_{\odot}$}
\def\s{$\sim$}
\def\Rp{$R_{\rm p}$}
\def\Ra{$R_{\rm a}$}
\def\Rm{$R_{\rm m}$}
\def\Ulsr{$U_{\rm LSR}$}
\def\Vlsr{$V_{\rm LSR}$}
\def\Wlsr{$W_{\rm LSR}$}
\def\asn{$\langle$\,S/N\,$\rangle$}
\def\mxfe{$\langle$[X/Fe]$\rangle$}
\def\mxh{$\langle$[X/H]$\rangle$}
\def\tc{$T_{\rm C}$}

\titlerunning{Chemo-chronological analysis of solar-type stars}
\authorrunning{R. da Silva et al.}

\title{Accurate and homogeneous abundance patterns in solar-type stars of
the solar neighbourhood: a chemo-chronological analysis
\thanks{Based on observations collected at the Cerro Tololo
Inter-American Observatory, Chile.}}

\author{R. da Silva\inst{1}
\and G.F. Porto de Mello\inst{2}
\and A.C. Milone\inst{1}
\and L. da Silva\inst{3}
\and L.S. Ribeiro\inst{1}
\and H.J. Rocha-Pinto\inst{2}}

\offprints{R. da Silva,\\
\email{dasilvr2@gmail.com}}

\institute{INPE, Divis\~ao de Astrof\'\i sica, Av. dos Astronautas, 1758
S\~ao Jos\'e dos Campos, 12201-970 Brazil
\and
UFRJ, Observat\'orio do Valongo, Ladeira do Pedro Ant\^onio, 43 Rio de
Janeiro, 20080-090 Brazil
\and
Observat\'orio Nacional, Rua Gal. Jos\'e Cristino 77, S\~ao Cristov\~ao,
Rio de Janeiro, 20921-400 Brazil}

\date{Received / accepted}


\abstract
{}
{We report the derivation of abundances of C, Na, Mg, Si, Ca, Sc, Ti, V, Cr,
Mn, Fe, Co, Ni, Cu, Zn, Sr, Y, Zr, Ba, Ce, Nd, and Sm in a sample of 25
solar-type stars of the solar neighbourhood, correlating the abundances with
the stellar ages, kinematics, and orbital parameters.}
{The spectroscopic analysis, based on data of high resolution and high
signal-to-noise ratio, was differential to the Sun and applied to atomic
line equivalent widths supplemented by the spectral synthesis of C and \C2\ 
features. We also performed a statistical study by using the method of tree
clustering analysis, searching for groups of stars sharing similar elemental
abundance patterns. We derived the stellar parameters from various criteria,
with average errors of 30~K, 0.13~dex, and 0.05~dex, respectively, for
\teff, \logg, and [Fe/H]. The average error of the [X/Fe] abundance ratios
is 0.06~dex. Ages were derived from theoretical HR diagrams and membership
of the stars in known kinematical moving groups.}
{We identified four stellar groups: one having, on average, over-solar
abundances (\mxh\ = +0.26~dex), another with under-solar abundances (\mxh\ =
$-$0.24~dex), and two with intermediate values (\mxh\ = $-$0.06 and
+0.06~dex) but with distinct chemical patterns. Stars sharing solar
metallicity, age, and Galactic orbit possibly have non-solar abundance
ratios, a possible effect either of chemical heterogeneity in their natal
clouds or migration. A trend of [Cu/Fe] with [Ba/Fe] seems to exist, in
agreement with previous claims in the literature, and maybe also of [Sm/Fe]
with [Ba/Fe]. No such correlation involving C, Na, Mn, and Zn is observed.
The [X/Fe] ratios of various elements show significant correlations with
age. [Mg/Fe], [Sc/Fe], and [Ti/Fe] increase with age. [Mn/Fe] and [Cu/Fe]
display a more complex behaviour, first increasing towards younger stars up
to the solar age, and then decreasing, a result we interpret as possibly
related to time-varying yields of SN\,Ia and the weak s-process in massive
stars. The steepest negative age relation is due to [Ba/Fe], but only for
stars younger than the Sun, and a similar though less significant behaviour
is seen for Zr, Ce, and Nd. [Sr/Fe] and [Y/Fe] show a linearly increasing
trend towards younger stars. The [Cu/Ba] and [Sm/Ba] therefore decrease for
younger stars. We found that [Ba/Mg], [Ba/Zn], and [Sr,Y,Ba/Sm] increase but
only for stars younger than the Sun, whereas the [Sr/Mg], [Y/Mg], [Sr/Zn],
and [Y/Zn] ratios increase linearly towards younger stars over the whole age
range.}
{}


\maketitle


\section{Introduction}

The Galactic chemical and dynamical history can be well framed by a series
of average ``laws'' \citep[see e.g.][]{Edvardssonetal1993,McWilliam1997},
namely: the age-metallicity relation (which in principle is accessible by
more than one chemical element), the [element/element] abundance ratios, the
stellar metallicity frequency distribution, the Galactic metallicity
gradient, and the star formation history, besides their mutual relationships
as a function of space and time. The success of the Galactic chemical
evolution models is to be judged by their ability to reproduce these
constraints \citep[see][]{AllenPortodeMello2011}.

In recent years there has been growing recognition that, even though such
average laws are meaningful and fundamental, there may be considerable
underlying complexity in the real Galaxy that has gone at least partly
unappreciated. In the present age of very large spectroscopic databases and
precise abundances for numerous chemical elements, a successful model must
harmonise stellar evolution inputs such as the initial mass function, the
star-formation rate, and mass loss processes, connecting these to the
specific properties of Galactic components, and a large diversity of spatial
and temporal structures, differing timescales for stellar nucleosynthetic
yields and their sensitivities to differing metallicities.

During the evolution of the Galactic disc, nucleosynthesis in successive
generations of stars occurs together with dynamical interactions with the
interstellar gas. The states of the Galaxy in past periods of its evolution
are still preserved in the abundance distributions of solar-type stars,
which constitute an ideal population to study the chemical evolution. These
stars have an age dispersion comparable to the age of the Galaxy. They are
similar to the Sun in many physical parameters, allowing the application of
a differential analysis and the consequent minimisation of theoretical
shortcomings of atmospheric models and systematic errors. In addition, their
chemical composition does not change in consequence of the mixing processes
in their surfaces, which means that the present abundance of a given element
is the same as in the time of their formation (an exception are the
abundances of Li, Be, and B, but these elements are not considered in this
work). Therefore, the chemical abundance of solar-type stars, combined with
kinematical, orbital, and evolutionary parameters (mass and age), provide a
powerful tool to investigate the chemical and dynamical evolution of the
Galaxy.

Over the last decade, several works have analysed the composition of disc
dwarf stars of spectral types F and G \citep{Chenetal2000,Reddyetal2003,
Bensbyetal2005,Chenetal2008,Nevesetal2009}. The metallicity ranges have
become wider, the number of stars and elements studied has become larger,
stars with and without detected planets have been compared, and the chemical
distinction between thin and thick stellar population has been refined. Here
we specifically ask, when regarded in as high a level of detail as possible
with present techniques, to what extent the relative abundances of chemical
elements can be traced as a function of age, the nature of underlying
nucleosynthetic processes, and whether these properties can be statistically
grouped for the nearby solar-type stars, defining ``snapshots'' relevant to
the chemo-chronological evolution of the Galaxy.

In this work we present a multi-elemental spectroscopic analysis of a sample
of 25 solar-type stars in the solar neighbourhood, all members of the thin
disc stellar population (excepting one star in the transition between thin
and thick discs). We have performed the determination of atmospheric
parameters (effective temperature, metallicity, surface gravity, and
micro-turbulence velocity), mass, age, kinematical and orbital parameters,
and elemental abundances based on equivalent widths or spectral synthesis.
Three different criteria were used to pin down the stellar effective
temperatures, and they showed excellent internal consistency. We have also
performed a statistical study of our abundance results using the method of
tree clustering analysis \citep{Everittetal2001}, through which we looked
for stellar groups that share similar abundances in [X/H], where X
represents one given element. Four groups were identified and then analysed
in terms of their relations with [Fe/H], age, [Ba/Fe], kinematics, and
Galactic orbits.

Despite the small range in metallicity, our sample stars cover a broad range
in age, and possible trends with age were traced. The relation between
[X/Fe] and [Ba/Fe] for a few elements has also been considered given the
Ba-rich nature of some of our stars. In particular, we investigated previous
correlations with Na and Cu suggested by \citet{Castroetal1999}. Finally,
considering the results of \citet{RochaPintoetal2006} that, on average,
metal-poor and old stars tend to have larger $\vert$\Rm$-R_\odot\vert$
(where \Rm\ is the mean orbital distance from the Galactic centre), we have
looked for any relations involving the stellar groups of the clustering
analysis and the kinematic and orbital parameters of the sample.

Though limited in size, our sample was carefully built up to undergo an
homogeneous and detailed analysis, based on spectra with high resolution and
high signal-to-noise ratio, in order to achieve a precision as high as
possible in our determinations. Particular care was exercised to derive the
stellar atmospheric parameters from different and independent criteria, in
an attempt to limit the abundance uncertainties as much as possible.

In Sect.~\ref{observ} we describe the observations and the reduction
process. In Sect.~\ref{analysis} we present the methods used to derive the
atmospheric parameters and the chemical abundances. The stellar evolution,
kinematics, and orbits are presented in Sect.~\ref{evol_kin_orb}. The tree
clustering method is described in Sect.~\ref{classtree}, and all the results
are discussed in Sect.~\ref{results}. Finally, we present our conclusions in
Sect.~\ref{concl}.


\section{Observations and data reduction}
\label{observ}

The sample stars were selected from the Bright Star
\citep{HoffleitJaschek1982} and Hipparcos \citep{ESA1997} catalogues
according to the following conditions:

\begin{itemize}

\item[\it i)] Solar neighbourhood stars in a distance $\leq$ 40~pc;

\item[\it ii)] Stars brighter than $V$ = 6.5 and with declination
$< +20\,\degr$;

\item[\it iii)] Stars with effective temperature and metallicity distributed
over about $5500 \leq$ \teff $\leq 6100$~K and $-0.3 \leq$ [Fe/H] $\leq
0.3$~dex, respectively, which represents the colour index range $0.52 \leq$
(\bv) $\leq 0.78$ from the (\bv) calibration described in
Sect.~\ref{phot_temp} (Eq.~\ref{cal_bv}); and

\item[\it iv)] Stars with no information of duplicity (capable of
significantly affecting the spectroscopy) available in the astrometric and
spectroscopic binary catalogues of \citet{HoffleitJaschek1982},
\citet{WarrenHoffleit1987}, \citet{Battenetal1989}, and
\citet{DuquennoyMayor1991}; the possibility of duplicity was afterward
revised in the Washington Double Star Catalogue \citep{Masonetal2001} and in
the survey of \citet{Raghavanetal2010}, and no close-in companions that
could affect our analysis were found.

\end{itemize}

Based on these criteria, 99 F, G, and K dwarfs and subgiants were selected,
out of which 25 stars were observed and analysed in this work. Our sample,
shown in Table~\ref{sample}, contains F and G dwarf and subgiants stars from
the thin disc stellar population, excepting the star HD\,50806, which is
probably in the transition between thin and thick discs.

The observations were carried out at the Cerro Tololo Inter-American
Observatory (CTIO, Chile) in two different runs:
$i)$ 15 stars were observed in March 25-26, 1994 using the Cassegrain
{\it \'echelle} spectrograph mounted on the 4~m telescope, with the red
camera, 140~$\mu$m slit, and Tek CCD detector of $1024 \times 1024$ pixels
($24 \times 24~\mu$m pixel size), and with a gain of 1~e$^-$/ADU; the
spectra have resolution $R \sim 29\,000$ and cover the wavelength range
4370$-$6870~\AA\ divided into 46 orders; and
$ii)$ 10 stars were observed in November 8-15, 1997 using the
bench-mounted {\it \'echelle} spectrograph and a $750~$mm folded Schmidt
camera attached to the 1.5~m telescope; the same CCD was used; the spectra
have resolution $R \sim 46\,000$ and cover the wavelength range
4550$-$6520~\AA\ divided into 37 orders.

\begin{table}
\centering
\caption[]{The 25 sample stars. The mean values of S/N ratios measured in
	   continuum windows around $\lambda\lambda$ 4500$-$5000~\AA\ are
	   listed for the stars and for the two sunlight spectra
	   (Ganymede).}
\label{sample}
\begin{tabular}{l r@{}l c c}
\hline\hline\noalign{\smallskip}
Object &
\multicolumn{2}{c}{$V$} &
Sp. type &
\asn \\
\noalign{\smallskip}\hline\noalign{\smallskip}
\multicolumn{5}{c}{\centering First run (March 26-25, 1994):} \\
\hline\noalign{\smallskip}
Ganymede   &    5&.10 & G2\,V    & 450 \\
HD\,20807  &	5&.24 & G0\,V	 & 430 \\
HD\,43834  &	5&.08 & G7\,V	 & 370 \\
HD\,84117  &	4&.93 & F8\,V	 & 410 \\
HD\,102365 &	4&.89 & G2\,V	 & 370 \\
HD\,112164 &	5&.89 & F9\,V	 & 400 \\
HD\,114613 &	4&.85 & G3\,V	 & 400 \\
HD\,115383 &	5&.19 & G0\,V	 & 320 \\
HD\,115617 &	4&.74 & G7\,V	 & 370 \\
HD\,117176 &	4&.97 & G5\,V	 & 410 \\
HD\,128620 & $-$0&.01 & G2\,V	 & 450 \\
HD\,141004 &	4&.42 & G0\,V	 & 520 \\
HD\,146233 &	5&.49 & G2\,V	 & 420 \\
HD\,147513 &	5&.37 & G5\,V	 & 290 \\
HD\,160691 &	5&.12 & G3\,IV-V & 400 \\
HD\,188376 &	4&.70 & G5\,IV   & 310 \\
\hline\noalign{\smallskip}
\multicolumn{5}{c}{\centering Second run (November 8-15, 1997):} \\
\hline\noalign{\smallskip}
Ganymede   &    5&.10 & G2\,V    & 410 \\
HD\,1835   &	6&.39 & G3\,V	 & 230 \\
HD\,26491  &	6&.37 & G1\,V	 & 310 \\
HD\,33021  &	6&.15 & G1\,IV   & 240 \\
HD\,39587  &	4&.39 & G0\,V	 & 510 \\
HD\,50806  &	6&.05 & G5\,V	 & 300 \\
HD\,53705  &	5&.56 & G0\,V	 & 370 \\
HD\,177565 &	6&.15 & G6\,V	 & 290 \\
HD\,181321 &	6&.48 & G2\,V	 & 220 \\
HD\,189567 &	6&.07 & G2\,V	 & 260 \\
HD\,196761 &	6&.36 & G8\,V	 & 400 \\
\hline
\end{tabular}
\end{table}

The two subsamples, although observed in different conditions, were both
selected based on the same criteria and will be treated as an homogeneous
single sample. The spectra collected in the first run have, on average,
signal-to-noise ratio (S/N = 395 $\pm$ 60) slightly larger than those of the
second one (S/N = 320 $\pm$ 90) and, despite having smaller resolution, may
provide smaller uncertainties in some parameters estimated here.
Nevertheless, all spectra have S/N $>$ 200 in the blue region, which
warranties spectral line profiles good enough to the equivalent width
measurements. Any differences in the error estimates are discussed
throughout the paper whenever needed.

Two spectra of the sunlight reflected by Ganymede were also observed, one in
each run. The S/N ratios were estimated using continuum windows in the
spectra selected by inspection of the solar flux atlas of
\citet{Kuruczetal1984} (hereafter the Solar Flux Atlas) and the solar line
identifications catalogue of \citet{Mooreetal1966} (hereafter the Solar
Lines Catalogue). The mean values \asn\ measured in the wavelength range
4500$-$5000~\AA\ are listed in Table~\ref{sample}. For larger wavelengths
the S/N ratios are even higher, approaching twice that for $\lambda$4500 in
the range 6000$-$6500~\AA.

The spectra were reduced using IRAF\footnote{{\it Image Reduction and
Analysis Facility}, distributed by the National Optical Astronomy
Observatories (NOAO), USA.} routines for order identification and
extraction, background subtraction (including bias and scattered light),
flat-field correction, wavelength calibration, radial-velocity shift
correction, and flux normalisation. The wavelength calibration was performed
onto the stellar spectra themselves using lines selected by inspection of
the Solar Flux Atlas and the Solar Lines Catalogue. The normalisation of the
continuum is a very delicate and relevant step in the analysis procedure,
since the accuracy of the equivalent width measurements is very sensitive to
a faulty determination of the continuum level. Therefore, continuum windows
free from telluric or photospheric lines were carefully selected also based
on the Solar Flux Atlas and the Solar Lines Catalogue.


\section{Spectroscopic analysis}
\label{analysis}

A differential spectroscopic analysis relative to the Sun was performed to
determine the atmospheric parameters and the chemical abundance of several
elements in our sample. The analysis was based on the equivalent widths of
atomic lines measured in the spectra, and on the spectral synthesis of
carbon atomic and molecular lines. The two groups of stars (15 observed in
the first and 10 in the second run) were treated in comparison with the
Ganymede spectrum of their respective run.


\subsection{Equivalent widths and atomic line parameters}
\label{ew}

Atomic lines of the elements Na, Mg, Si, Ca, Sc, Ti, V, Cr, Mn, Fe, Co, Ni,
Cu, Zn, Sr, Y, Zr, Ba, Ce, Nd, and Sm were selected throughout the spectral
range for equivalent width ($EW$) measurements. The lines were chosen based
on the Solar Flux Atlas and the Solar Lines Catalogue, selecting only those
for which the profiles were sufficiently clean from blends in order to
provide reliable measurements. For both solar and stellar spectra, the $EW$
values of more than 7500 lines of these elements (about 300 lines per star)
were measured by hand by Gaussian function fit using IRAF routines.

\begin{figure*}
\centering
\begin{minipage}[b]{0.49\textwidth}
\centering
\resizebox{0.8\hsize}{!}{\includegraphics{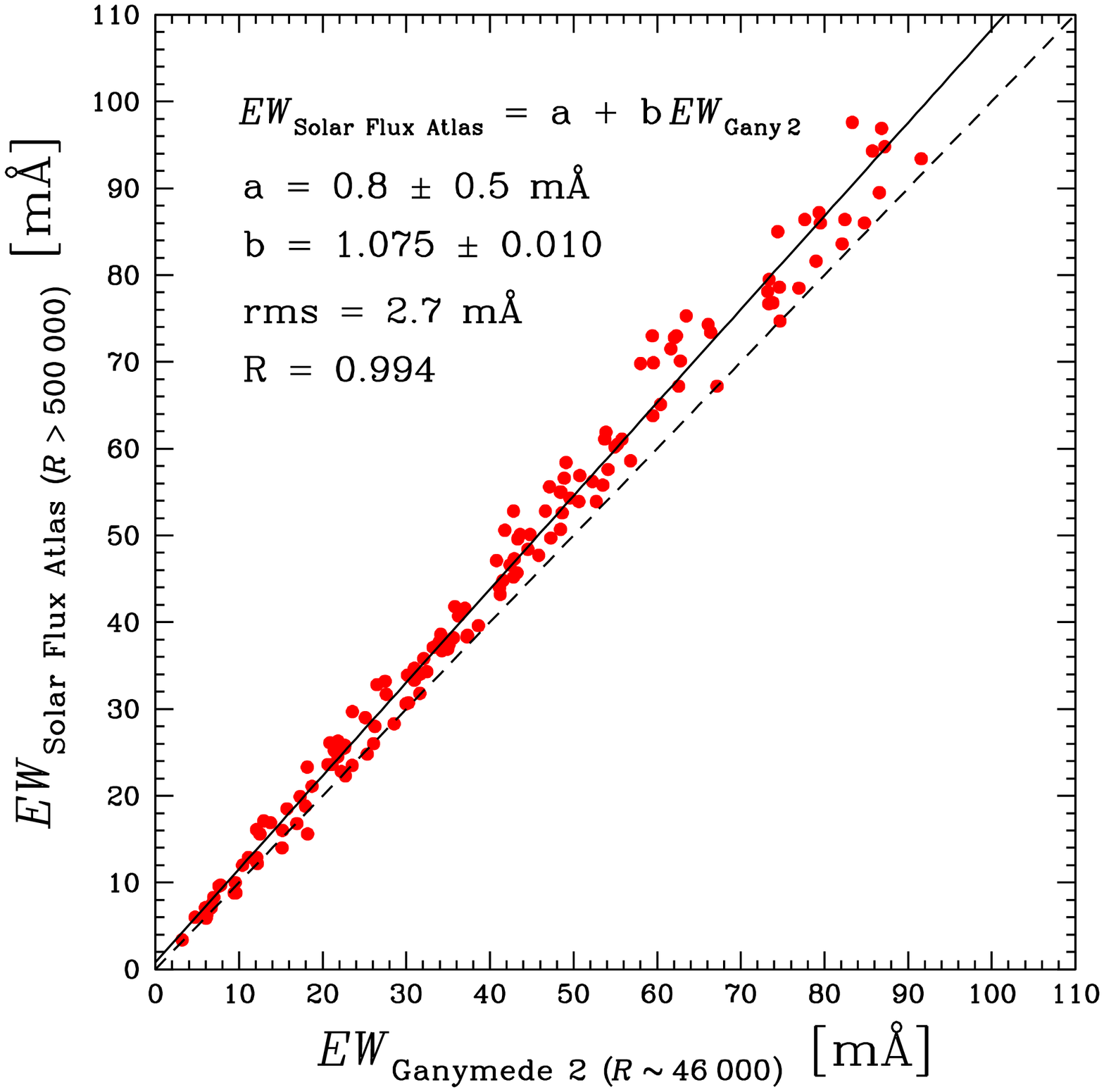}}
\end{minipage}
\begin{minipage}[b]{0.49\textwidth}
\centering
\resizebox{0.8\hsize}{!}{\includegraphics{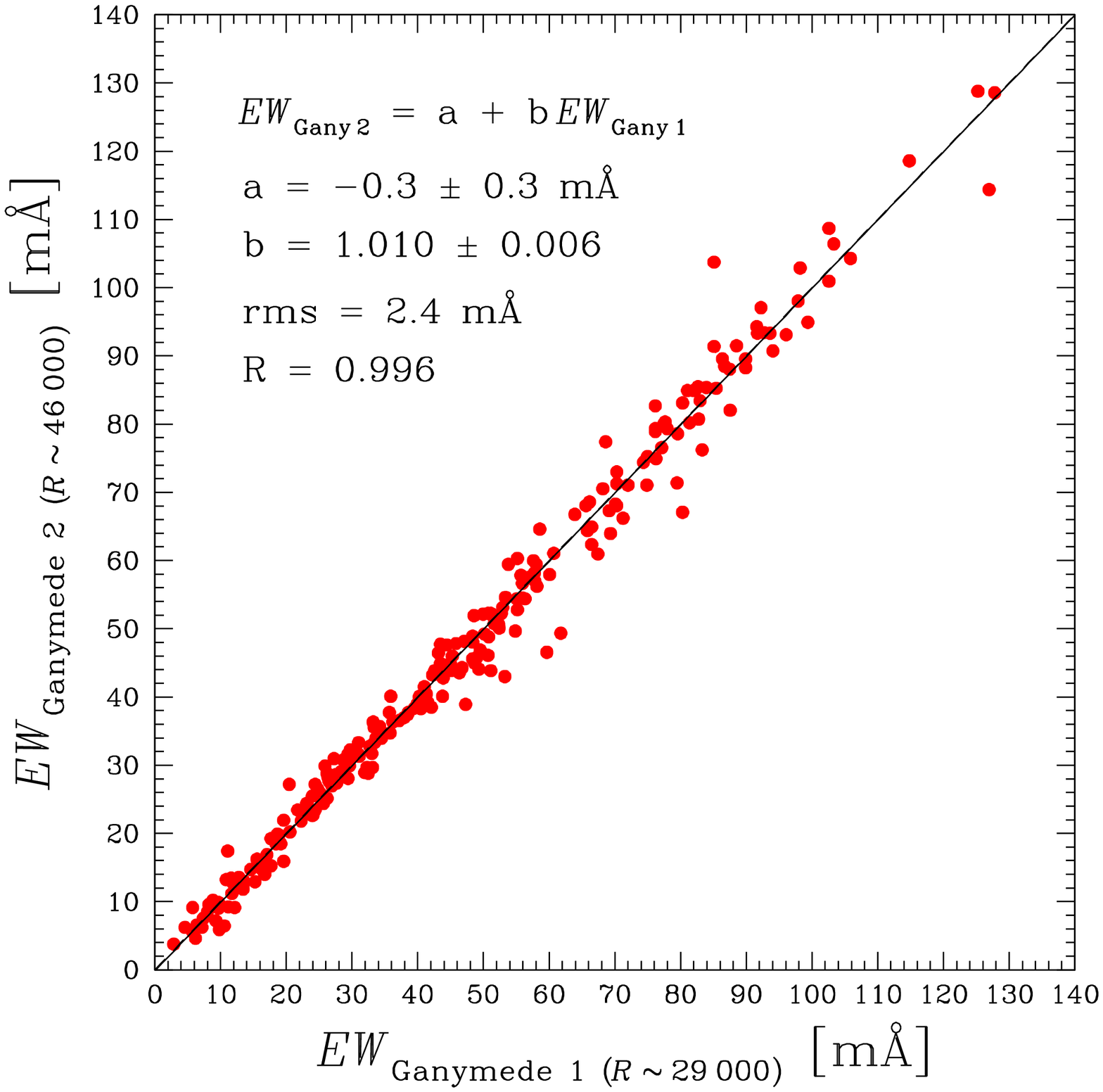}}
\end{minipage}
\caption{{\it Left panel}: comparison of the EWs measured in this work in
	 the Ganymede spectrum (second observation run) by Gaussian function
	 fit to those measured by \citet{Meylanetal1993} in the Solar Flux
	 Atlas by Voigt function fit; filled circles represent 145 lines of
	 several elements; the linear regression (solid line) is the same
	 expressed by Eq.~\ref{fit2}.
	 {\it Right panel}: comparison of the EWs listed in
	 Table~\ref{linelist} of all lines in the Ganymede spectra in common
	 to both runs after the conversions according to Eq.~\ref{fit1} and
	 \ref{fit2}; the coefficients of the linear regression (solid line)
	 are shown.}
\label{Gauss_Voigt}
\end{figure*}

Strong line profiles are better described by Voigt functions than by
Gaussian functions. We show in Fig.~\ref{Gauss_Voigt} (left panel) a
comparison of the $EW$s measured in this work in the Ganymede spectrum of
the second observation run (Gany 2) by Gaussian function fit to those
measured in the Solar Flux Atlas ($R > 500\,000$ and S/N $\sim$3000) by
Voigt function fit \citep{Meylanetal1993}. A similar diagram was obtained
using the Ganymede spectrum of the first run (Gany 1) and the following
relations represent the linear least square regressions fitted to both
diagrams:
\begin{equation}
  \label{fit1}
  EW_{\rm Solar\,Flux\,Atlas} = (0.3 \pm 0.6) + (1.065 \pm 0.012)\,EW_{\rm Gany\,1}
\end{equation}
\begin{equation}
  \label{fit2}
  EW_{\rm Solar\,Flux\,Atlas} = (0.8 \pm 0.5) + (1.075 \pm 0.010)\,EW_{\rm Gany\,2}
\end{equation}
where $EW$ is given in m\AA. The standard deviations and the
cross-correlation coefficients are, respectively, $\sigma$ = 3.4~m\AA\ and
$r$ = 0.991 for Eq.~\ref{fit1}, and $\sigma$ = 2.7~m\AA\ and $r$ = 0.994 for
Eq.~\ref{fit2}. Therefore, to reduce possible systematic uncertainties and
provide direct comparison with other works, all our $EW$s were transformed
to a common system using the regression coefficients of these equations (the
constant terms have no statistical significance within 2$\sigma$). The
regressions were derived in order to have a direct transformation to the
Solar Flux Atlas system. We also did the converse (using $EW_{\rm Ganymede}$
vs. $EW_{\rm Solar\,Flux\,Atlas}$ diagrams) for comparison, and the
resulting regressions are comparable with those of Eq.~\ref{fit1} and
\ref{fit2}, within 1$\sigma$.

The wavelength and lower-level excitation potential ($\chi$) of the atomic
lines used were taken from the Solar Lines Catalogue. The oscillator
strengths ($gf$) were computed using a solar model atmosphere applied to the
$EW$s of Ganymede (converted using Eq.~\ref{fit1} and \ref{fit2}) in order
to provide the standard solar abundances of \citet{GrevesseNoels1993}. The
adopted solar abundances are of course inconsequential in a differential
analysis.

The solar and stellar model atmospheres were computed with a code kindly
supplied by Dr. Monique Spite (Meudon Observatory, Paris) that interpolates
the model-atmosphere grid from \citet{Edvardssonetal1993}. We used an
updated version of the original code from \citet{Spite1967}. The fundamental
atmospheric parameters (effective temperature \teff, metallicity [Fe/H],
surface gravity \logg, and micro-turbulence velocity $\xi$) and the
population ratio of helium and hydrogen atoms ($n_{\rm He}/n_{\rm H}$) are
taken as input. For the Sun, the adopted parameters are \teff = 5777~K,
\logg = 4.44, $\xi$ = 1.3~\kms, $n_{\rm He}/n_{\rm H}$ = 0.1, and
$\log{\epsilon}_{\,\odot}$ = 7.50 (the solar Fe abundance).

The spectral lines used and their parameters are listed in
Table~\ref{linelist}, in which the $EW$s are the raw values (before the
conversion). We do not list the $EW$s of the other stars but they are
available upon request.

The atomic lines of the elements Mg, Sc, V, Mn, Co, and Cu have important
hyperfine structure (HFS). Their $gf$ values, listed in Table~\ref{hfs},
were taken from \citet{Steffen1985} and also revised according to the $EW$s
of Ganymede and the standard solar abundances of \citet{GrevesseNoels1993},
as done for the other elements. For the elements Zn, Sr, Y, Zr, Ce, Ba, and
Nd, for which theoretical HFS exist, either their effects are negligible or
the spectral lines used are too weak to depend on the HFS assumption. For
the elements without HFS data listed in \citet{Steffen1985}, values of
neighbouring multiplets were adopted. The only exception was the Mg line at
$\lambda$5785.285, for which no HSF information was available. Its $gf$
value, listed in Table~\ref{linelist}, was obtained in the same way as those
lines without HFS. This is not a strong line ($EW < 60$ m\AA) so that the
error induced in the Mg abundance determination is not important.
\citet{delPelosoetal2005} have recently shown that for Mn and Co the
differences in their abundances computed using different values of HFS are
not greater than 0.10~dex.

The HFS of Ba, and also its isotopic splitting, are of some importance only
for the line at $\lambda$6496.9, and can be neglected for the lines at
$\lambda$5853.7 and $\lambda$6141.7 \citep[see][]{Korotinetal2011}.
However, for the 25 stars of our sample we have found a good agreement among
the abundances yielded by the three lines, with a mean standard deviation of
0.07 dex. Moreover, a test performed using only $\lambda$5853.7 and
$\lambda$6141.7 indicated that the global behaviour and trends found in the
abundance diagrams, as well as all our conclusions involving the Ba results
would not change if only these two lines were used.

Concerning the fact that we used two spectrographs with different spectral
resolutions, we performed a test in which we degraded the spectrum of
Ganymede and of the metal-rich star HD\,1835, both observed in the second
observation run, to match the resolution of the first run. New values of EWs
were then obtained, and no systematic differences were found when they are
compared with the original measurements. Moreover, a comparison of the EWs of
Ganymede listed in Table~\ref{linelist} and converted according to
Eq.~\ref{fit1} and \ref{fit2} (see Fig.~\ref{Gauss_Voigt}, right panel)
demonstrates that the equivalent widths of the two observation runs were
properly transformed to a common system, reinforcing our assumption of an
homogeneous analysis.


\subsection{Derivation of atmospheric parameters}
\label{atmpar_sec}

In order to determine the fundamental atmospheric parameters we developed a
code that iteratively calculates these parameters for a given star based on
initial input values. The so called excitation effective temperature (\texc)
was calculated through the excitation equilibrium of neutral iron by
removing any dependence in a [\ion{Fe}{I}/H] vs. $\chi$ diagram. The
micro-turbulence velocity was obtained by removing the dependence of
[\ion{Fe}{I}/H] on $EW$, and the ionisation surface gravity (\loggi) was
computed through the ionisation equilibrium between \ion{Fe}{I} and
\ion{Fe}{II}. Finally, the metallicity was yielded by the $EW$ of
\ion{Fe}{I} lines.

The temperature used in our abundance analyses was the excitation effective
temperature, which is a better representation of the temperature
stratification of the line forming layers. In order to compute the stellar
luminosity with more reliability by also accounting for any consequence due
to small LTE departures, we also considered two other temperature
indicators, which are described in Sect.~\ref{phot_temp}.

Table~\ref{atmpar_tab} lists the spectroscopic atmospheric parameters of the
program stars. An estimate of their uncertainties was performed based on the
analysis of HD\,146233 and HD\,26491, which are representative stars in our
sample (from the first and second runs, respectively) with regard to their
parameters and quality of the spectroscopic data. For both stars we have
similar errors and they were obtained as follows:

\begin{table}
\centering
\caption[]{Atmospheric parameters from our spectroscopic analysis. For
           comparison, the evolutionary surface gravity computed as
	   described in Sect.~\ref{evol_par} (Eq.~\ref{log_g}) is also
	   shown.}
\label{atmpar_tab}
\begin{tabular}{l c c c c r@{}l}
\hline\hline\noalign{\smallskip}
Star &
\parbox[c]{0.6cm}{\centering \texc\ {\tiny [K]}} &
\loggi & \logge &
\parbox[c]{1.0cm}{\centering $\xi$ {\tiny [\kms]}} &
\multicolumn{2}{c}{[Fe/H]} \\
\noalign{\smallskip}\hline\noalign{\smallskip}
Sun        & 5777 & 4.44 & 4.44 & 1.30 &    0&.00 \\
HD\,1835   & 5890 & 4.52 & 4.49 & 1.66 &    0&.21 \\
HD\,20807  & 5878 & 4.51 & 4.44 & 1.33 & $-$0&.22 \\
HD\,26491  & 5820 & 4.38 & 4.28 & 1.38 & $-$0&.09 \\
HD\,33021  & 5750 & 4.14 & 4.05 & 1.40 & $-$0&.20 \\
HD\,39587  & 6000 & 4.52 & 4.48 & 1.72 &    0&.00 \\
HD\,43834  & 5630 & 4.47 & 4.44 & 1.23 &    0&.11 \\
HD\,50806  & 5610 & 4.12 & 4.03 & 1.36 &    0&.02 \\
HD\,53705  & 5810 & 4.32 & 4.26 & 1.34 & $-$0&.22 \\
HD\,84117  & 6074 & 4.20 & 4.27 & 1.56 & $-$0&.06 \\
HD\,102365 & 5643 & 4.47 & 4.40 & 1.04 & $-$0&.28 \\
HD\,112164 & 6031 & 4.05 & 3.87 & 1.79 &    0&.32 \\
HD\,114613 & 5706 & 3.97 & 3.87 & 1.55 &    0&.15 \\
HD\,115383 & 6126 & 4.43 & 4.26 & 1.61 &    0&.23 \\
HD\,115617 & 5587 & 4.41 & 4.42 & 1.22 &    0&.00 \\
HD\,117176 & 5587 & 4.13 & 3.91 & 1.36 & $-$0&.04 \\
HD\,128620 & 5857 & 4.44 & 4.31 & 1.45 &    0&.23 \\
HD\,141004 & 5926 & 4.28 & 4.18 & 1.49 &    0&.03 \\
HD\,146233 & 5817 & 4.45 & 4.42 & 1.32 &    0&.05 \\
HD\,147513 & 5891 & 4.63 & 4.48 & 1.41 &    0&.04 \\
HD\,160691 & 5777 & 4.32 & 4.19 & 1.40 &    0&.27 \\
HD\,177565 & 5630 & 4.42 & 4.43 & 1.24 &    0&.08 \\
HD\,181321 & 5810 & 4.34 & 4.56 & 2.30 & $-$0&.06 \\
HD\,188376 & 5514 & 3.71 & 3.61 & 1.55 &    0&.00 \\
HD\,189567 & 5700 & 4.44 & 4.32 & 1.22 & $-$0&.27 \\
HD\,196761 & 5410 & 4.44 & 4.49 & 1.08 & $-$0&.32 \\
\hline
\end{tabular}
\end{table}

\begin{itemize}

\item[\it i)]
The uncertainty in \texc\ is related to the standard error of the angular
coefficient of the linear regression fitted to the [\ion{Fe}{I}/H] vs.
$\chi$ diagram. The temperature is changed until this coefficient is of the
same order of its error. The difference between the best value and the
previous one from the last iteration provides the uncertainty
$\sigma$(\texc) = 30~K;

\item[\it ii)] The uncertainty in metallicity is the standard deviation of
the abundance yielded by individual \ion{Fe}{I} lines, which is
$\sigma$([Fe/H]) = 0.05~dex;

\item[\it iii)] To estimate the uncertainty in \loggi, its value is changed
until the difference between the averaged abundance yielded by \ion{Fe}{I}
and \ion{Fe}{II} lines is of the order of their internal errors
($\sim$0.05~dex), which led to an uncertainty $\sigma$(\loggi) = 0.13~dex;

\item[\it iv)] The uncertainty in the micro-turbulence velocity is estimated
regarding the [\ion{Fe}{I}/H] vs. $EW$ diagram. The $\xi$ value is modified
until the angular coefficient of the regression is of the same order of its
error. An uncertainty $\sigma$($\xi$) = 0.04~\kms\ was found for this
parameter.

\end{itemize}

The spectroscopic atmospheric parameters were used to compute the model
atmospheres, which in turn are required in the abundance determination. In
our analysis, we used the model-atmosphere grid derived by
\citet{Edvardssonetal1993} for stars with effective temperatures from 5250
to 6000~K, surface gravity from 2.5 to 5.0~dex, and metallicity from $-$2.3
to +0.3~dex (with small extrapolations when needed). These are 1D,
plane-parallel, constant flux, line-blanketed, and LTE models computed over
45 layers.

The model atmospheres are, essentially, subject to errors in the atmospheric
parameters, in the LTE simplifications, and in the thermal homogeneity
assumption. However, the effects of non-LTE and thermal inhomogeneities are
hopefully minor for the elements and the stellar types studied here, being
more important for low metallicity and low surface gravity stars
\citep{Edvardssonetal1993,Asplund2005}. Possible errors induced by such
simplified assumptions are dominated by other sources of uncertainties. In
addition, in a differential analysis, the errors in the stellar atmospheric
structure are of second order.

We also investigated what would be the effects on the derived abundances if
another model-atmosphere grid were used. We compared Edvardsson and Kurucz
models and, using the same equivalent widths, gf values, and atmospheric
parameters obtained from a solar spectrum, we found that the differences in
abundance [X/Fe] for most of the elements are of the order of 0.03~dex or
smaller, achieving a maximum of 0.07~dex. We note that these values are for
the Sun and represent the differences between the two sets of model
atmospheres. The effects of these differences when computing the stellar
abundances relative to the Sun are minimised in a differential approach.


\subsection{Abundance determination and their uncertainties}
\label{abund}

The abundance of the elements studied were determined using an adapted
version of a code also supplied by Dr. Monique Spite. The code takes into
account the solar $gf$ values and the stellar model atmospheres (computed
using the atmospheric parameters of each star) to calculate the abundances
that fits the equivalent widths measured in the spectra (transformed
according to the procedure described in Sect.~\ref{ew}). The results of this
abundance determination are presented and discussed in Sect.~\ref{results}.

\begin{table}
\centering
\caption[]{A comparison between the estimated abundance errors
           $\sigma_{\rm est}$ (see Sect.~\ref{abund} and \ref{synth}) and
	   the dispersions of these abundances around the mean
	   $\sigma_{\rm disp}$ for elements with N = 5 or more lines
	   measured. For each element of each observation run the larger
	   value was adopted to represent the errors $\sigma$([X/Fe]).}
\label{ab_err_comp}
\begin{tabular}{l c c c c c c}
\hline\hline\noalign{\smallskip}
\multirow{2}{*}{[X/Fe]} &
\multicolumn{3}{c}{HD\,146233} &
\multicolumn{3}{c}{HD\,26491} \\[0.1cm]
& $\sigma_{\rm est}$ & $\sigma_{\rm disp}$ & N &
$\sigma_{\rm est}$ & $\sigma_{\rm disp}$ & N \\[-0.1cm]
\noalign{\smallskip}\hline\noalign{\smallskip}
C  & 0.07 & --   & 2  & 0.07 & --   & 2  \\
Na & 0.03 & --   & 2  & 0.06 & --   & 2  \\
Mg & 0.03 & --   & 4  & 0.06 & --   & 4  \\
Si & 0.05 & 0.06 & 11 & 0.06 & 0.03 & 17 \\
Ca & 0.03 & 0.04 & 6  & 0.05 & 0.05 & 13 \\
Sc & 0.06 & 0.03 & 6  & 0.09 & 0.03 & 13 \\
Ti & 0.07 & 0.04 & 24 & 0.10 & 0.04 & 38 \\
V  & 0.07 & 0.07 & 8  & 0.11 & 0.04 & 11 \\
Cr & 0.06 & 0.04 & 14 & 0.07 & 0.04 & 29 \\
Mn & 0.04 & 0.05 & 8  & 0.06 & 0.04 & 11 \\
Co & 0.07 & 0.05 & 9  & 0.11 & 0.04 & 12 \\
Ni & 0.04 & 0.03 & 23 & 0.07 & 0.04 & 26 \\
Cu & 0.04 & --   & 3  & 0.07 & --   & 3  \\
Zn & 0.05 & --   & 1  & 0.06 & --   & 1  \\
Sr & 0.06 & --   & 1  & 0.07 & --   & 1  \\
Y  & 0.07 & 0.05 & 5  & 0.09 & 0.05 & 5  \\
Zr & 0.09 & --   & 3  & 0.08 & --   & 3  \\
Ba & 0.06 & --   & 3  & 0.08 & --   & 3  \\
Ce & 0.07 & 0.07 & 5  & 0.12 & --   & 3  \\
Nd & 0.16 & --   & 2  & 0.12 & --   & 2  \\
Sm & 0.09 & --   & 1  & 0.15 & --   & 1  \\
\hline
\end{tabular}
\end{table}

The main sources of uncertainties in the abundance determination come from
the errors in the $EW$s (the most important), the $gf$ values, the
atmospheric parameters, and the adopted model atmospheres (these two latter
are discussed in Sect.~\ref{atmpar_sec}).

\begin{table*}
\centering
\caption[]{Colour indices and photometric effective temperatures (given in
           K). The values of (\bv) and (\btvt) were taken from the Hipparcos
	   Catalogue . The H$\alpha$ effective temperatures from
	   \citet{LyraPortodeMello2005} are also listed. For the star
	   HD\,128620 ($\alpha$\,Cen\,A), the photometric determination of
	   \teff\ was not performed (see discussion in
	   Sect.~\ref{phot_temp}). }
\label{colour}
\begin{tabular}{l c c r@{}l r@{}l c c c c c c}
\hline\hline\noalign{\smallskip}
Star &
\bv &
\btvt &
\multicolumn{2}{c}{\by$^{\,1}$} &
\multicolumn{2}{c}{$\beta^{\,1}$} &
\parbox[c]{0.9cm}{\centering $T_{\rm eff}^{\rm phot}$ {\tiny (\bv)}} &
\parbox[c]{1.1cm}{\centering $T_{\rm eff}^{\rm phot}$ {\tiny (\btvt)}} &
\parbox[c]{0.8cm}{\centering $T_{\rm eff}^{\rm phot}$ {\tiny (\by)}} &
\parbox[c]{0.8cm}{\centering $T_{\rm eff}^{\rm phot}$ {\tiny ($\beta$)}} &
\parbox[c]{0.8cm}{\centering $T_{\rm eff}^{\rm phot}$ {\tiny (mean)}} &
\parbox[c]{0.8cm}{\centering \tHa} \\
\noalign{\smallskip}\hline\noalign{\smallskip}
HD\,1835   & 0.659 & 0.758 & 0&.420  & 2&.606  & 5822 & 5804 & 5796 & 5908 & 5826 & 5846 \\
HD\,20807  & 0.600 &	-- & 0&.380  & 2&.592  & 5878 &   -- & 5956 & 5745 & 5876 & 5860 \\
HD\,26491  & 0.636 & 0.697 & 0&.404  & 2&.587  & 5803 & 5844 & 5827 & 5685 & 5797 & 5774 \\
HD\,33021  & 0.625 & 0.682 & 0&.402  & 2&.590  & 5805 & 5844 & 5814 & 5721 & 5800 & 5823 \\
HD\,39587  & 0.594 & 0.659 & 0&.376  & 2&.599  & 5955 & 5965 & 6031 & 5827 & 5956 & 5966 \\ 
HD\,43834  & 0.720 & 0.829 & 0&.442  & 2&.601  & 5611 & 5600 & 5629 & 5850 & 5662 & 5614 \\
HD\,50806  & 0.708 & 0.800 & 0&.437  & \md{--} & 5618 & 5634 & 5639 &   -- & 5631 & 5636 \\
HD\,53705  & 0.624 & 0.685 & 0&.396  & 2&.595  & 5803 & 5830 & 5850 & 5780 & 5820 & 5821 \\
HD\,84117  & 0.530 & 0.581 & 0&.339  & 2&.622  & 6134 & 6136 & 6260 & 6089 & 6166 & 6188 \\
HD\,102365 & 0.660 &	-- & 0&.408  & 2&.588  & 5673 &   -- & 5755 & 5697 & 5714 & 5644 \\
HD\,112164 & 0.640 & 0.696 & 0&.392  & 2&.632  & 5909 & 5984 & 6000 & 6200 & 6015 & 5965 \\
HD\,114613 & 0.700 & 0.796 & 0&.441  & \md{--} & 5683 & 5693 & 5646 &   -- & 5671 & 5732 \\
HD\,115383 & 0.590 & 0.644 & 0&.371  & 2&.615  & 6029 & 6073 & 6114 & 6011 & 6064 & 5952 \\
HD\,115617 & 0.710 &	-- & 0&.434  & 2&.582  & 5606 &   -- & 5653 & 5625 & 5631 & 5562 \\
HD\,117176 & 0.710 & 0.804 & 0&.445  & 2&.576  & 5593 & 5601 & 5571 & 5552 & 5580 & 5493 \\
HD\,128620 &	-- &	-- & \md{--} & \md{--} &   -- &   -- &   -- &   -- &   -- & 5820 \\
HD\,141004 & 0.600 & 0.672 & 0&.382  & 2&.606  & 5946 & 5944 & 5999 & 5908 & 5955 & 5869 \\
HD\,146233 & 0.650 & 0.736 & 0&.398  & 2&.596  & 5801 & 5798 & 5899 & 5792 & 5830 & 5790 \\
HD\,147513 & 0.620 & 0.703 & 0&.391  & 2&.609  & 5888 & 5873 & 5942 & 5942 & 5913 & 5840 \\
HD\,160691 & 0.700 & 0.786 & 0&.432  & \md{--} & 5721 & 5762 & 5734 &   -- & 5739 & 5678 \\
HD\,177565 & 0.705 & 0.803 & 0&.436  & 2&.584  & 5646 & 5650 & 5660 & 5649 & 5652 & 5673 \\
HD\,181321 & 0.628 & 0.694 & 0&.396  & \md{--} & 5836 & 5861 & 5887 &   -- & 5864 & 5845 \\
HD\,188376 & 0.750 &	-- & 0&.458  & \md{--} & 5485 &   -- & 5497 &   -- & 5492 & 5436 \\
HD\,189567 & 0.648 & 0.718 & 0&.410  & 2&.583  & 5713 & 5730 & 5745 & 5637 & 5712 & 5697 \\
HD\,196761 & 0.719 & 0.828 & 0&.441  & \md{--} & 5474 & 5431 & 5525 &   -- & 5482 & 5544 \\
\hline
\end{tabular}
\begin{flushleft}
{$^1$}
\citet{Crawford1975},
\citet{FabregatReglero1990},
\citet{Ferroetal1990},
\citet{GronbechOlsen1976,GronbechOlsen1977},
\citet{Olsen1977,Olsen1983,Olsen1993,Olsen1994a,Olsen1994b}, \\
\citet{OlsenPerry1984},
\citet{Perryetal1987},
\citet{RegleroFabregat1991},
\citet{SchusterNissen1988},
\citet{Twarog1980}.
\end{flushleft}
\end{table*}

The uncertainties in the $EW$s were estimated as follows: by plotting the
diagram $EW_{\rm HD\,26491}$ vs. $EW_{\rm Ganymede}$ and computing the
standard deviation of the linear regression, we obtained $\sigma$ =
2.9~m\AA. The solar $EW$s were measured in the surrogate spectrum of the Sun
collected under the same circumstances as for the program stars. Therefore,
we assumed that $\sigma$ is a quadratic sum of the errors in $EW$ of both
objects and that they are similar to each other. Thus, for the star
HD\,26491 the value of $\sigma$($EW$) is $\sigma/\sqrt{2}$ = 2.1~m\AA.
Similarly, for the star HD\,146233 we obtained $\sigma$($EW$) = 1.7~m\AA.
These values were adopted to represent the uncertainties in $EW$ of the two
observation runs. Because the solar $gf$ values were computed to reproduce
solar the equivalent widths, the errors in $EW$ contribute twice to the
total uncertainty, with approximately the same magnitude.

Each one in turn, $EW$, \texc, [Fe/H], $\xi$, and \loggi\ are changed by
1$\sigma$ in the sense of increasing the abundance ratios and new abundances
are computed for each element. The differences between new and previous
abundance values provide the errors induced by each parameter and a
quadratic sum of these errors yields the total uncertainty in the elemental
abundance ratios. The estimated errors ($\sigma_{\rm est}$) are listed in
Table~\ref{ab_err_comp} for both HD\,146233 and HD\,26491, and they are
compared to the dispersions around the mean ($\sigma_{\rm disp}$) for
elements with at least five lines measured in the spectra. For these
elements, the larger values were adopted to represent the errors
$\sigma$([X/Fe]) in each observation run. Otherwise, $\sigma_{\rm est}$ was
adopted.


\subsection{Photometric and H$\alpha$ effective temperatures}
\label{phot_temp}

The effective temperature of the sample stars were also obtained using some
photometric calibrations, providing the photometric effective temperature
(\tphot). These calibrations, derived by \citet{PortodeMelloetal2011} for
the (\bv), ($B_{\rm T}-V_{\rm T}$), (\by), and $\beta$ colour indices, are
given by the following equations:
\begin{equation}
\label{cal_bv}
T_{{\rm eff}\,(B-V)}^{\rm phot} = 7747 - 3016\,(B-V)\,\big\{1-0.15{\rm\,[Fe/H]}\big\}
\end{equation}
\begin{equation}
\label{cal_btvt}
T_{{\rm eff}\,(B_{\rm T}-V_{\rm T})}^{\rm phot} = 7551 - 2406\,(B_{\rm T}-V_{\rm T})\,\big\{1-0.2{\rm\,[Fe/H]}\big\}
\end{equation}
\begin{equation}
\label{cal_by}
T_{{\rm eff}\,(b-y)}^{\rm phot} = 8481 - 6516\,(b-y)\,\big\{1-0.09{\rm\,[Fe/H]}\big\}
\end{equation}
\begin{equation}
\label{cal_beta}
T_{{\rm eff}\,(\beta)}^{\rm phot} = 11654\,\big\{\,\beta - 2.349\big\}^{0.5}
\end{equation}
for \teff\ given in K. The standard deviations of these calibrations are
$\sigma$ = 65, 64, 55, and 70~K, respectively.

The (\bv) and (\btvt) colour indices of our stars were taken from the
Hipparcos Catalogue, and (\by) and $\beta$ from the literature (see
Table~\ref{colour}), when available. For the star HD\,33021, the $\beta$
values adopted are only from \citet{Perryetal1987} because these authors
made 41 measurements of this index. For the star HD\,50806, only one
reference for the $\beta$ index was found, and the effective temperature
from this index strongly disagrees with that obtained from the other colours
and we thus discarded it.

Table~\ref{colour} lists the colour indices used and the photometric
effective temperature (\tphot) obtained, which is a mean of the temperatures
computed using the four calibrations, weighted by their variances. The
references for (\by) and $\beta$ are also listed. The values of (\by) from
\citet{GronbechOlsen1976}, \citet{Olsen1983}, \citet{Twarog1980}, and
\citet{SchusterNissen1988} were converted to the \citet{Olsen1993} system
according to equations provided by the latter author.

\begin{figure*}
\centering
\begin{minipage}[b]{0.245\textwidth}
\centering
\resizebox{\hsize}{!}{\includegraphics{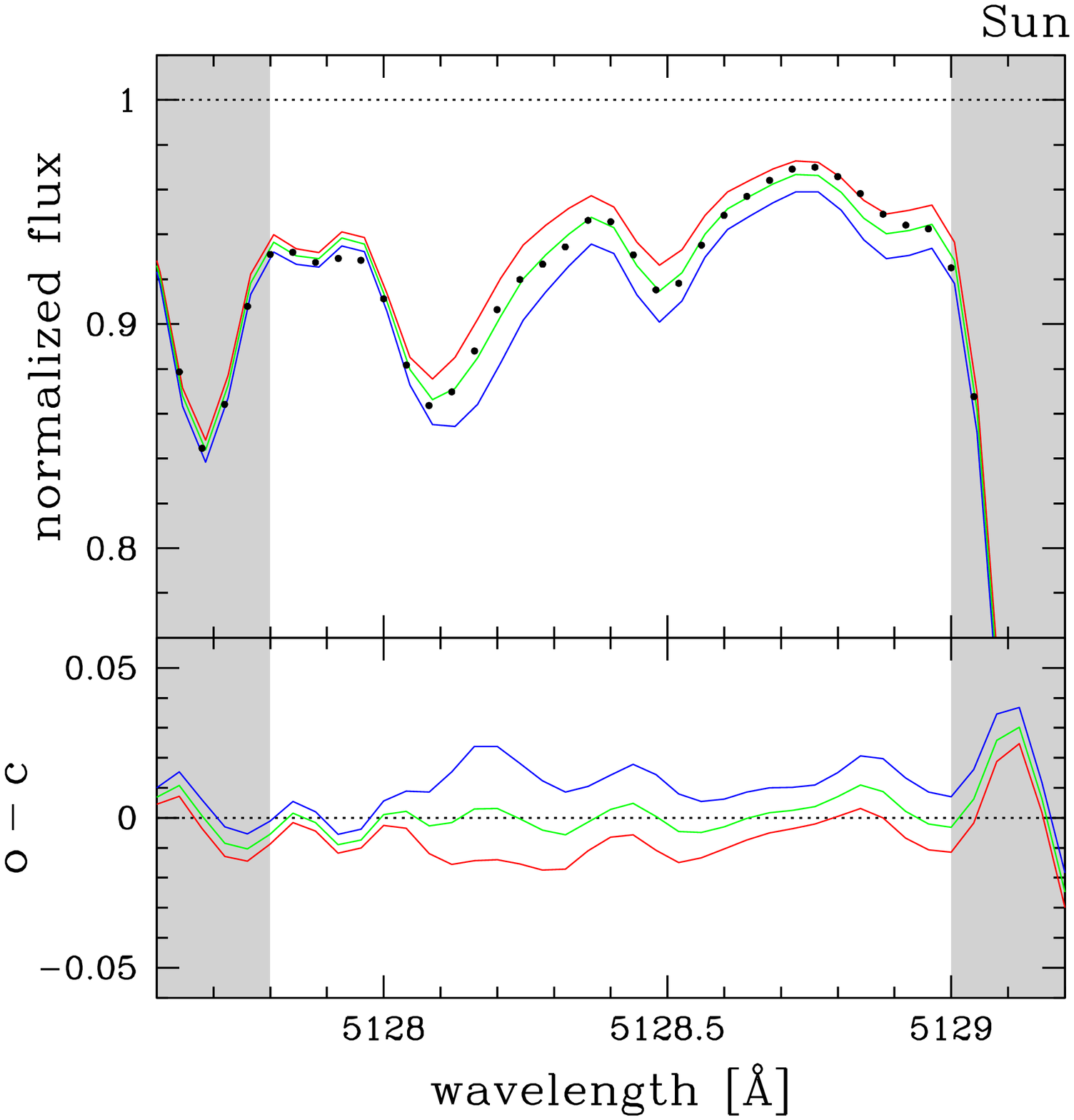}}
\end{minipage}
\begin{minipage}[b]{0.245\textwidth}
\centering
\resizebox{\hsize}{!}{\includegraphics{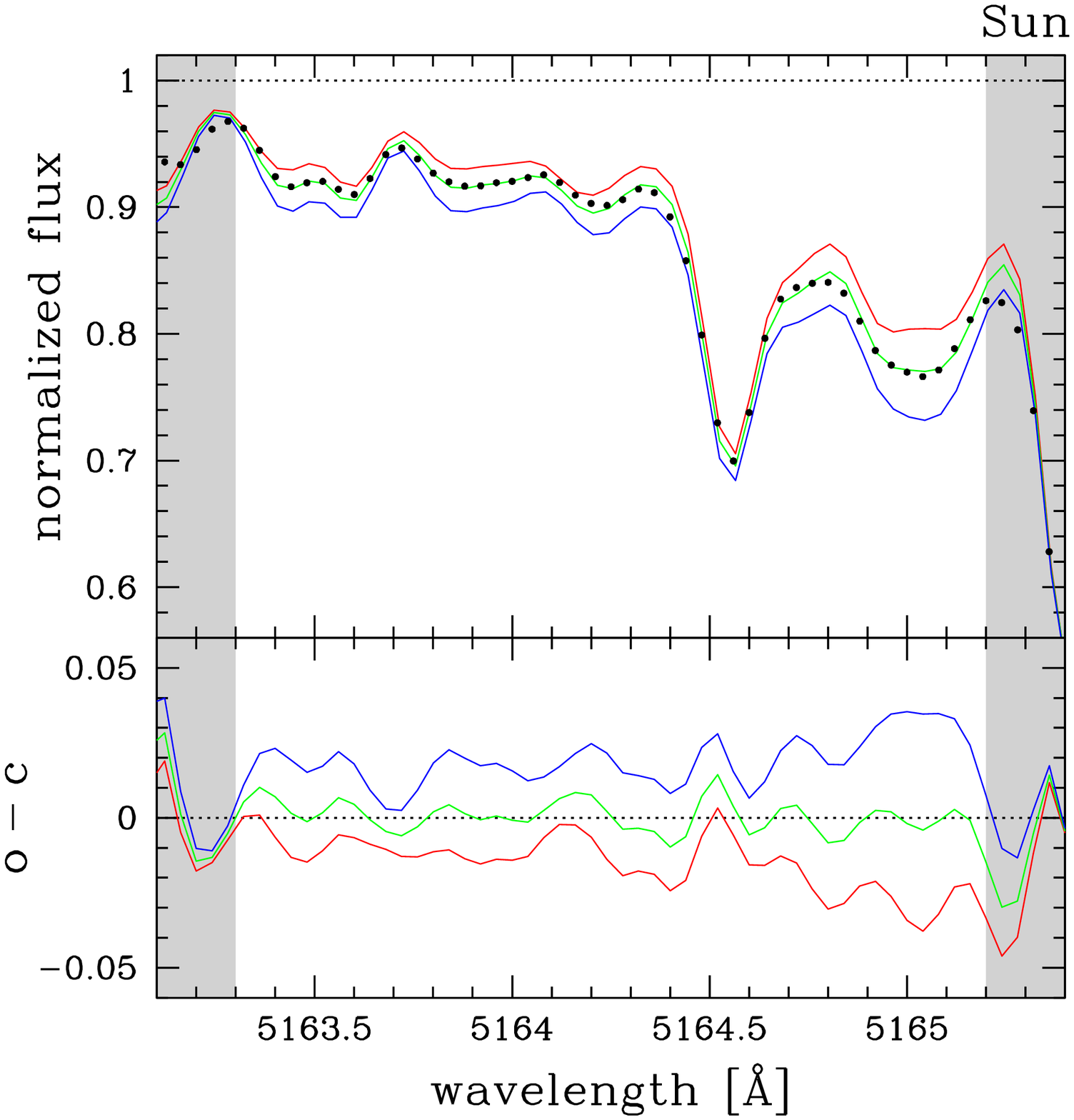}}
\end{minipage}
\begin{minipage}[b]{0.245\textwidth}
\centering
\resizebox{\hsize}{!}{\includegraphics{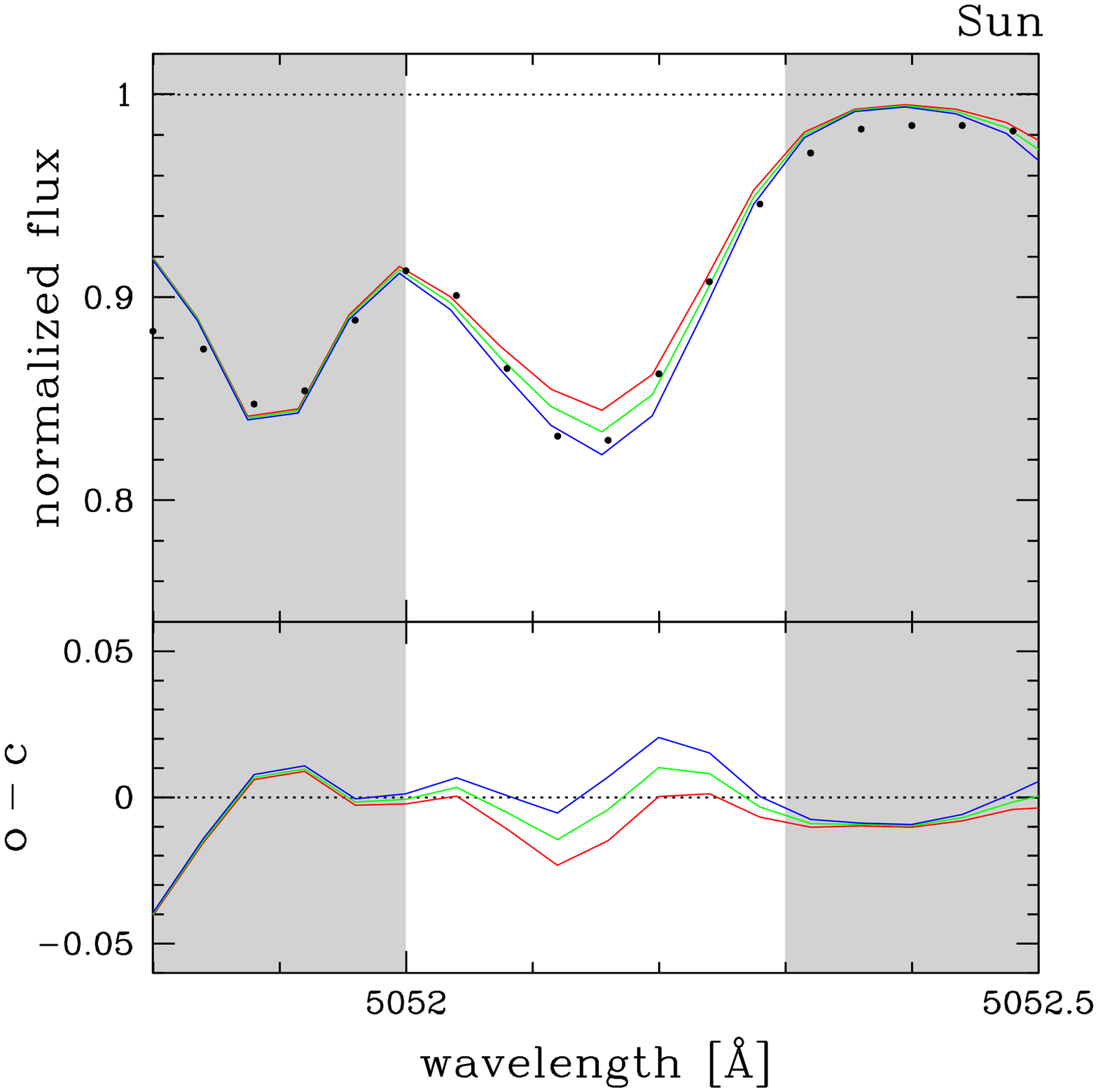}}
\end{minipage}
\begin{minipage}[b]{0.245\textwidth}
\centering
\resizebox{\hsize}{!}{\includegraphics{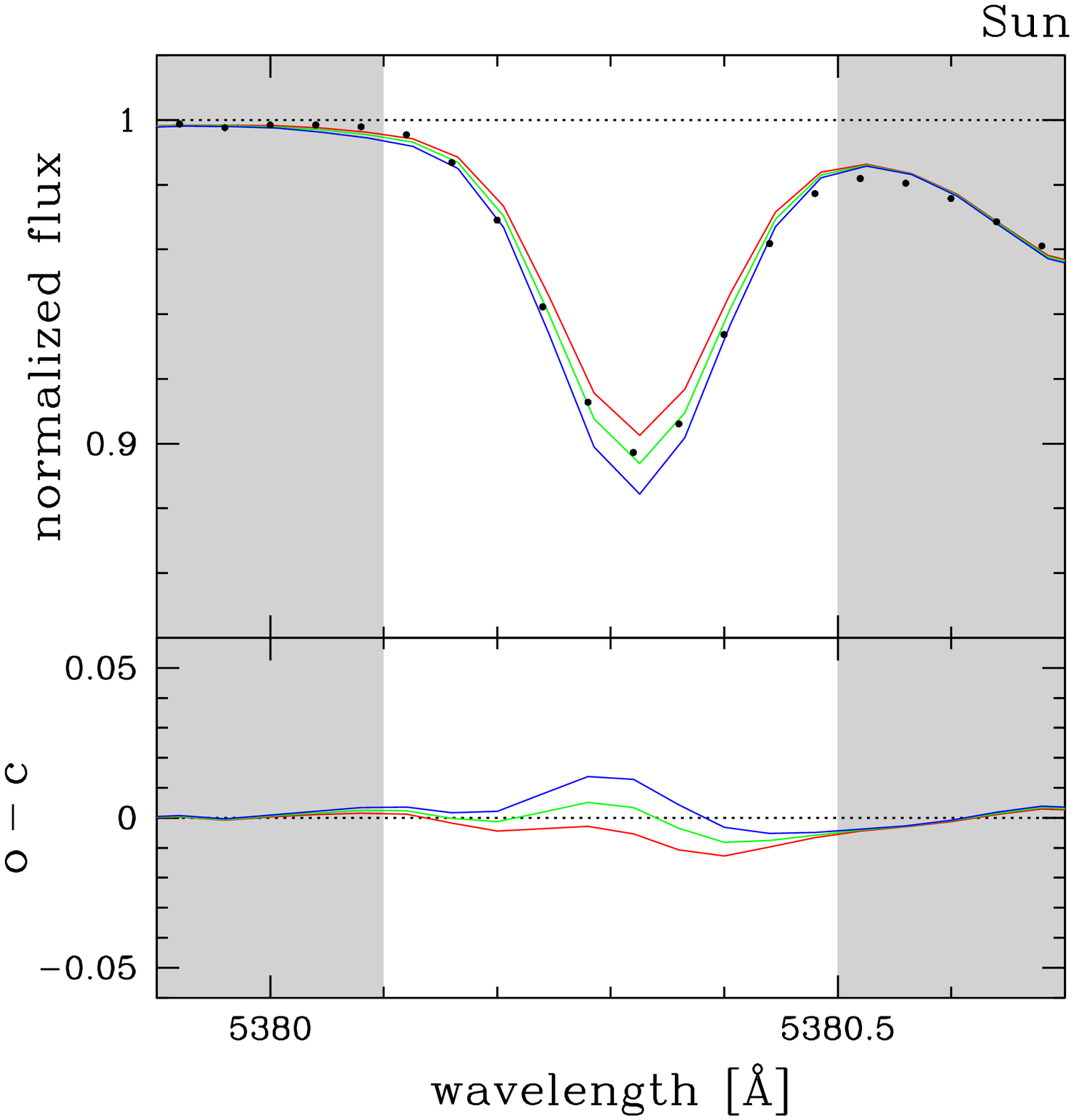}}
\end{minipage} \\
\begin{minipage}[b]{0.245\textwidth}
\centering
\resizebox{\hsize}{!}{\includegraphics{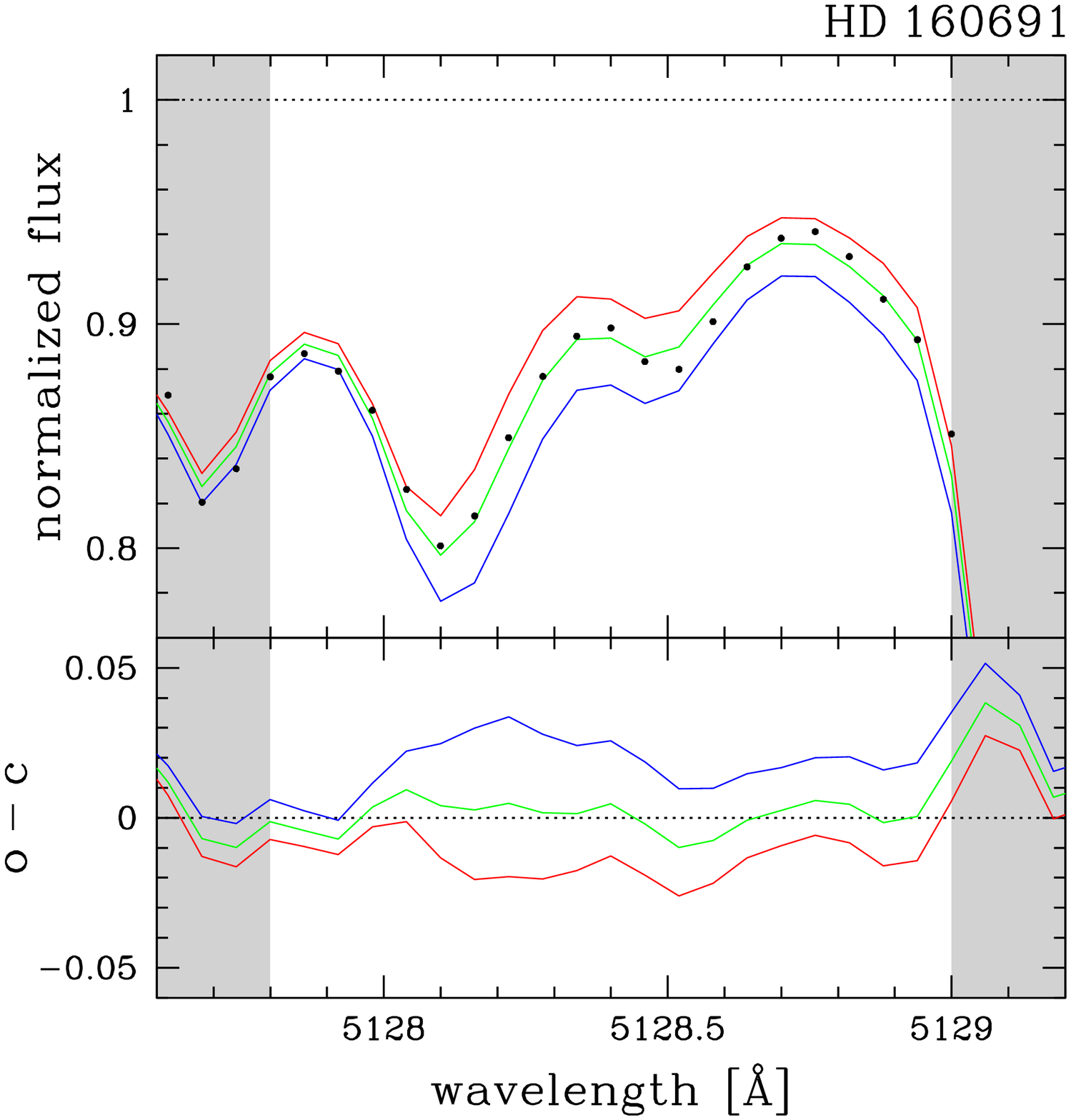}}
\end{minipage}
\begin{minipage}[b]{0.245\textwidth}
\centering
\resizebox{\hsize}{!}{\includegraphics{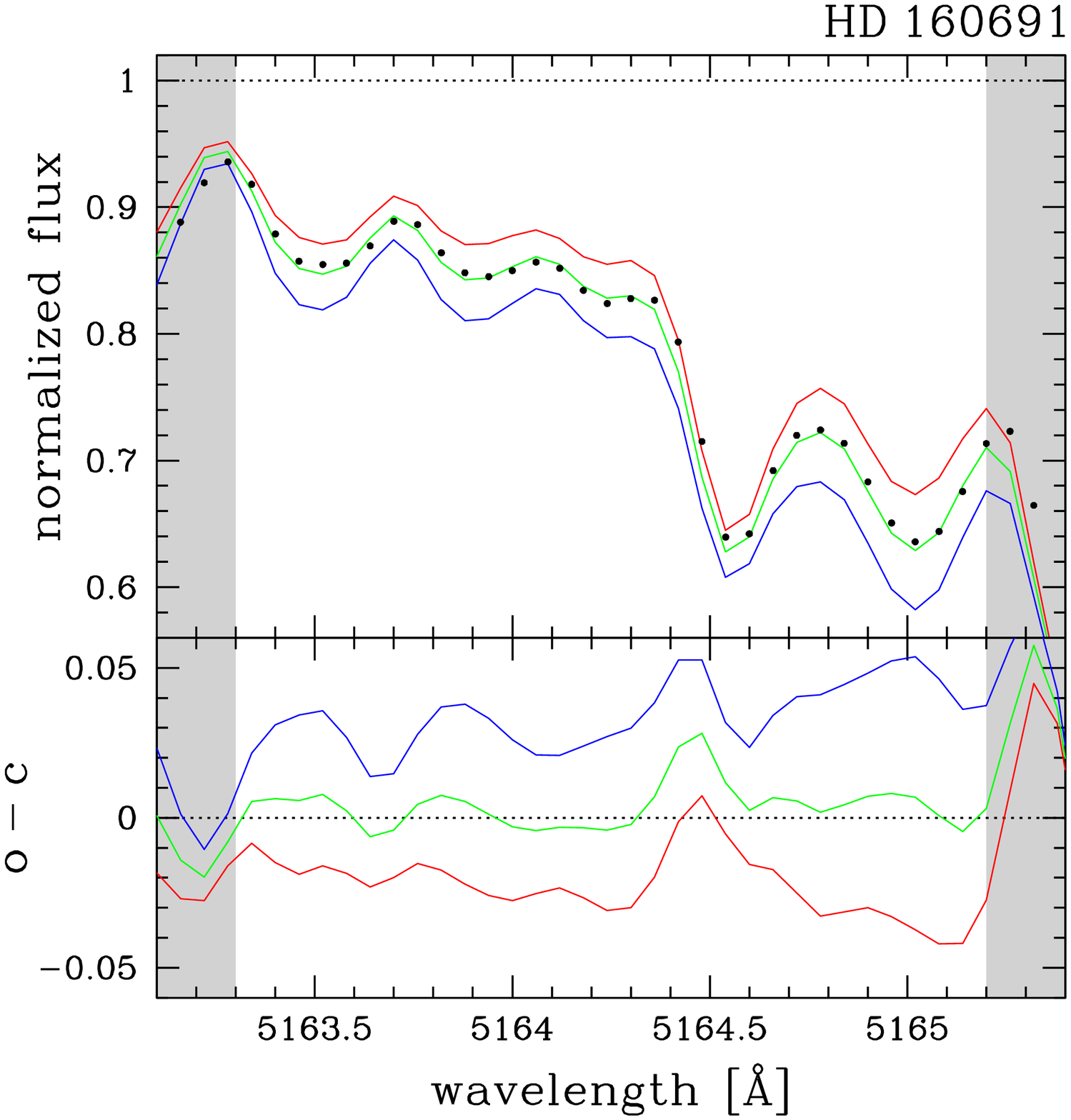}}
\end{minipage}
\begin{minipage}[b]{0.245\textwidth}
\centering
\resizebox{\hsize}{!}{\includegraphics{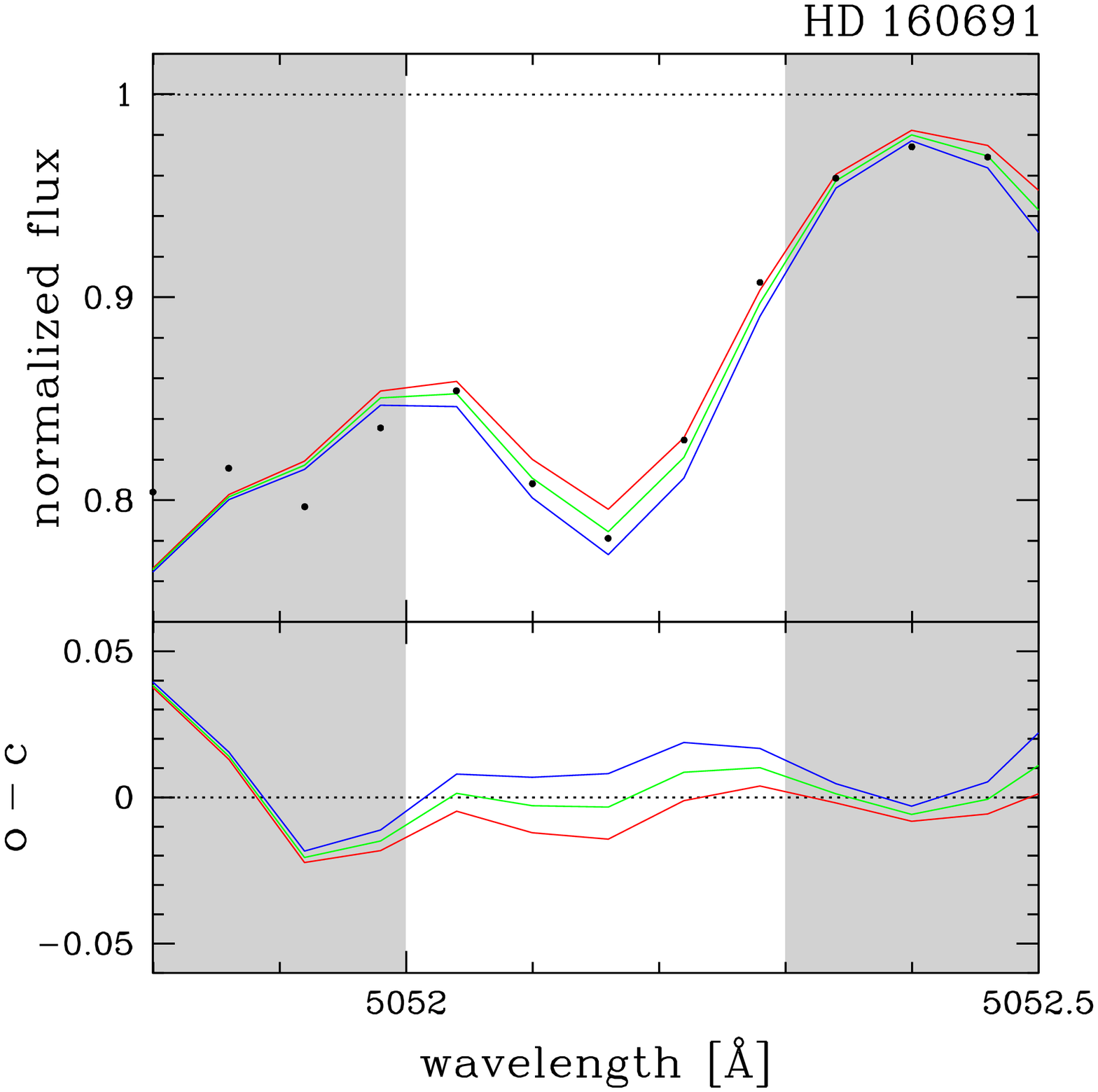}}
\end{minipage}
\begin{minipage}[b]{0.245\textwidth}
\centering
\resizebox{\hsize}{!}{\includegraphics{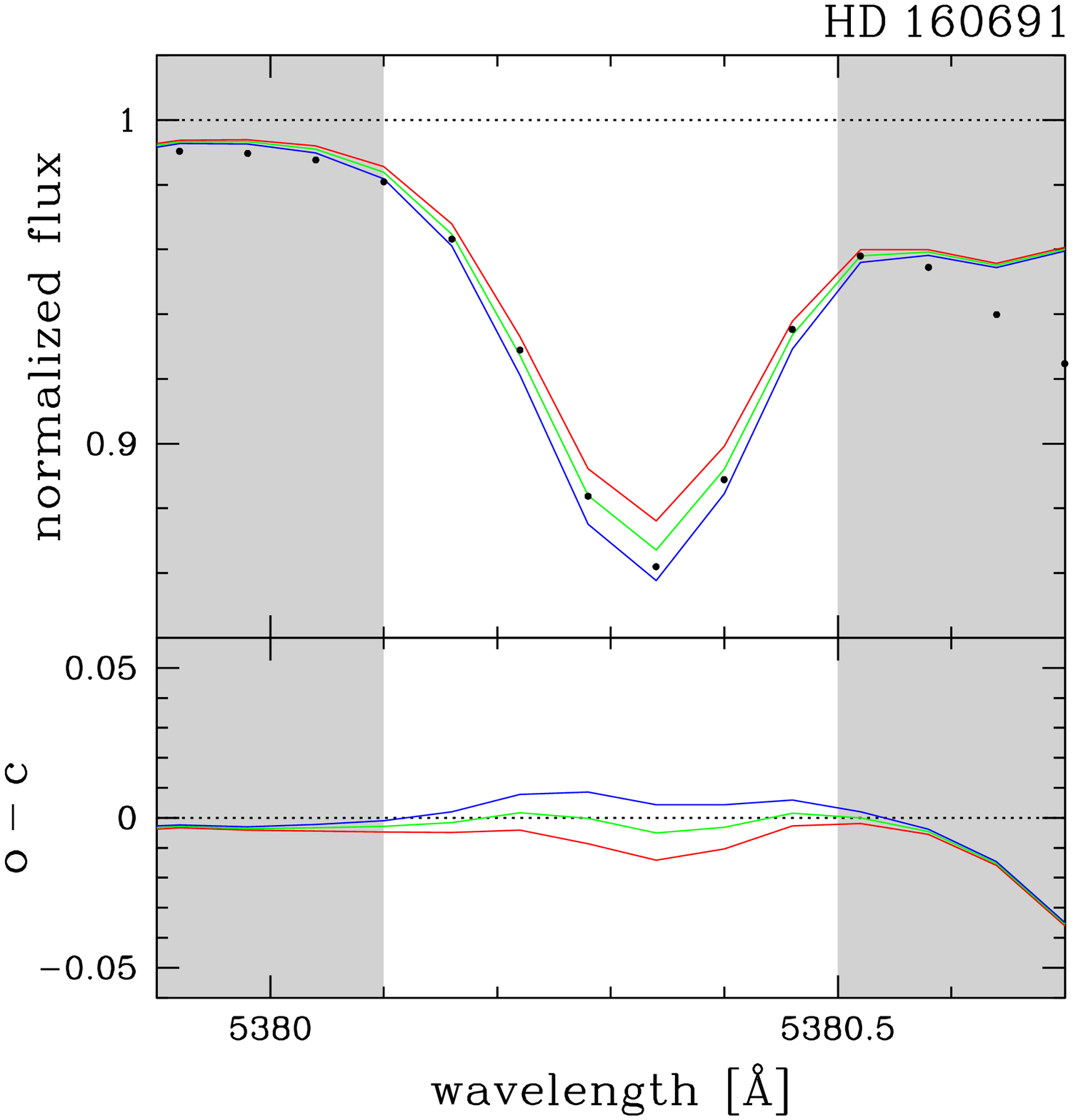}}
\end{minipage}
\caption{Spectral synthesis of the ${\rm C}_2$ molecular band regions
         ($\lambda$5128 and $\lambda$5165) and of the C atomic lines
	 ($\lambda$5052.2 and $\lambda$5380.3) for Ganymede of the second
	 observation run (top panels) and for one metal-rich star of the
	 first run, HD\,160691 (bottom panels). The solid lines represent
	 three models with different values of [C/Fe]: $-$0.05 (red), 0.0
	 (green), and +0.05~dex (blue). The differences between observed
	 (dots) and computed spectra (o$-$c) are also shown in the bottom of
	 each panel.}
\label{synt_gany}
\end{figure*}

The star HD\,128620 is the primary component ($V$ = $-$0.01) of the
$\alpha$\,Cen triple system. The \teff\ determination for very bright stars
using photometric colours is normally considered risky due to systematic
effects that may affect the results (such as non-linearity and detector dead
time) and also, in the case of this system, due to a possible contamination
by the companion. For this reason, we preferred do not include this star in
our \teff\ estimates based on the photometric indices. Nevertheless, our
spectroscopic determination for $\alpha$\,Cen\,A, \texc\ = 5857 $\pm$ 30~K,
is  in good agreement with the photometric determination performed by
\citet{PortodeMelloetal2008}, \teff\ = 5794 $\pm$ 34~K.

Concerning the uncertainties in \tphot, on the one hand, the internal error
of the weighted mean, computed using the standard deviations in the four
photometric calibrations, is 31~K. On the other hand, the mean value of the
standard deviations around \tphot\  (i.e., the dispersion of the four values
of temperature around the weighted mean) is 41~K. Therefore, in this work we
adopted $\sigma$(\tphot) = 40~K as the internal uncertainty in our
photometric effective temperatures. This uncertainty intrinsically takes
into account the errors in the colour indices themselves.

The stellar effective temperature can also be estimated by modelling the
wing profile of the H$\alpha$ line, which is very sensitive to changes in
this parameter. \citet{LyraPortodeMello2005} applied this method to solar
neighbourhood stars, and values of \tHa\ for our sample were used as a third
\teff\ indicator, with an uncertainty $\sigma$(\tHa) = 50~K.


\subsection{Carbon abundance from spectral synthesis}
\label{synth}

The carbon abundance was derived using the spectral synthesis method applied
to molecular lines of electronic-vibrational band heads of the ${\rm C}_2$
Swan System at $\lambda$5128 and $\lambda$5165, and also to C atomic lines
at $\lambda$5052.2 and $\lambda$5380.3. To reproduce the atomic and
molecular absorption lines in the observed spectra of the sample stars, the
MOOG spectral synthesis
code\footnote{http://www.as.utexas.edu/$\sim$chris/moog.html}, developed by
Chris Sneden (University of Texas, USA), was used. The synthetic spectra
were computed in steps of 0.02 {\AA}, also taking into account the continuum
opacity contribution in ranges of 0.5~{\AA}. The Uns\"old approximation
multiplied by 6.3 was adopted in the calculations of the line damping
parameters.

The model atmospheres are the same used in the spectroscopic analysis. They
also include the micro-turbulence velocity and the elemental abundances of
each star, both assumed to be constant in all layers. For any element X for
which no abundance was determined in this work, we adopted the metallicity
of the respective star to set the [X/H] ratio.

The atomic and molecular line parameters used to compute the synthetic
spectra are: the central wavelength, the $gf$ values, the lower-level
excitation potential, and the constant of dissociation energy $D_0$ (only
for molecular features). Atomic and molecular data were taken, respectively,
from the {\it Vienna Atomic Line Database} -- VALD \citep{Kupkaetal1999,
Kupkaetal2000,Piskunovetal1995,Ryabchikovaetal1997} and from
\citet{Kurucz1992}. In addition to \C2, the spectral regions studied also
include MgH molecular features that may contribute to the continuum
formation. The oscillator strengths of \C2\ and MgH lines were revised
according to the normalisation of the H\"onl-London factors
\citep{WhitingNicholls1974}.

To account for the spectral line broadening, the synthetic spectra were
computed by means of the convolution with three input parameters:
$i)$ the spectroscopic instrumental broadening;
$ii)$ the limb darkening of the stellar disc; and
$iii)$ a composite of velocity fields, such as rotation velocity and
macro-turbulence broadening, named \Vbroad.
The instrumental broadening was estimated by means of the FWHM of thorium
lines present in Th-Ar spectra observed at the CTIO. The linear
limb-darkening coefficient (on average $u \sim 0.7$ for all the sample
stars) was individually estimated by interpolation of \texc\ and \loggi\ in
Table~1 of \citet{DiazCordovesetal1995}. As a first estimate of \Vbroad, the
projected rotation velocity (\vsini) of the stars was used, which was
computed based on the profile of four isolated \ion{Fe}{I} lines
($\lambda$5852.2, $\lambda$5855.1, $\lambda$5856.1, and $\lambda$5859.6) in
the spectra. Small corrections in \Vbroad\ were applied when needed
according to an eye-trained inspection of the synthetic spectra. The final
values are listed in Table~\ref{ev_cin_par}, where they can be compared to
the stellar age and the chromospheric activity level.

Figure~\ref{synt_gany} shows two examples of synthetic spectra of the \C2\ 
molecular band regions around $\lambda$5128 and $\lambda$5165, and of the C
atomic lines at $\lambda$5052.2 and $\lambda$5380.3 for the sunlight
spectrum reflected by Ganymede (second observation run) and for the
metal-rich star HD\,160691. The spectral synthesis was first applied to the
Ganymede spectra of both runs, then the $gf$ values of some atomic and
molecular lines were revised when needed, and finally the synthesis was
applied to the other stars, treated according to their observation runs. For
each case, the best fit was obtained through the minimisation of the $rms$
between observed and synthetic spectra.

\begin{figure}
\centering
\resizebox{\hsize}{!}{\includegraphics{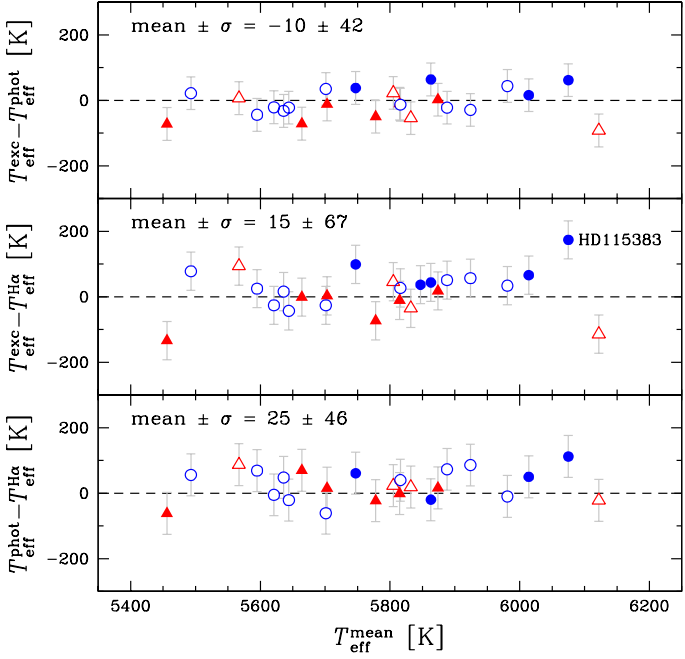}}
\caption{Comparisons of the three effective temperature indicators: the
         excitation vs. photometric (top panel), excitation vs. H$\alpha$
	 (middle panel), and photometric vs. H$\alpha$ (bottom panel)
	 temperatures. The symbols represent the stellar groups defined in
	 Sect.~\ref{classtree} (see Fig.~\ref{dendrogram}).}
\label{teff_comp}
\end{figure}

In order to estimate the uncertainties in the [C/Fe] abundance
determination, we performed a spectral synthesis of the most prominent
molecular band used ($\lambda$5165) adopting model atmospheres perturbed by
the errors estimated for the atmospheric parameters. This procedure resulted
in:
$\pm$0.03~dex due to the error in \texc;
$\pm$0.01~dex due to the error in [Fe/H];
$\pm$0.02~dex due to the error in $\xi$; and
$\pm$0.03~dex due to the error in \loggi.
The uncertainties related to errors in \Vbroad\  ($\sim$1.0~\kms\ or
smaller) and in the limb darkening coefficient are negligible. The quadratic
sum of the individual contributions (also including a global error of
0.05~dex estimated based on the $rms$ minimisation of the solar spectrum)
yields a total uncertainty $\sigma$([C/Fe]) = 0.07~dex.


\section{Evolutionary, Kinematic, and orbital parameters}
\label{evol_kin_orb}


\subsection{Mass and age determination}
\label{evol_par}

The evolutionary parameters mass and age were obtained by interpolation in
the Yonsei-Yale (${\rm Y}^2$) evolutionary tracks and isochrones
\citep{Yietal2001,Kimetal2002} drawn on the HR diagram and computed for
different values of metallicity.

The luminosity used in the diagrams were calculated with parallaxes taken
from the new reduction of the Hipparcos data \citep{vanLeeuwen2007},
bolometric corrections ($BC$) from \citet{Flower1996}, and an absolute
bolometric magnitude for the Sun $M_{\rm bol}^\odot = 4.75$ for
$M_{\rm v}^\odot = 4.82$. We estimated, for these nearby stars with precise
parallaxes, a mean uncertainty of 0.01~dex in $\log(L/L_\odot)$.

\begin{figure*}
\centering
\begin{minipage}[b]{0.49\textwidth}
\centering
\resizebox{0.88\hsize}{!}{\includegraphics[angle=-90]{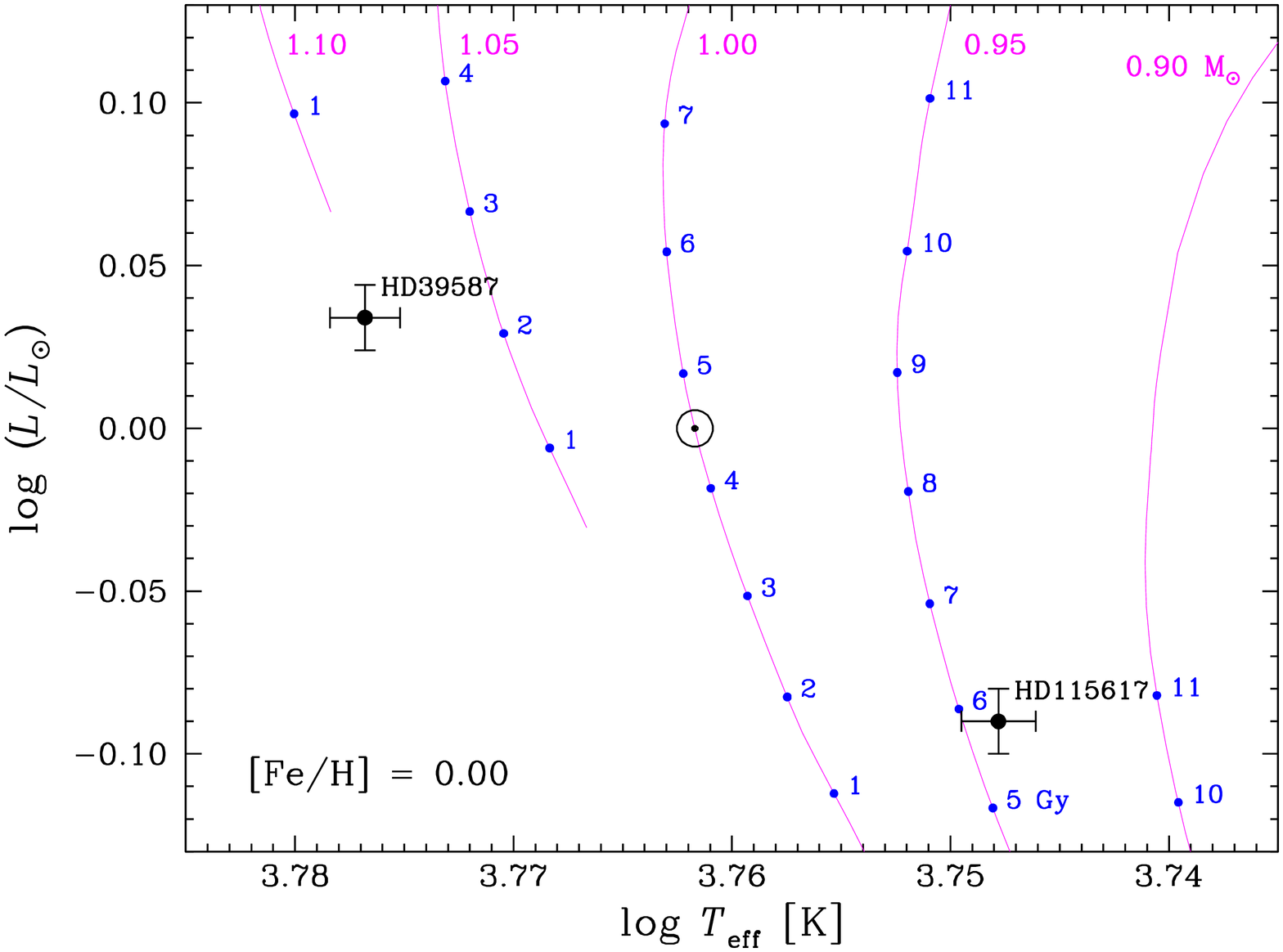}}
\end{minipage}
\begin{minipage}[b]{0.49\textwidth}
\centering
\resizebox{0.88\hsize}{!}{\includegraphics[angle=-90]{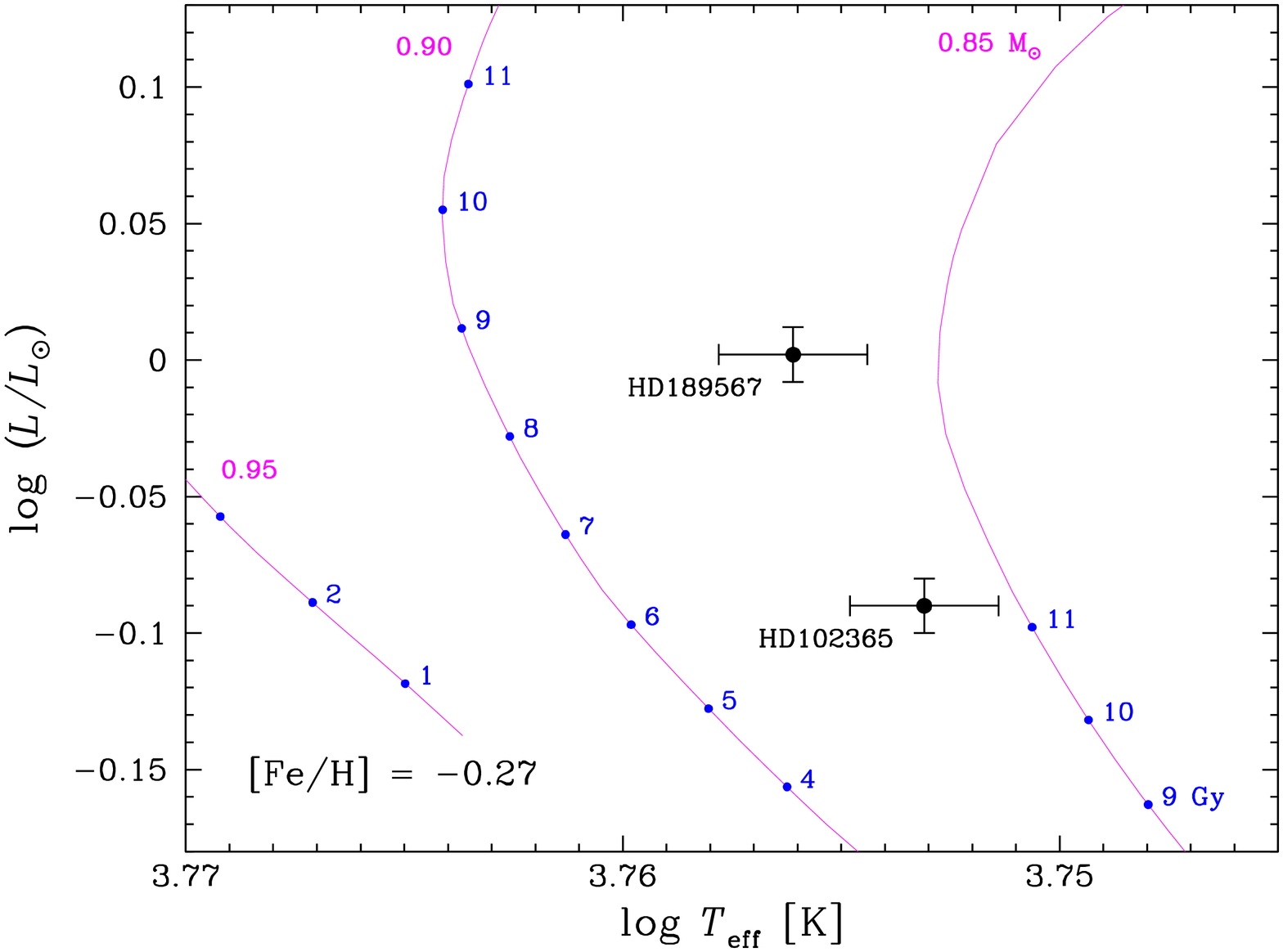}}
\end{minipage} \\
\begin{minipage}[b]{0.49\textwidth}
\centering
\resizebox{0.88\hsize}{!}{\includegraphics[angle=-90]{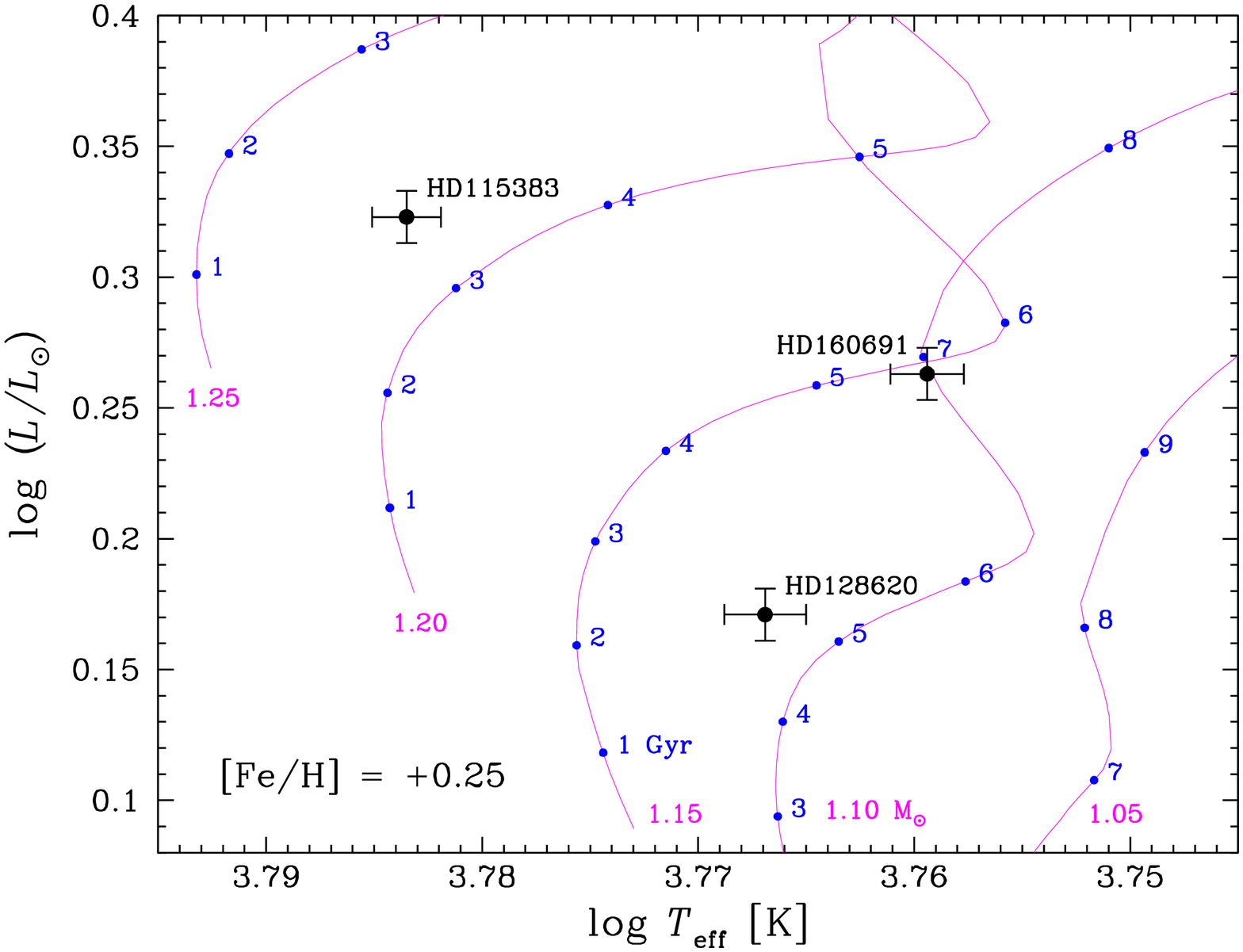}}
\end{minipage}
\begin{minipage}[b]{0.49\textwidth}
\centering
\resizebox{0.88\hsize}{!}{\includegraphics[angle=-90]{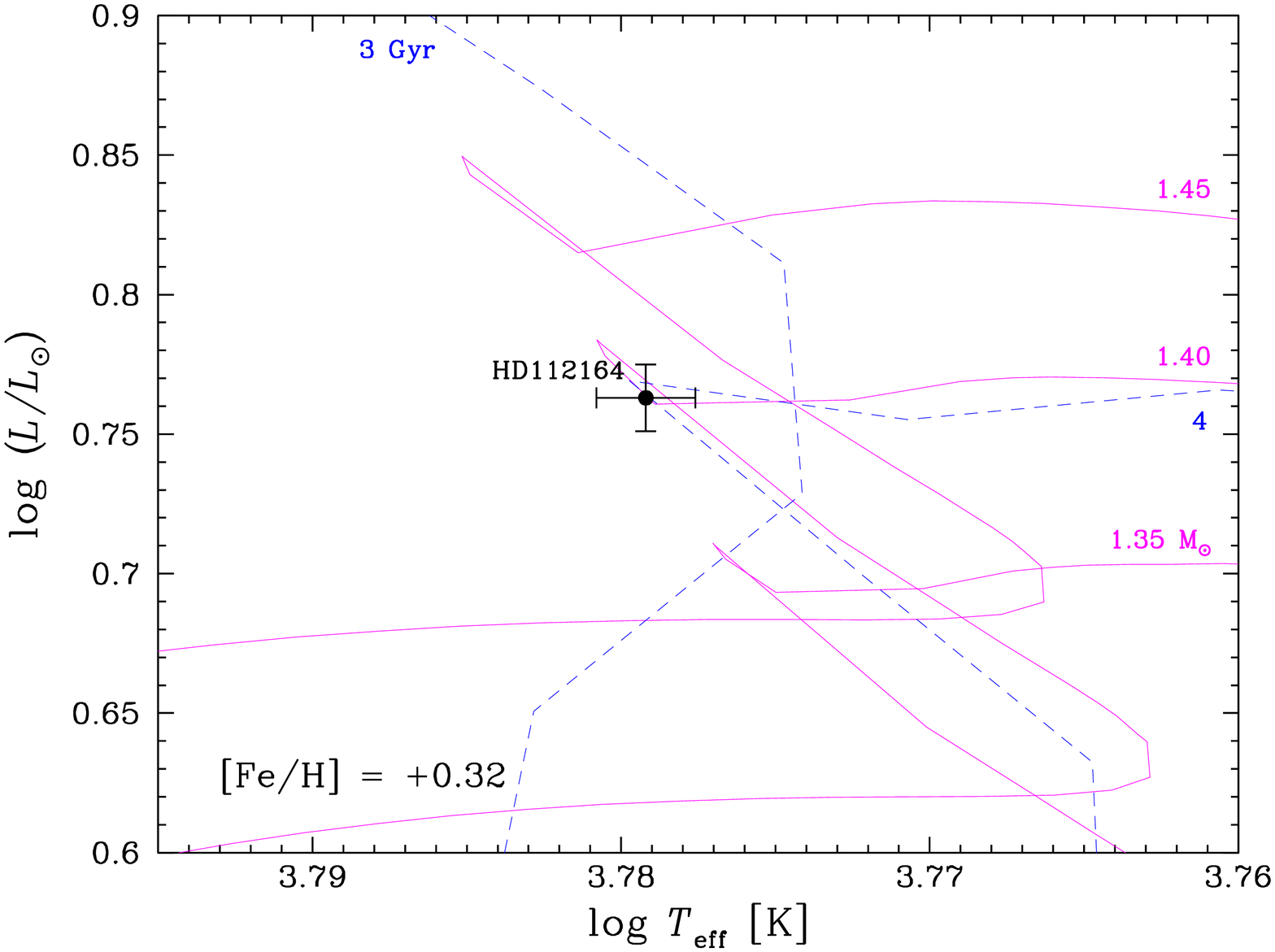}}
\end{minipage}
\caption{Evolutionary tracks (magenta solid lines) and isochrones (blue dots
         and dashed lines) from Yonsei-Yale, calculated for different values
	 of metallicity, showing how to derive the stellar masses and ages.
	 The ages in Gyr are indicated alongside the evolutionary tracks,
	 for which the masses are indicated in \msun. The Sun's position for
	 \teff\ = 5777~K is also shown in the solar metallicity panel
	 (top-left corner).}
\label{ev_diag}
\end{figure*}

The effective temperature is a weighted mean (\tmean) of the excitation,
photometric, and H$\alpha$ temperatures. The weights are given by
$1/\sigma^2$ for $\sigma$(\texc) = 30~K, $\sigma$(\tphot) = 40~K, and
$\sigma$(\tHa) = 50~K, obtained as described in Sect.~\ref{atmpar_sec} and
\ref{phot_temp}. The uncertainty of the weighted mean is 22~K, calculated
using these three values of $\sigma$. On the other hand, the standard
deviations of the three values of temperature around the weighted mean is
29~K. Therefore, we conclude that our estimates of effective temperature
based on the three indicators agree with each other very well (see the
comparison in Fig.~\ref{teff_comp}) and that the mean value has a mean
internal error $\sigma(T^{\rm mean}_{\rm eff})$ = 30~K.

\citet{PortodeMelloetal2008} determined the effective temperature of
$\alpha$\,Cen\,A (HD\,128620) and B also using the excitation, photometric,
and H$\alpha$ approaches. They found a good agreement for $\alpha$\,Cen\,A,
a solar temperature star. However, for $\alpha$\,Cen\,B (\teff\
$\sim$5200~K), the excitation effective temperature is about 100-150~K
higher than the photometric and the H$\alpha$ counterparts, which the
authors attributed to non-LTE effects. Although the agreement for the
coolest and hottest stars in our sample is not that good, especially in the
comparison of excitation and H$\alpha$ temperatures, Fig.~\ref{teff_comp}
does not show any systematic difference among the three indicators, and the
differences are nonetheless within 2$\sigma$ for all the \teff\ range (only
HD\,115383 has \texc\ larger than \tHa\ by 3$\sigma$). This confirms that 1D
LTE model atmospheres may adequately represent solar-type stars, at least in
a differential analysis relative to the Sun.

We remind that the temperature used in our abundance analyses (based on the
equivalent widths or synthesis of spectral features) was the excitation
effective temperature, which better characterises the temperature radial
profile in the stellar photosphere and the formation of absorption lines in
the emergent spectrum. On the other hand, to better represent the luminosity
of a star and to account for any possible effect due to small deviations
from LTE, we adopted the weighted mean of the three temperature indicators.

The stars were grouped according to their values of metallicity (12 groups
from [Fe/H] = $-$0.32 to +0.32~dex) and then their masses and ages were
computed using evolutionary tracks and isochrones for each stellar group.
The difference in metallicity between each star and its respective HR
diagram is not greater than 0.02~dex. A few examples for some metallicities
are shown in Fig.~\ref{ev_diag}. To reproduce the Sun's position in the
diagrams, adopting \teff\ = 5777~K and $age$ = 4.53~Gy
\citep{GuentherDemarque1997}, the evolutionary tracks and isochrones were
displaced in $\log(T_{\rm eff})$ and $\log(L/L_\odot)$ by 0.001628
($\sim$22~K in \teff) and 0.011, respectively. These values are, at any
rate, of the same order or smaller than the uncertainties on these
parameters.

As an independent check, we calculated the evolutionary surface gravity
using the values of mass and effective temperature obtained, which we called
$g_{\rm evol}$, using the following equation:
\begin{equation}
\label{log_g}
\log \frac{g_{\rm evol}}{g_\odot} = \log \frac{M}{M_\odot} +
4\log \frac{T^{\rm mean}_{\rm eff}}{T_{\rm eff}^\odot} +
0.4(M_{\rm bol} - M_{\rm bol}^\odot)
\end{equation}
where $M_{\rm bol}$ is the absolute bolometric magnitude for the stars. The
values of \logge\ are listed in Table~\ref{atmpar_tab} together with the
ionisation surface gravity. They are in very good agreement, having a
dispersion of only 0.09~dex, smaller than the uncertainty of 0.13~dex
estimated for \loggi.


\subsection{Galactic velocities, distance, and eccentricity}

\begin{figure*}
\centering
\begin{minipage}[b]{0.33\textwidth}
\centering
\resizebox{\hsize}{!}{\includegraphics{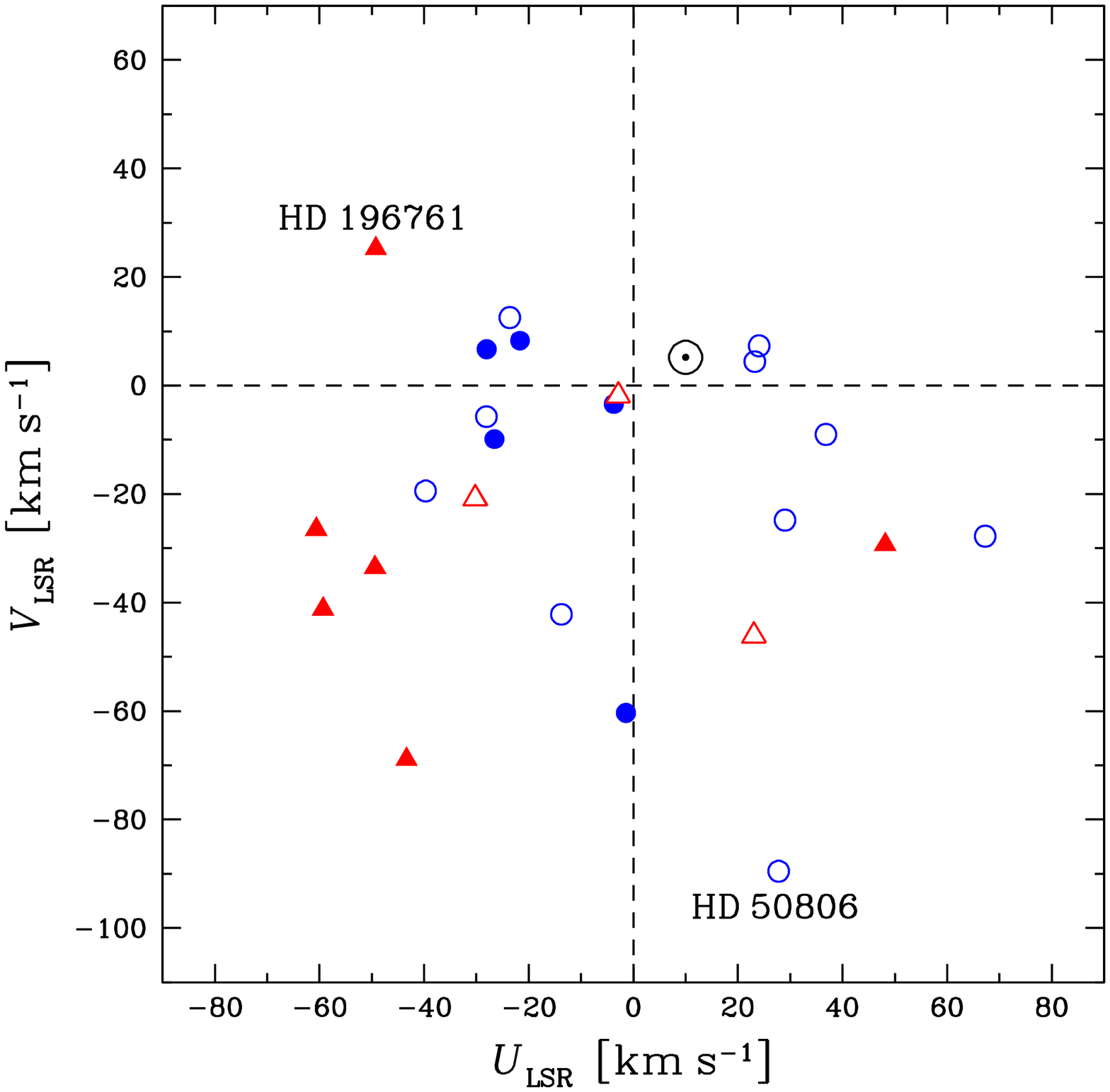}}
\end{minipage}
\begin{minipage}[b]{0.33\textwidth}
\centering
\resizebox{\hsize}{!}{\includegraphics{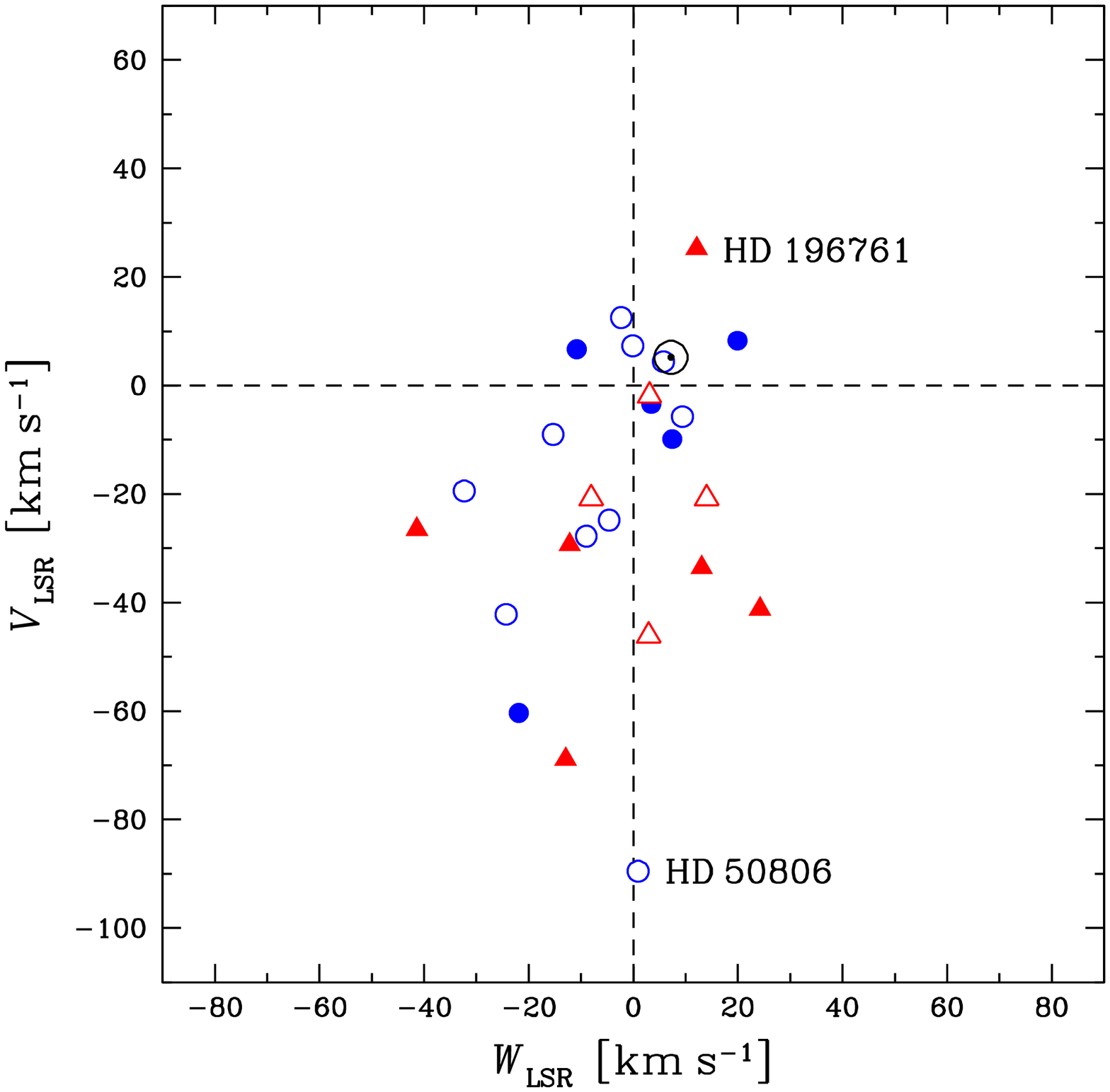}}
\end{minipage}
\begin{minipage}[b]{0.33\textwidth}
\centering
\resizebox{\hsize}{!}{\includegraphics{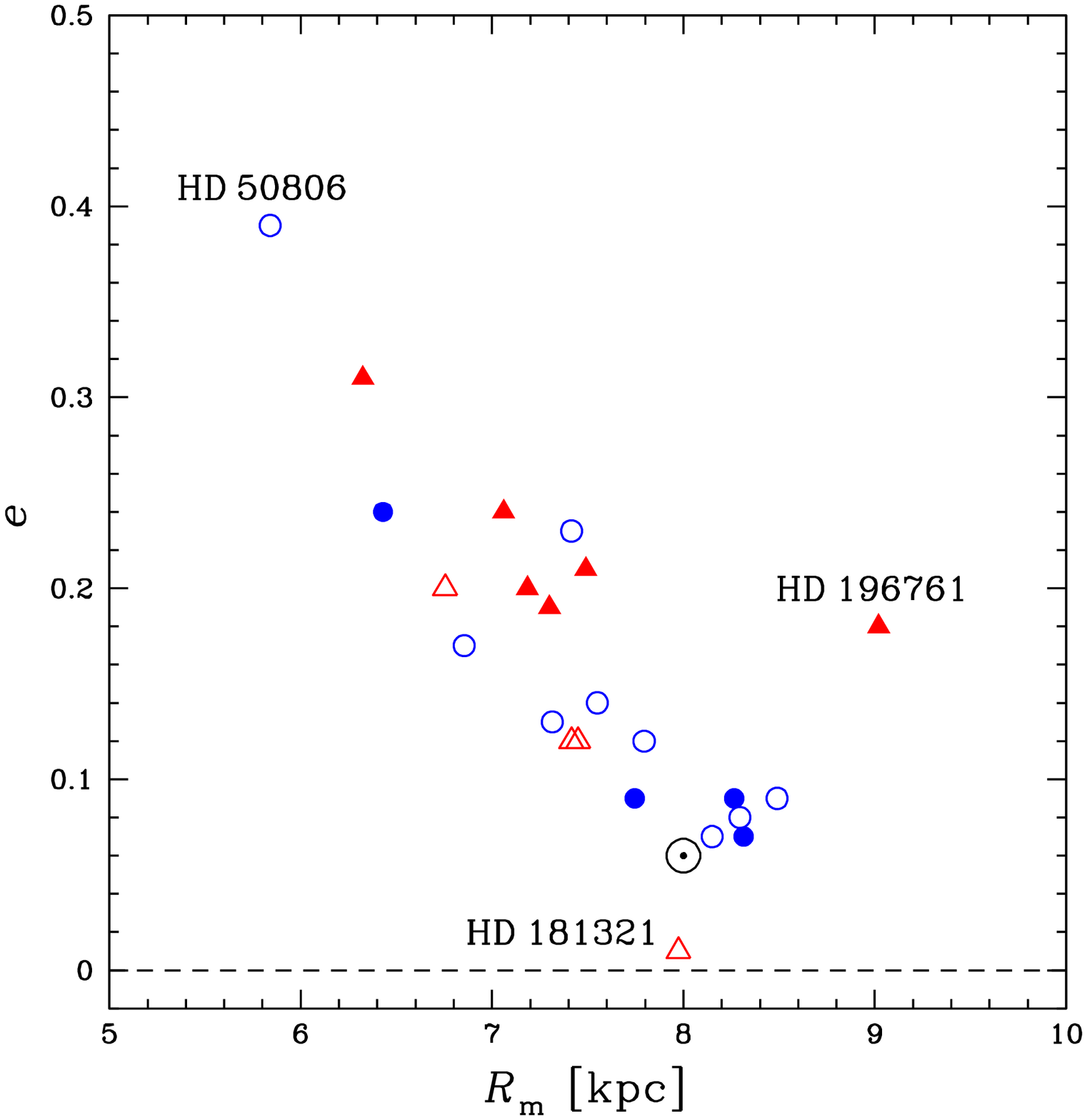}}
\end{minipage}
\caption{Galactic velocity components (left and middle panels) and orbital
         eccentricity as a function of the mean orbital distance from the
	 Galactic centre (right panel) for the sample stars and the Sun. The
	 symbols represent the stellar groups defined in
	 Sect.~\ref{classtree}.}
\label{cin_diag}
\end{figure*}

The kinematic properties of our sample were investigated by computing the
Galactic velocity components \Ulsr, \Vlsr, and \Wlsr\  (see
Fig.~\ref{cin_diag}) with respect to the Local Standard of Rest (LSR). We
developed a code that uses equations of \citet{JohnsonSoderblom1987},
parallaxes and proper motions both from the new reduction of the Hipparcos
data, and radial velocities from \citet{Holmbergetal2007},
\citet{Torresetal2006} for HD\,114613, and \citet{Santosetal2004} for
HD\,160691. For the Sun, the adopted values of \Ulsr, \Vlsr, and \Wlsr\ are
10.0, 5.3, and 7.2~\kms, respectively \citep{DehnenBinney1998}.

The mean orbital distance from the Galactic centre (\Rm) and the orbital
eccentricity ($e$) were also considered in our analysis (see
Fig.~\ref{cin_diag}), where $e$ = (\Ra $-$ \Rp)/(\Ra + \Rp) and \Rm\ =
(\Ra + \Rp)/2 were computed using the perigalactic (\Rp) and the apogalactic
(\Ra) orbital distances from the Geneva-Copenhagen survey
\citep{Holmbergetal2009}. For the Sun, the adopted values are $e$ = 0.06 and
\Rm\ = 8~kpc.


\section{Tree clustering analysis}
\label{classtree}

We looked for statistically significant abundance groups in our sample using
a hierarchical clustering analysis. To avoid missing abundance values, the
analysis uses only those elements having abundances measured for all stars
and was applied to the [X/H] abundance space.

\begin{figure}
\centering
\begin{minipage}[b]{0.48\textwidth}
\centering
\resizebox{0.78\hsize}{!}{\includegraphics[angle=-90]{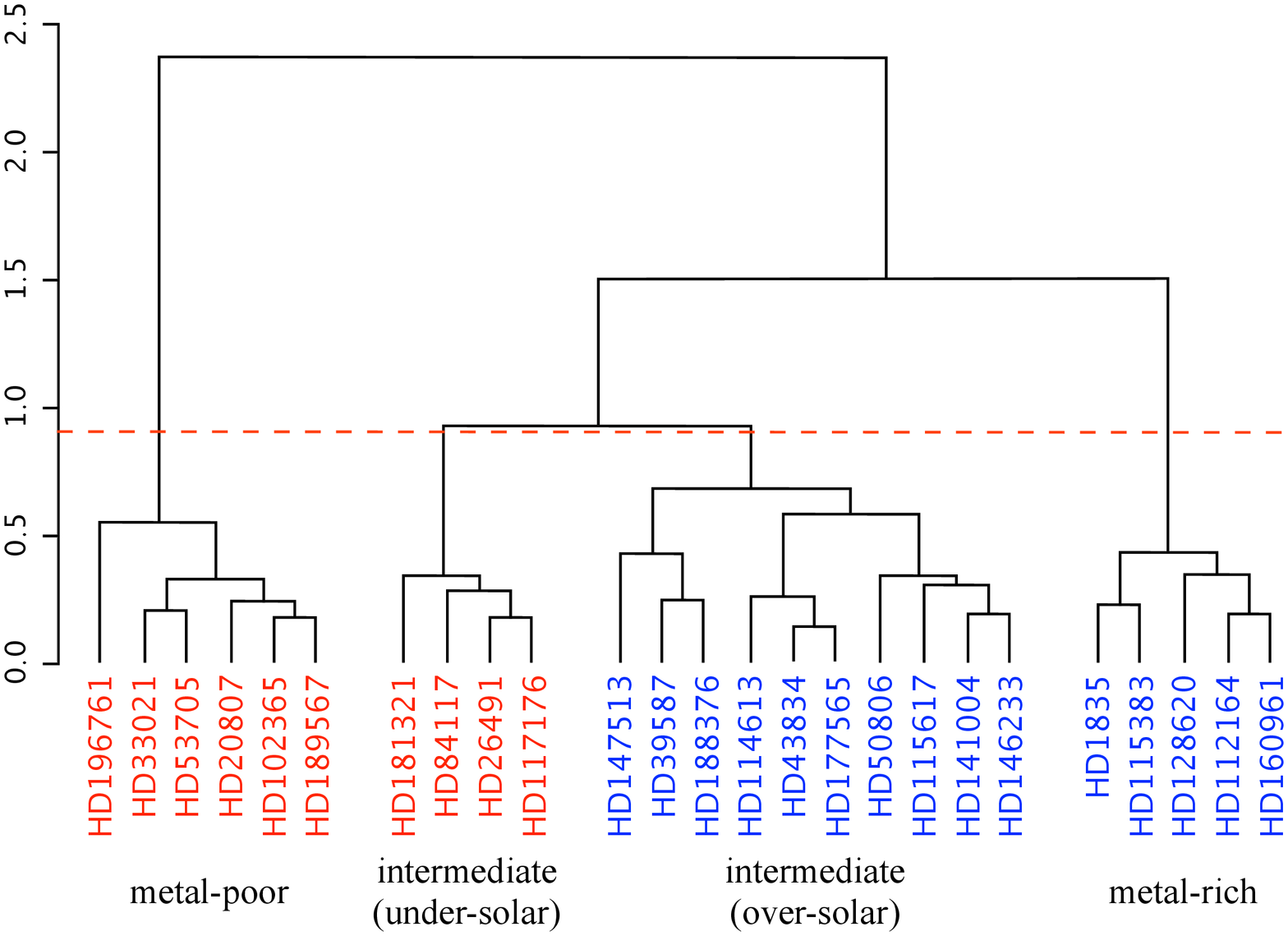}}
\end{minipage} \\[0.3cm]
\begin{minipage}[b]{0.48\textwidth}
\centering
\resizebox{0.88\hsize}{!}{\includegraphics[angle=-90]{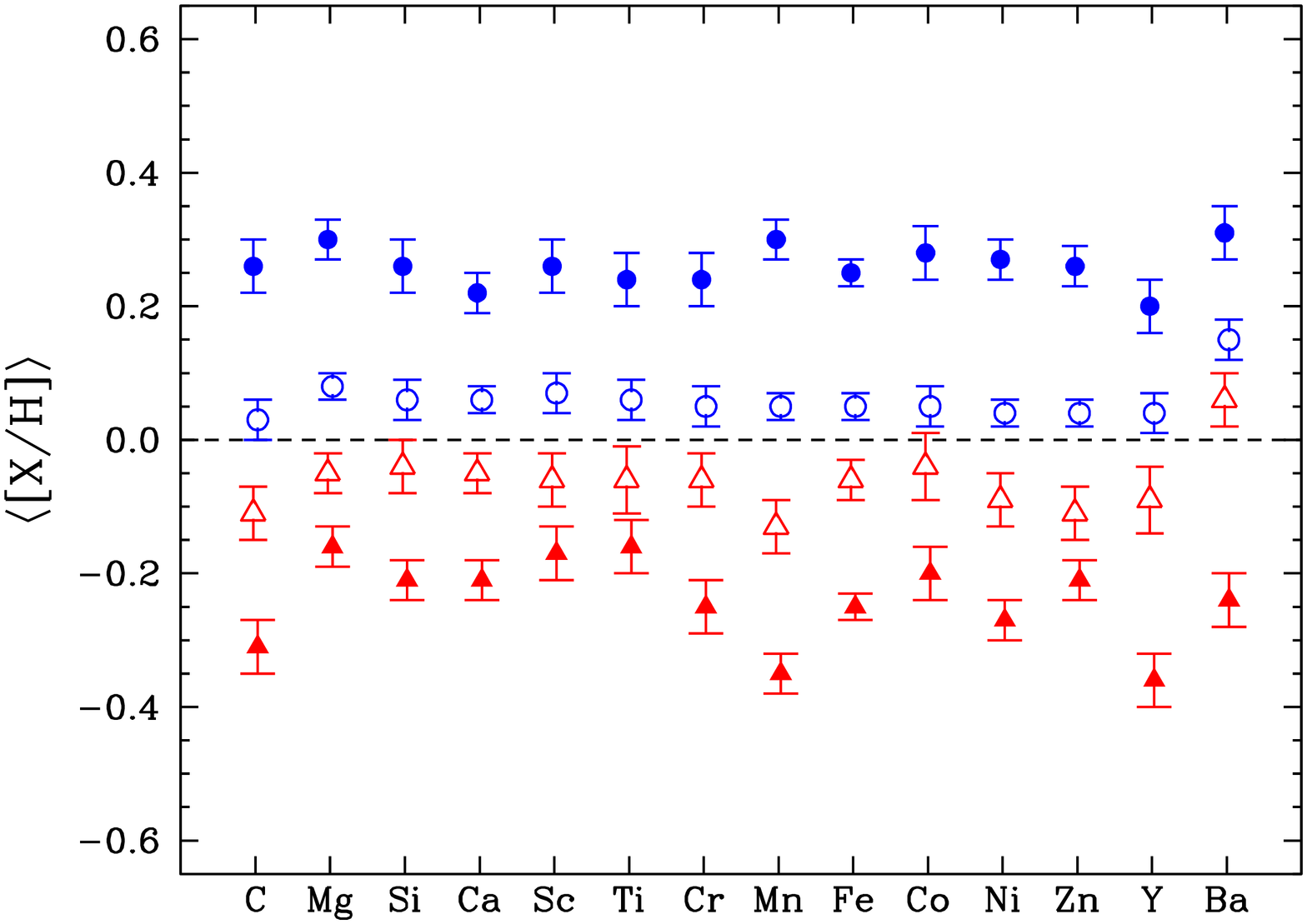}}
\end{minipage} \\
\begin{minipage}[b]{0.48\textwidth}
\centering
\resizebox{0.88\hsize}{!}{\includegraphics[angle=-90]{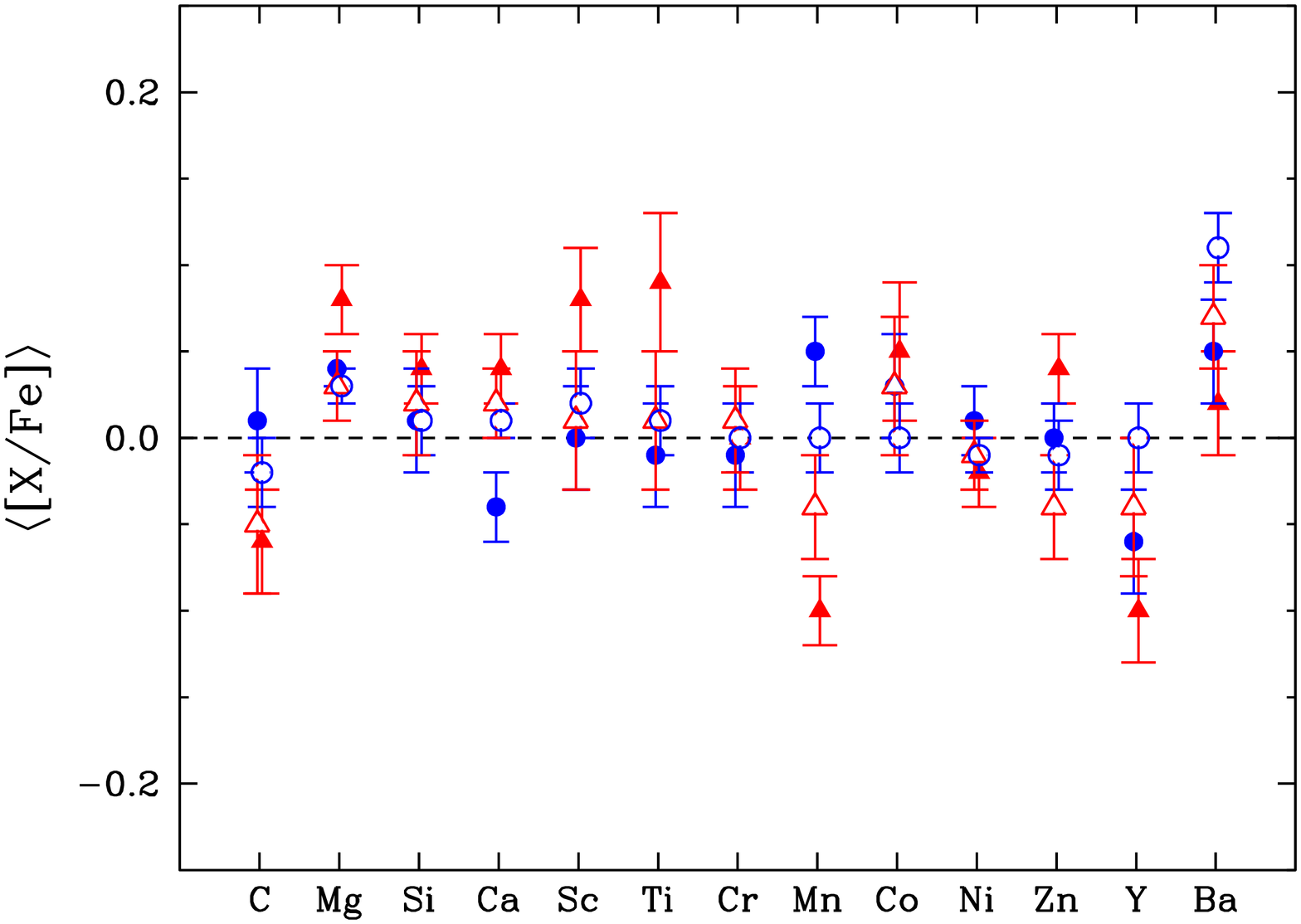}}
\end{minipage}
\caption{Dendrogram (upper panel) and mean abundance ratios \mxh\ (middle
	 panel) and \mxfe\ (bottom panel) for the four stellar groups from
	 the tree clustering analysis (Sect.~\ref{classtree}). They are
	 classified as metal-poor ({\color{red} $\blacktriangle$}),
	 intermediate abundance (under solar {\color{red} $\triangle$} or
	 over solar {\color{blue} \Large $\circ$}), or metal-rich
	 ({\color{blue} \Large $\bullet$}) stars, and the same symbolism is
	 adopted all over the paper. The dashed red line on the dendrogram
	 represents a dissimilarity number of 0.9 dex.}
\label{dendrogram}
\end{figure}

We used the complete linkage method for the hierarchical clustering
\citep{Everittetal2001} and euclidean distances as measures of
dissimilarities in this abundance space. A hierarchical clustering algorithm
works by joining similar objects in a hierarchical structure. Initially,
each object is assigned to its own cluster. The algorithm proceeds
iteratively, joining the two most similar clusters in each pass until there
is just a single cluster. The resulting hierarchy of clusters for our data
is shown in Fig.~\ref{dendrogram} (upper panel) as a dendrogram. In this
plot, the most similar objects are linked together in the bottom forming
clusters, which are then iteratively linked together in pairs by similarity.
The vertical axis in a dendrogram measures the dissimilarity between each
individual or cluster. Since we used euclidean distances in the [X/H]
abundance space, the units of this axis is dex, although it measures the
total dissimilarity in this abundance space and not in a single variable.

Clusters can be defined by specifying a reasonable total dissimilarity value
for pruning the dendrogram. There is no unambiguous or optimal way for
defining this pruning value, especially because the clusters found depend on
the clustering method and cluster shapes. Since our sample is quite small,
we arbitrarily decided to prune our dendrogram at the total dissimilarity of
0.9 dex, which is shown in Fig.~\ref{dendrogram} by the dashed red line.
This pruning value was chosen in order to have four more or less equally
populated clusters. Considering the size of our sample, less than four
clusters would simply limit our discussion to poor against rich stars, while
a larger number of clusters would make such an analysis meaningless.

The average \mxh\ behaviour of the four clusters for each element considered
in this analysis is shown in the middle panel of Fig.~\ref{dendrogram}. Two
of these clusters have over-solar abundances, with averages +0.26 and
+0.06~dex for all the elements, whereas the two others have under-solar
abundance values, with averages $-$0.06 and $-$0.24~dex. We can observe in
the figure that the elemental abundance patterns of the metal-poor and
metal-rich groups are distinct between each other. In particular, it seems
to exit a chemical distinction even between the two intermediate groups: the
under-solar intermediate group has an abundance pattern that roughly follows
the element by element pattern of the metal-poor group, whereas the pattern
of the over-solar intermediate group resembles the scaled-solar mixture.

The clustering analysis we have presented was tentatively based on
biological ideas of evolving species, in the sense that the material the
stars came from is continuously changing. For this reason we made use of
[X/H] abundances ratios instead of [X/Fe]. We implicitly need the time
evolution that [X/H] has, because we want a time hierarchy in the output
groups. A similar analysis in the [X/Fe] space can still show groups, but
the hierarchical relation between these groups in a dendrogram will not
necessarily show evolutionary trends, because this variable is only
indirectly linked with time. Notwithstanding, we have checked this, but the
output groups show no meaningful interpretation in terms of chemical
evolution or abundance ratio groups. The outcome could be different if the
sample were larger, but this needs to be verified with another sample, what
is beyond the scope of this paper. Nevertheless, we included in
Fig.~\ref{dendrogram} (bottom panel) the average \mxfe\ behaviour of the
stars clustered in the [X/H] parameter space. The small variation of \mxfe\ 
with respect to the solar values reinforces our point above: that for this
specific small sample, the [X/Fe] parameter space is dynamically very narrow
and does not favour a cluster analysis.


\section{Results and discussion}
\label{results}

Table~\ref{ev_cin_par} lists the evolutionary (mass and age), kinematic
(\Ulsr, \Vlsr, and \Wlsr\ velocities), and orbital (mean orbital distance
from the Galactic centre and orbital eccentricity) parameters computed for
the program stars. They are grouped following the tree clustering analysis
performed in Sect.~\ref{classtree}.

The uncertainties in the mass and age determination may vary widely
depending on the stellar position in the HR diagram. We made an estimate of
these errors for a few representative stars in our sample (cool and hot
dwarfs and subgiants). We took into account the errors estimated for 
$\log(L/L_\odot)$, $\log(T_{\rm eff})$, and also [Fe/H] considering that the
evolutionary tracks and isochrones are metallicity dependent. We found that
the uncertainties in mass stand between 0.02 and 0.08~\msun, whereas those
in age vary from about 0.5~Gyr (or smaller) for evolved stars up to about
2.5~Gyr for cool main-sequence stars.

For HD\,1835, HD\,39587, HD\,147513, and HD\,181321 we determined an
approximative value for their masses and an upper limit for their ages given
their position in the HR diagram (close to the Zero Age Main Sequence).
Indeed, these are very young stars: one of them, HD\,1835, is likely a
member of the Hyades star cluster ($\sim$600~Myr) according to
\citet{LopezSantiagoetal2010}; two others, HD\,39587
\citep{SoderblomMayor1993,Fuhrmann2004} and HD\,147513
\citep{SoderblomMayor1993,Montesetal2001} belong to the Ursa Major moving
group of $\sim$300~Myr \citep[see also][]{Castroetal1999}; and HD\,181321 is
a member of the Castor moving group ($\sim$200~Myr) according to
\citet{Montesetal2001}. For these four stars, we adopted the ages of their
respective moving group in our study.

The stars HD\,112164 and HD\,160691, indicated by asterisks (*) in
Table~\ref{ev_cin_par}, are located in a region of the HR diagram where
successive evolutionary tracks and isochrones are superposed (see example in
Fig.~\ref{ev_diag}). Therefore, their mass and age determination may yield
larger uncertainties: $\lesssim$ 0.12~\msun\ and $\lesssim$ 0.7~Gyr for
HD\,112164, and $\lesssim$ 0.04~\msun\ and $\lesssim$ 1.0~Gyr for
HD\,160691.

The \Ulsr, \Vlsr, and \Wlsr\ velocities have a typical uncertainty of
0.3~\kms\ or smaller. An exception is the star HD\,188376, for which the
large errors (2.5, 1.1, and 1.8~\kms, respectively) are due to a large
uncertainty in its parallax. The level of activity in the chromosphere of
the stars, which is related to their age, was also investigated. The table
lists the flux in the centre of the H$\alpha$ line ($F'_{{\rm H}\alpha}$),
computed by \citet{LyraPortodeMello2005} and used as a chromospheric
activity indicator (the larger the value of $F'_{{\rm H}\alpha}$, the higher
the level of chromospheric activity). The uncertainty for this parameter is
0.5$\times10^5$~erg\,cm$^{-2}$\,s$^{-1}$.

We note that a few stars in our sample have at least one planetary companion
detected. They are HD102365, HD115617, HD117176, HD147513, and HD160691 (see
The Extrasolar Planets Encyclopaedia: http://exoplanet.eu). Comparisons of
properties of stars with and without planets are frequently published. In
the analysis performed in this paper, however, no peculiar information
distinguishing the two populations has been found.


\subsection{Elemental abundances}

Table~\ref{ab_ratios} lists the chemical abundances relative to iron [X/Fe]
obtained for the elements studied. For some elements of some stars, the
abundance determination was not possible due to the poor quality or the
weakness of their spectral lines (empty fields in the table). The carbon
abundance ratios [C/Fe] are shown in Table~\ref{carb_ab}.

We also computed the mean abundance ratios \mxfe\ of the following groups of
elements:
$i)$ two groups of light metals: (Mg, Si) and (Ca, Sc, Ti);
$ii)$ two groups of the iron peak: (V, Cr, Co, Ni) and (Mn, Cu);
$iii)$ light elements from the s-process: (Sr, Y, Zr), to which we refer
as ls; and
$iv)$ heavy elements from the s-process: (Ba, Ce, Nd), referred to as hs
(see Table~\ref{mean_ab_err_comp}).
They were grouped either because they have possibly the same nucleosynthetic
origin or because they share a similar behaviour in the diagrams. The
abundance ratio between heavy and light elements from the s-process, [hs/ls]
= [hs/Fe] $-$ [ls/Fe], was also calculated.

Figure~\ref{ab_ratios_fig} shows diagrams with the abundance ratios of the
program stars for individual elements and nucleosynthetic groups. The
uncertainties are listed in Tables~\ref{ab_err_comp} and
\ref{mean_ab_err_comp}. As for the individual elements, the estimated errors
are compared to the dispersions around the mean for groups having at least
two elements. For each group of each observation run (represented by the
stars HD\,146233 and HD\,26491), the larger values were adopted to be the
uncertainties in the grouped abundance ratios. In Fig.~\ref{ab_ratios_fig},
the stars are represented by different symbols according to the
tree clustering analysis performed in Sect.~\ref{classtree}.

The star HD\,1835 is enriched in Ca, Sr, and Ba. The mean value of all
s-process elements also suggests an over-solar abundance. Sm, the only
r-process element analysed here, shows an under-solar abundance of
$-$0.3~dex, but with a large error. As already mentioned, this is a very
young star, a probable member of the Hyades star cluster of age
$\sim$600~Myr, which is in agreement with its high level of chromospheric
activity indicated by $F'_{{\rm H}\alpha}$ in Table~\ref{ev_cin_par}.

\begin{table*}
\centering
\caption[]{Evolutionary, kinematic, and orbital parameters, separating the
           stars according to the classification of Sect.~\ref{classtree}.
	   The H$\alpha$ fluxes ($F'_{{\rm H}\alpha}$), from
	   \citet{LyraPortodeMello2005}, are given in units of
	   $10^5$~erg\,cm$^{-2}$\,s$^{-1}$. The mean distance from the
	   Galactic centre (\Rm) and the orbital eccentricity ($e$) are from
	   \citet{Holmbergetal2009}. The asterisk (*) indicates stars
	   adjacent to superposed evolutionary tracks and isochrones on the
	   HR diagram.}
\label{ev_cin_par}
\begin{tabular}{l c r@{}l r@{}l r@{}l r@{}l r@{}l r@{}l r@{}l r@{}l r@{}l c c}
\hline\hline\noalign{\smallskip}
 &
\parbox[c]{0.8cm}{\centering \tmean\ {\tiny [K]}} &
\multicolumn{2}{c}{[Fe/H]} &
\multicolumn{2}{c}{$\log L/{L_\odot}$} &
\multicolumn{2}{c}{$M/{M_\odot}$} &
\multicolumn{2}{c}{\parbox[c]{0.7cm}{\centering $age$ {\tiny [Gyr]}}} &
\multicolumn{2}{c}{\parbox[c]{1.0cm}{\centering \Vbroad\ {\tiny [\kms]}}} &
\multicolumn{2}{c}{$F'_{{\rm H}\alpha}$} &
\multicolumn{2}{c}{\parbox[c]{1.0cm}{\centering \Ulsr\ {\tiny [\kms]}}} &
\multicolumn{2}{c}{\parbox[c]{1.0cm}{\centering \Vlsr\ {\tiny [\kms]}}} &
\multicolumn{2}{c}{\parbox[c]{1.0cm}{\centering \Wlsr\ {\tiny [\kms]}}} &
\parbox[c]{0.9cm}{\centering \Rm\ {\tiny [kpc]}} &
$e$ \\
\noalign{\smallskip}\hline\noalign{\smallskip}
Sun        & 5777 &    0&.00 &    0&.00 &   1&.00  &   4&.53 &   1&.8  &  3&.4 &    10&.0 &     5&.3 &     7&.2 & 8.00 & 0.06 \\
\hline\noalign{\smallskip}
\multicolumn{22}{l}{metal-poor stars:} \\
\noalign{\smallskip}
HD\,20807  & 5874 & $-$0&.22 & $-$0&.02 &   0&.96  &  4&.2  & $<$&\,2 &  3&.6 & $-$59&.3 & $-$41&.2 &	 24&.2 & 7.06 & 0.24 \\
HD\,33021  & 5778 & $-$0&.20 &    0&.35 &   0&.98  & 10&.1  &	4&.1  &  3&.8 &    48&.2 & $-$29&.3 & $-$12&.2 & 7.30 & 0.19 \\
HD\,53705  & 5815 & $-$0&.22 &    0&.14 &   0&.93  & 10&.1  &	4&.0  &  1&.9 & $-$43&.4 & $-$68&.9 & $-$12&.9 & 6.33 & 0.31 \\
HD\,102365 & 5664 & $-$0&.28 & $-$0&.09 &   0&.86  & 10&.1  & $<$&\,2 &  2&.1 & $-$49&.4 & $-$33&.5 &	 13&.1 & 7.18 & 0.20 \\
HD\,196761 & 5456 & $-$0&.32 & $-$0&.27 &   0&.80  & 10&.8  &	3&.3  &  4&.9 & $-$49&.2 &    25&.3 &	 12&.1 & 9.02 & 0.18 \\
HD\,189567 & 5703 & $-$0&.27 &    0&.00 &   0&.87  & 11&.9  &	4&.4  &  2&.9 & $-$60&.6 & $-$26&.5 & $-$41&.4 & 7.49 & 0.21 \\
\hline\noalign{\smallskip}
\multicolumn{22}{l}{intermediate abundance (under solar) stars:} \\
\noalign{\smallskip}
HD\,26491  & 5805 & $-$0&.09 &    0&.13 &   0&.97  &  8&.5  &	3&.5  &  2&.1 & $-$30&.2 & $-$20&.8 &  $-$8&.0 & 7.42 & 0.12 \\
HD\,84117  & 6122 & $-$0&.06 &    0&.29 &   1&.12  &  3&.9  &	5&.9  &  0&.0 & $-$30&.2 & $-$20&.8 &	 14&.1 & 7.45 & 0.12 \\
HD\,117176 & 5567 & $-$0&.04 &    0&.47 &   1&.08  &  8&.2  & $<$&\,2 &  1&.5 &    23&.1 & $-$46&.2 &	  3&.0 & 6.75 & 0.20 \\
HD\,181321 & 5832 & $-$0&.06 & $-$0&.12 & \s1&.02  &  0&.2  &  12&.5  & 14&.2 &  $-$2&.9 &  $-$1&.9 &	  3&.1 & 7.97 & 0.01 \\
\hline\noalign{\smallskip}
\multicolumn{22}{l}{intermediate abundance (over solar) stars:} \\
\noalign{\smallskip}
HD\,39587  & 5981 &    0&.00 &    0&.03 & \s1&.10  &  0&.3  &	9&.0  & 10&.3 &    24&.0 &     7&.3 &  $-$0&.1 & 8.29 & 0.08 \\
HD\,43834  & 5636 &    0&.11 & $-$0&.08 &   0&.99  &  3&.9  & $<$&\,2 &  2&.8 &    29&.0 & $-$24&.8 &  $-$4&.6 & 7.31 & 0.13 \\
HD\,50806  & 5621 &    0&.02 &    0&.35 &   1&.02  &  9&.9  &	2&.0  &  2&.8 &    27&.8 & $-$89&.5 &	  0&.9 & 5.84 & 0.39 \\
HD\,114613 & 5701 &    0&.15 &    0&.62 &   1&.26  &  5&.1  & $<$&\,2 &  1&.7 & $-$28&.1 &  $-$5&.7 &	  9&.4 & --   & --   \\
HD\,115617 & 5595 &    0&.00 & $-$0&.09 &   0&.94  &  6&.8  & $<$&\,2 &  4&.2 & $-$13&.8 & $-$42&.2 & $-$24&.3 & 6.85 & 0.17 \\
HD\,141004 & 5924 &    0&.03 &    0&.32 &   1&.10  &  6&.3  & $<$&\,2 &  1&.6 & $-$39&.7 & $-$19&.4 & $-$32&.3 & 7.55 & 0.14 \\
HD\,146233 & 5816 &    0&.05 &    0&.02 &   1&.03  &  3&.4  & $<$&\,2 &  2&.7 &    36&.8 &  $-$9&.0 & $-$15&.3 & 7.79 & 0.12 \\
HD\,147513 & 5888 &    0&.04 & $-$0&.02 & \s1&.05  &  0&.3  & $<$&\,2 &  7&.0 &    23&.2 &     4&.4 &	  5&.8 & 8.15 & 0.07 \\
HD\,177565 & 5644 &    0&.08 & $-$0&.06 &   1&.00  &  4&.2  &	2&.3  &  3&.8 &    67&.3 & $-$27&.8 &  $-$9&.0 & 7.41 & 0.23 \\
HD\,188376 & 5493 &    0&.00 &    0&.90 &   1&.53  &  2&.7  & $<$&\,2 &  0&.5 & $-$23&.6 &    12&.5 &  $-$2&.3 & 8.49 & 0.09 \\
\hline\noalign{\smallskip}
\multicolumn{22}{l}{metal-rich stars:} \\
\noalign{\smallskip}
HD\,1835   & 5863 &    0&.21 &    0&.00 & \s1&.15  &  0&.6  &	5&.9  &  7&.7 & $-$26&.5 &  $-$9&.9 &	  7&.4 & 7.74 & 0.09 \\
HD\,112164 & 6014 &    0&.32 &    0&.76 &   1&.40* &  3&.5* &	4&.6  &  1&.4 &  $-$1&.4 & $-$60&.4 & $-$21&.9 & 6.43 & 0.24 \\
HD\,115383 & 6075 &    0&.23 &    0&.32 &   1&.22  &  2&.9  &	7&.0  &  8&.0 & $-$28&.1 &     6&.7 & $-$10&.8 & 8.27 & 0.09 \\
HD\,128620 & 5847 &    0&.23 &    0&.17 &   1&.11  &  4&.4  & $<$&\,2 &  4&.7 & $-$21&.7 &     8&.3 &	 19&.9 & 8.31 & 0.07 \\
HD\,160691 & 5747 &    0&.27 &    0&.26 &   1&.12* &  6&.2* & $<$&\,2 &  2&.4 &  $-$3&.7 &  $-$3&.4 &	  3&.4 & --   & --   \\
\hline						     
\end{tabular}
\end{table*}

Two other very young stars are HD\,39587 and HD\,147513, both members of the
kinematic Ursa Major group. They are clearly overabundant in the s-process
elements, especially Ba, and underabundant in C, which is in agreement with
the results of \citet{PortodeMellodaSilva1997a} and \citet{Castroetal1999}.
These stars were proposed by \citet{PortodeMellodaSilva1997a} to be
{\it barium stars}, originated in a phenomenon in which the more massive
component of a binary system evolves as a thermally pulsing asymptotic giant
branch (TP-AGB) star and the material produced in the He-burning envelope,
enriched in s-process elements, is dredged-up to the surface and then
accreted by its companion by wind mass transfer. The initially more massive
star is now a white dwarf whereas the companion has become the primary
barium star. At present, HD\,39587 is a single-lined spectroscopic and
astrometric binary, with a low mass companion of 0.15~\msun\ 
\citep{Konigetal2002}, and HD\,147513 has a common proper motion companion,
a DA2 white dwarf, at an angular separation of 345$\arcsec$
\citep{Holbergetal2002}. The barium-star scenario was not supported by
\citet{Castroetal1999}, who proposed that the two stars simply have usual Ba
abundances for their age and that probably all the Ursa Major group members
are Ba-enriched, either due to a primordial origin or because they are
young (see discussion in Sect.~\ref{ab_age_sect}).

HD\,181321 and HD\,188376 are two other Ba-rich stars. HD\,188376 is the
most evolved and massive star analysed here, clearly in the evolutionary
stage of a subgiant. The other star, HD\,181321, is the youngest and has the
highest level of chromospheric activity in our sample. Indeed, our
determination for \Vbroad\ is 12.5~\kms, indicating a fast-rotating star. It
has solar atmospheric parameters, excepting a high value of micro-turbulence
velocity ($\xi$ = 2.3~\kms). The kinematic and orbital parameters are also
very close to the solar values. In other words, this star has, on the one
hand, about the same effective temperature, metallicity, surface gravity,
mass, Galactic orbit, and space velocities as the Sun. On the other hand, it
is very young and significantly enriched in Ba, strengthening the relation
between Ba abundance and age (see Sect.~\ref{ab_age_sect}).

The high microturbulence velocity of the star HD\,181321 is probably
prompted by the strengthened convection and turbulence in its upper
photosphere, which is subjected to large non-thermal energy influxes from
the chromosphere. The UV radiation excess from the chromosphere of an active
star can scape to the photosphere and cause departures from LTE due to an
ionisation imbalance. The induced overionisation is commonly manifested by
differences either in excitation and photometric effective temperatures, or
in ionisation and evolutionary surface gravities
\citep[see][]{PortodeMelloetal2008,Ribasetal2010}. For HD\,181321, our
determination of \texc\ and \tphot\ are in very good agreement with each
other. Therefore, only the difference in surface gravity and the large value
of microturbulence velocity are possible signs that an overionisation is
taking place in the photosphere of this active star. A full non-LTE analysis
and a photospheric and chromospheric modelling would probably settle the
issue, but this goes beyond the scope of this paper.

\begin{figure*}
\centering
\begin{minipage}[b]{0.33\textwidth}
\centering
\resizebox{\hsize}{!}{\includegraphics[angle=-90]{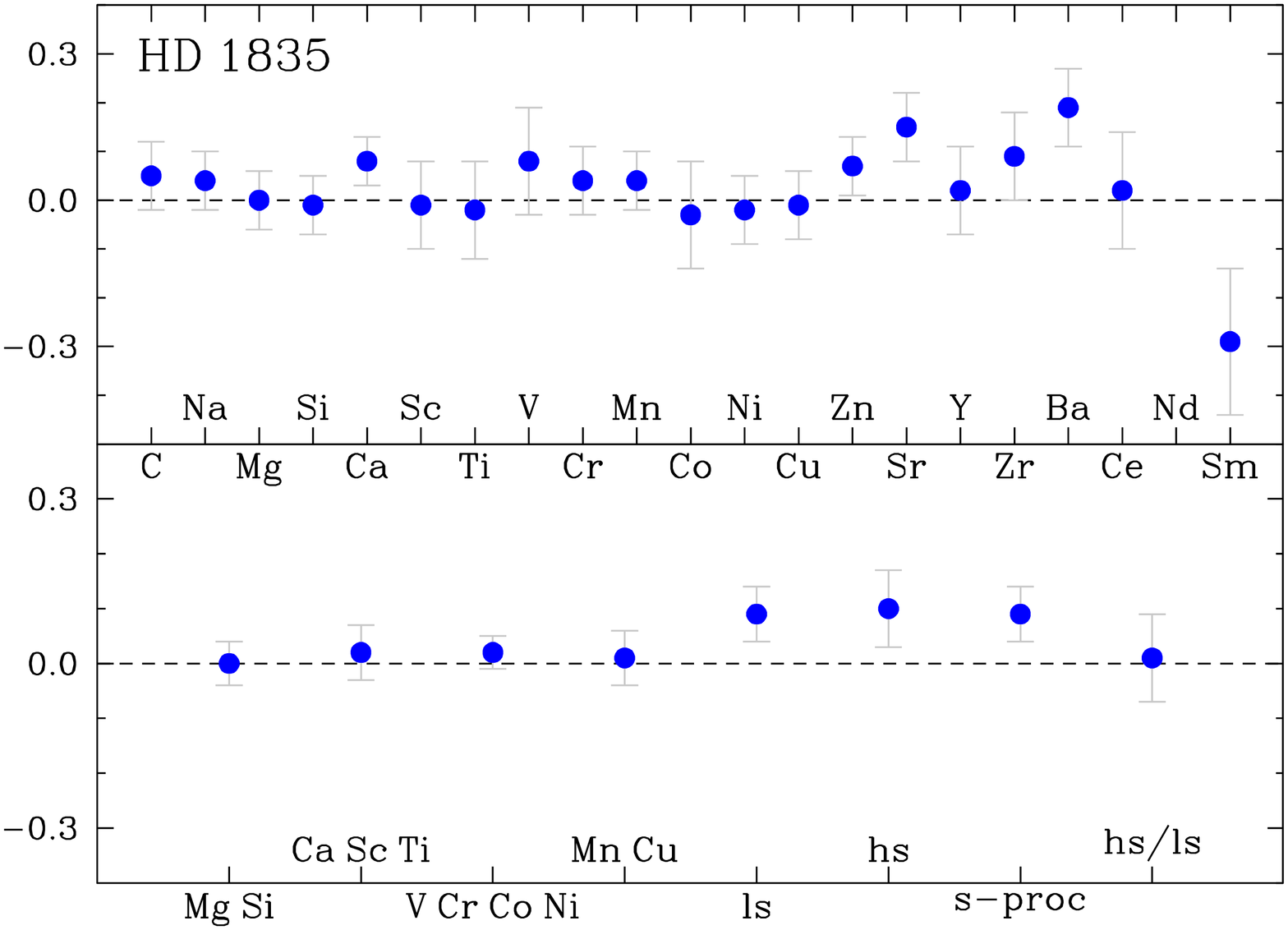}}
\end{minipage}
\begin{minipage}[b]{0.33\textwidth}
\centering
\resizebox{\hsize}{!}{\includegraphics[angle=-90]{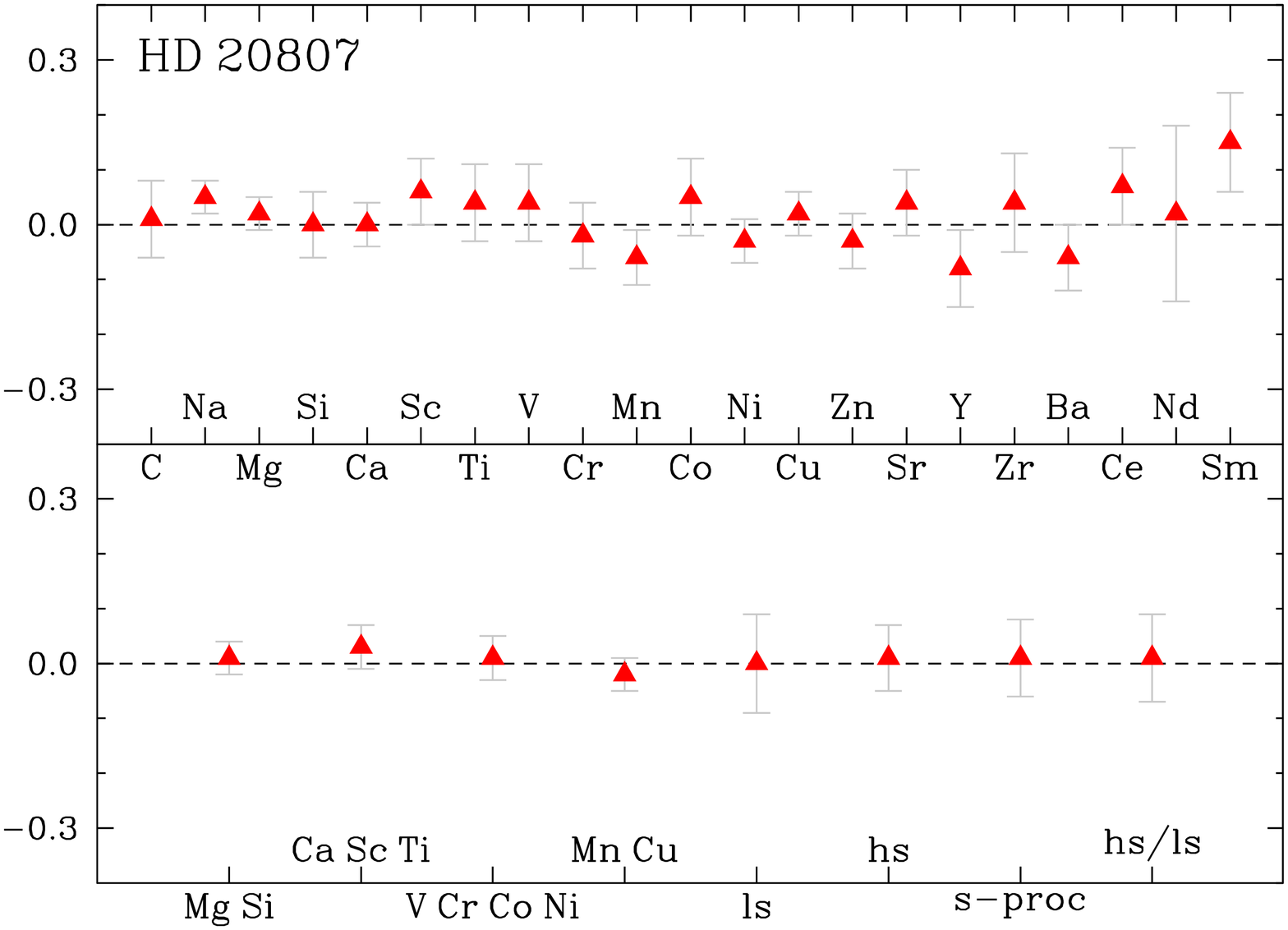}}
\end{minipage}
\begin{minipage}[b]{0.33\textwidth}
\centering
\resizebox{\hsize}{!}{\includegraphics[angle=-90]{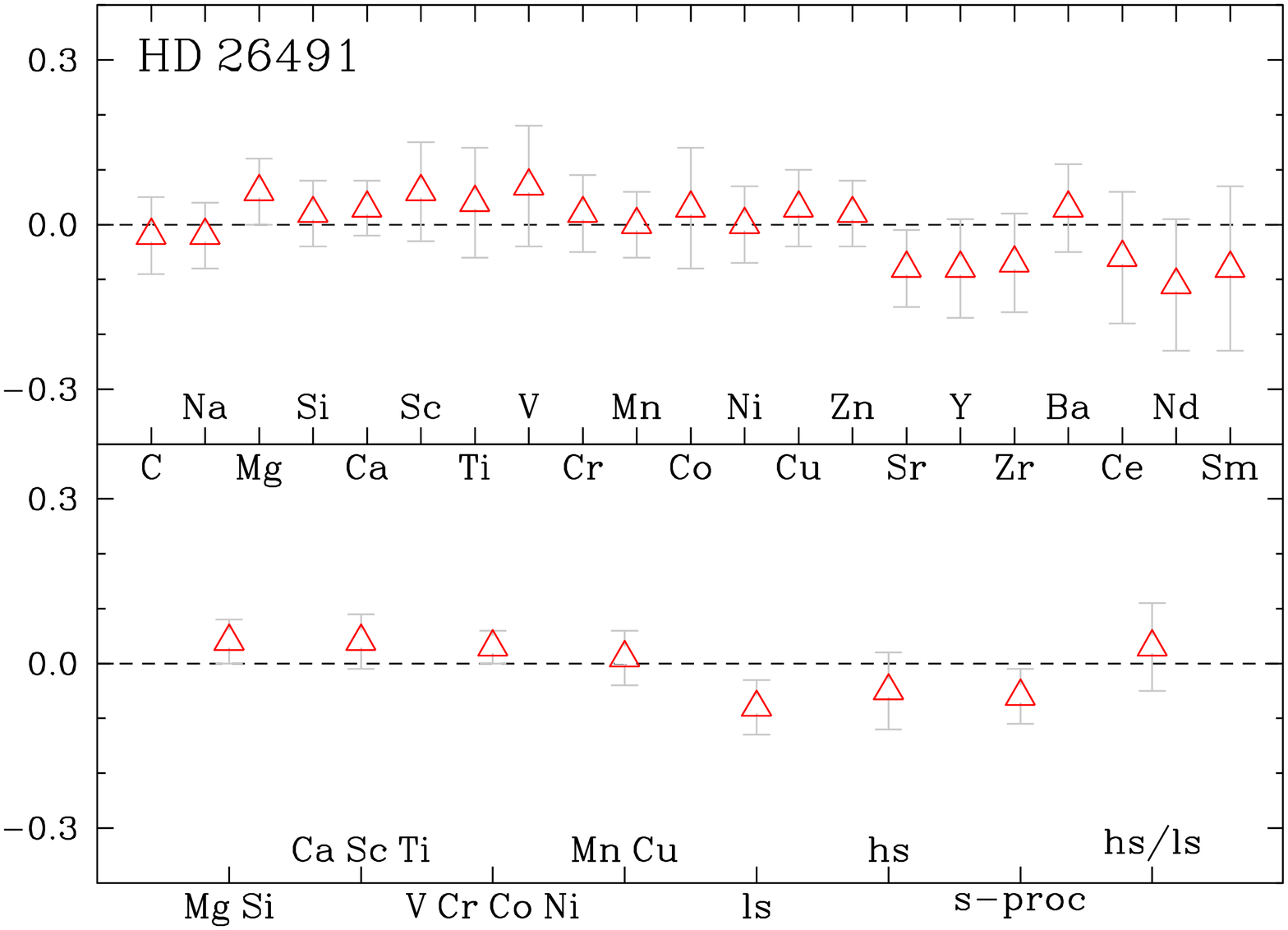}}
\end{minipage} \\
\begin{minipage}[b]{0.33\textwidth}
\centering
\resizebox{\hsize}{!}{\includegraphics[angle=-90]{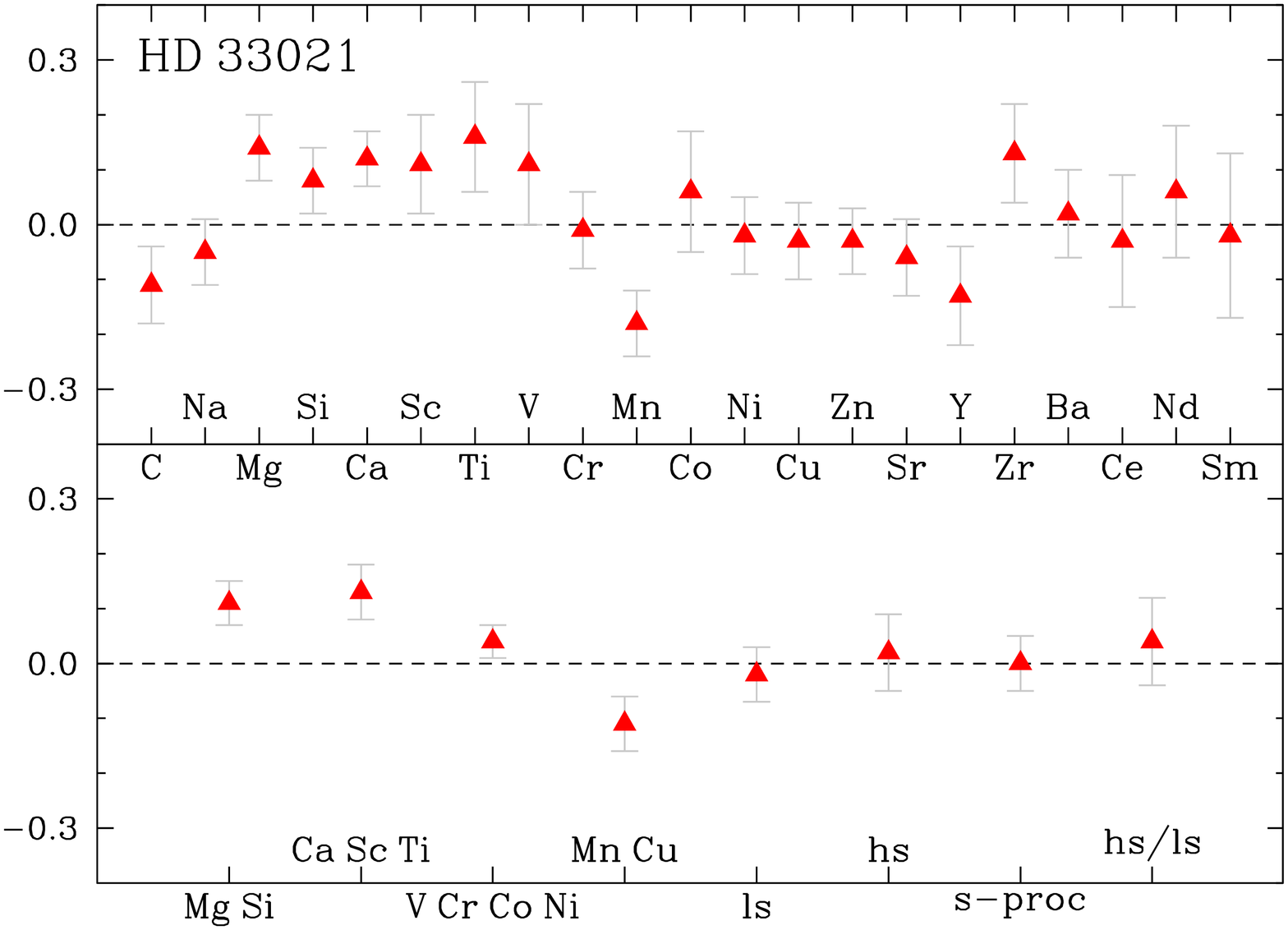}}
\end{minipage}
\begin{minipage}[b]{0.33\textwidth}
\centering
\resizebox{\hsize}{!}{\includegraphics[angle=-90]{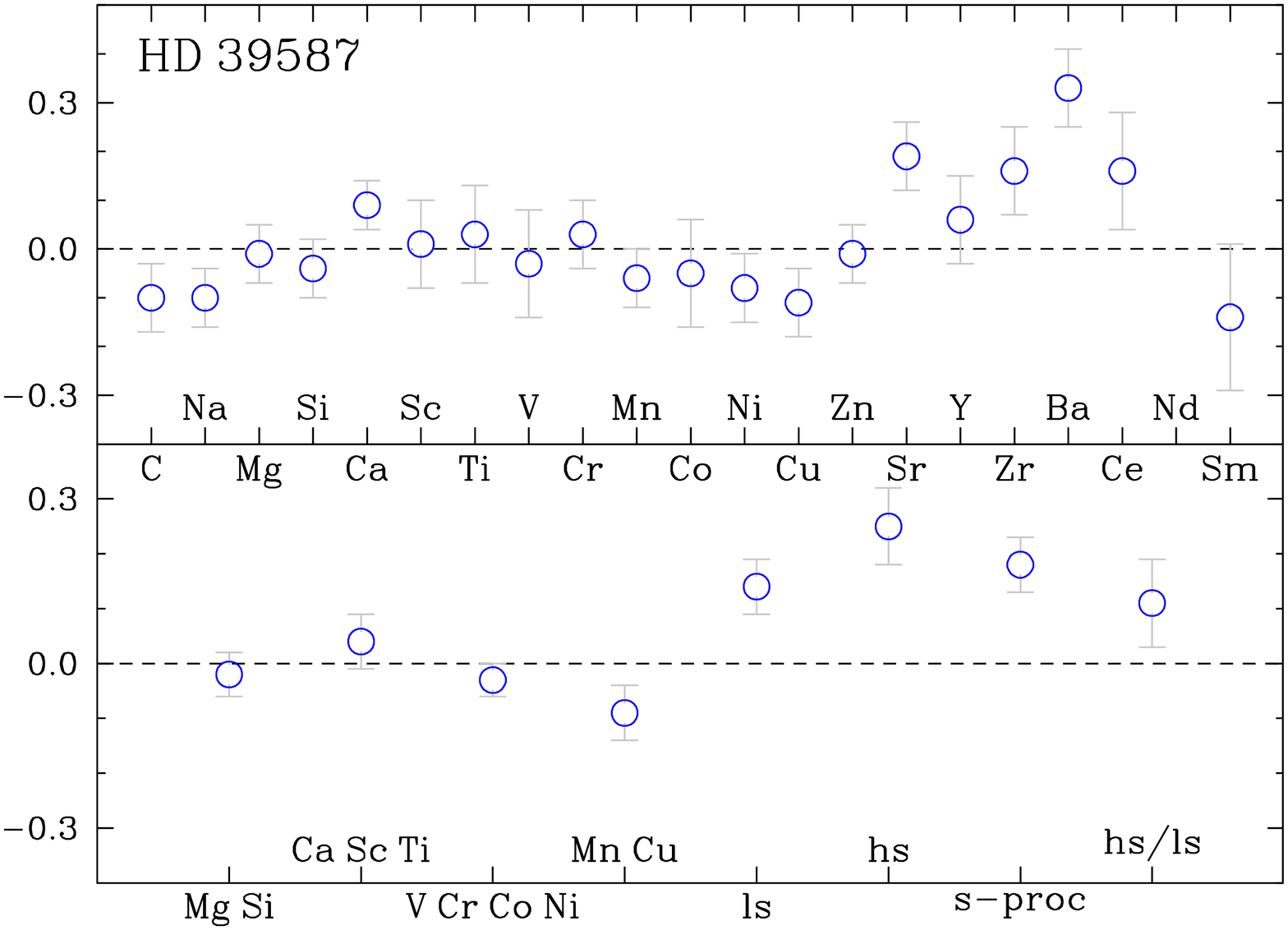}}
\end{minipage}
\begin{minipage}[b]{0.33\textwidth}
\centering
\resizebox{\hsize}{!}{\includegraphics[angle=-90]{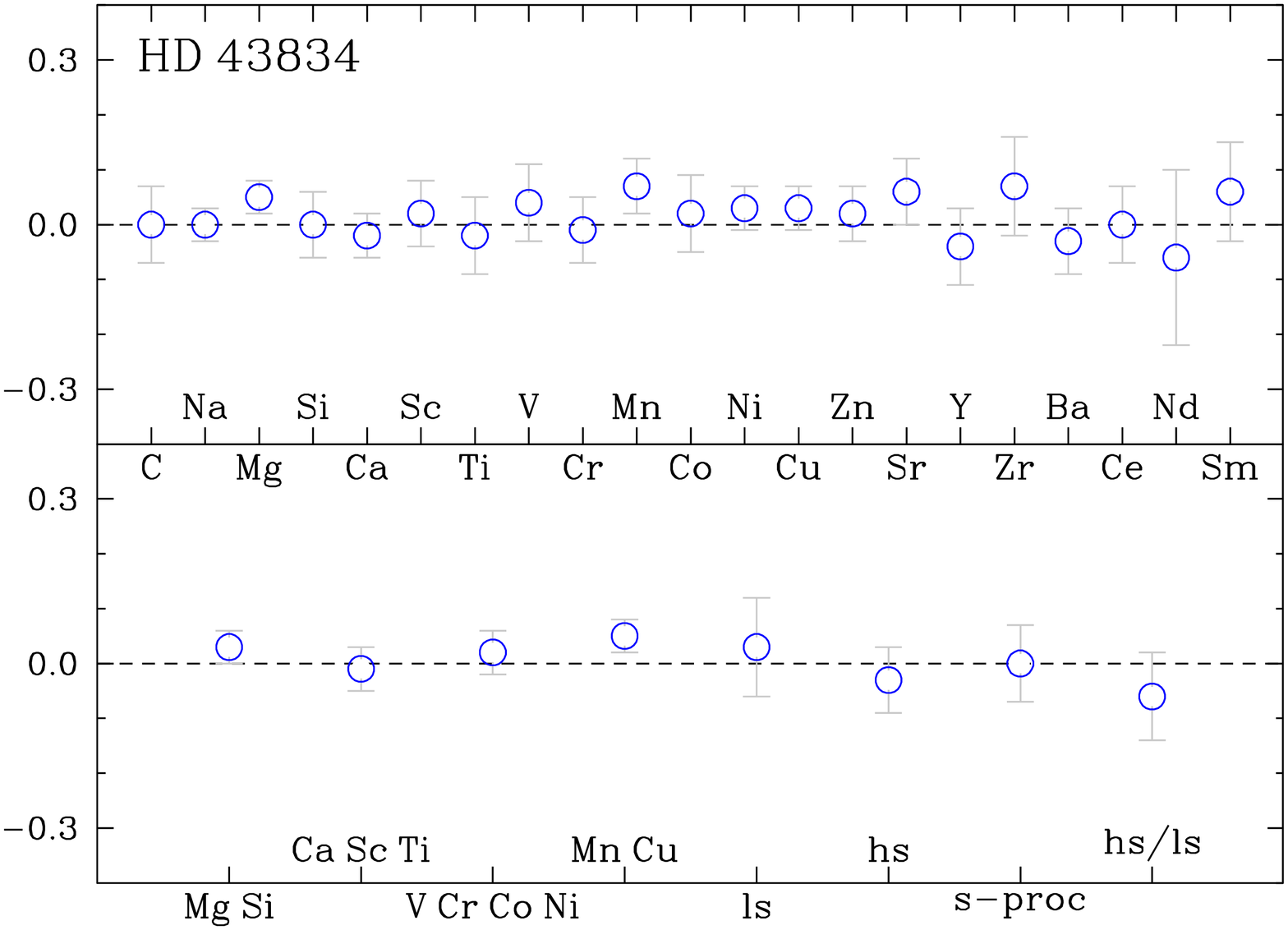}}
\end{minipage} \\
\begin{minipage}[b]{0.33\textwidth}
\centering
\resizebox{\hsize}{!}{\includegraphics[angle=-90]{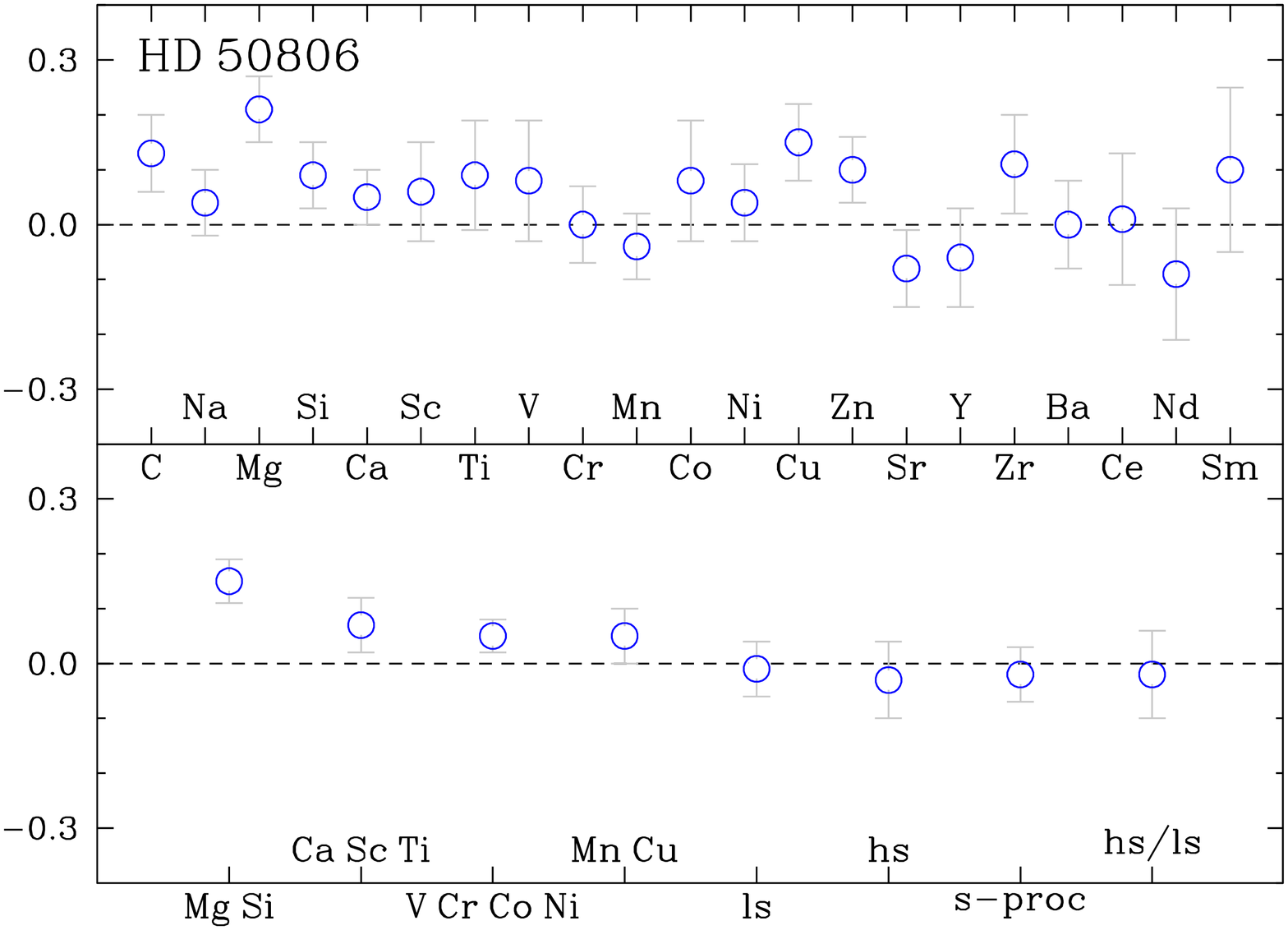}}
\end{minipage}
\begin{minipage}[b]{0.33\textwidth}
\centering
\resizebox{\hsize}{!}{\includegraphics[angle=-90]{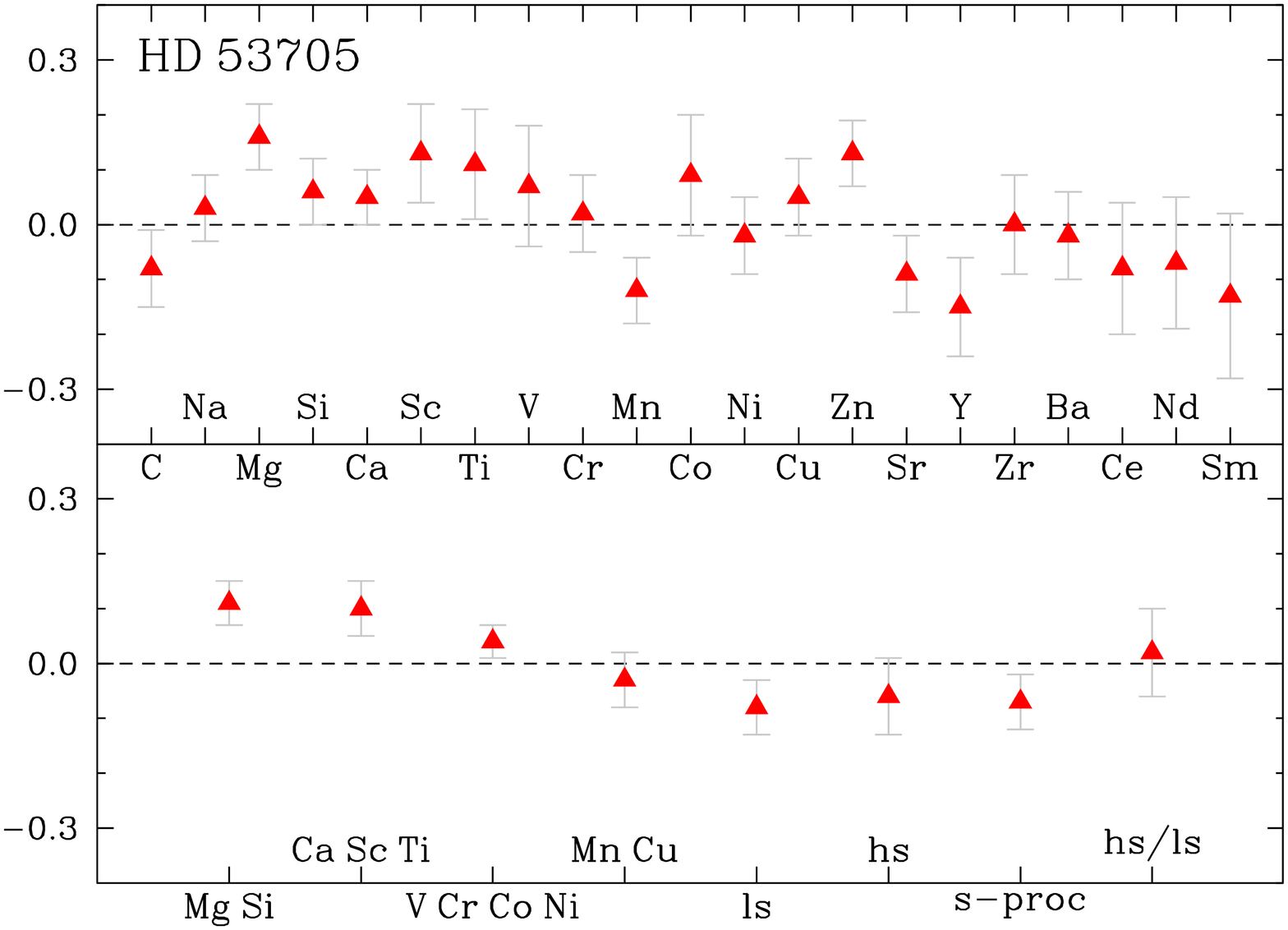}}
\end{minipage}
\begin{minipage}[b]{0.33\textwidth}
\centering
\resizebox{\hsize}{!}{\includegraphics[angle=-90]{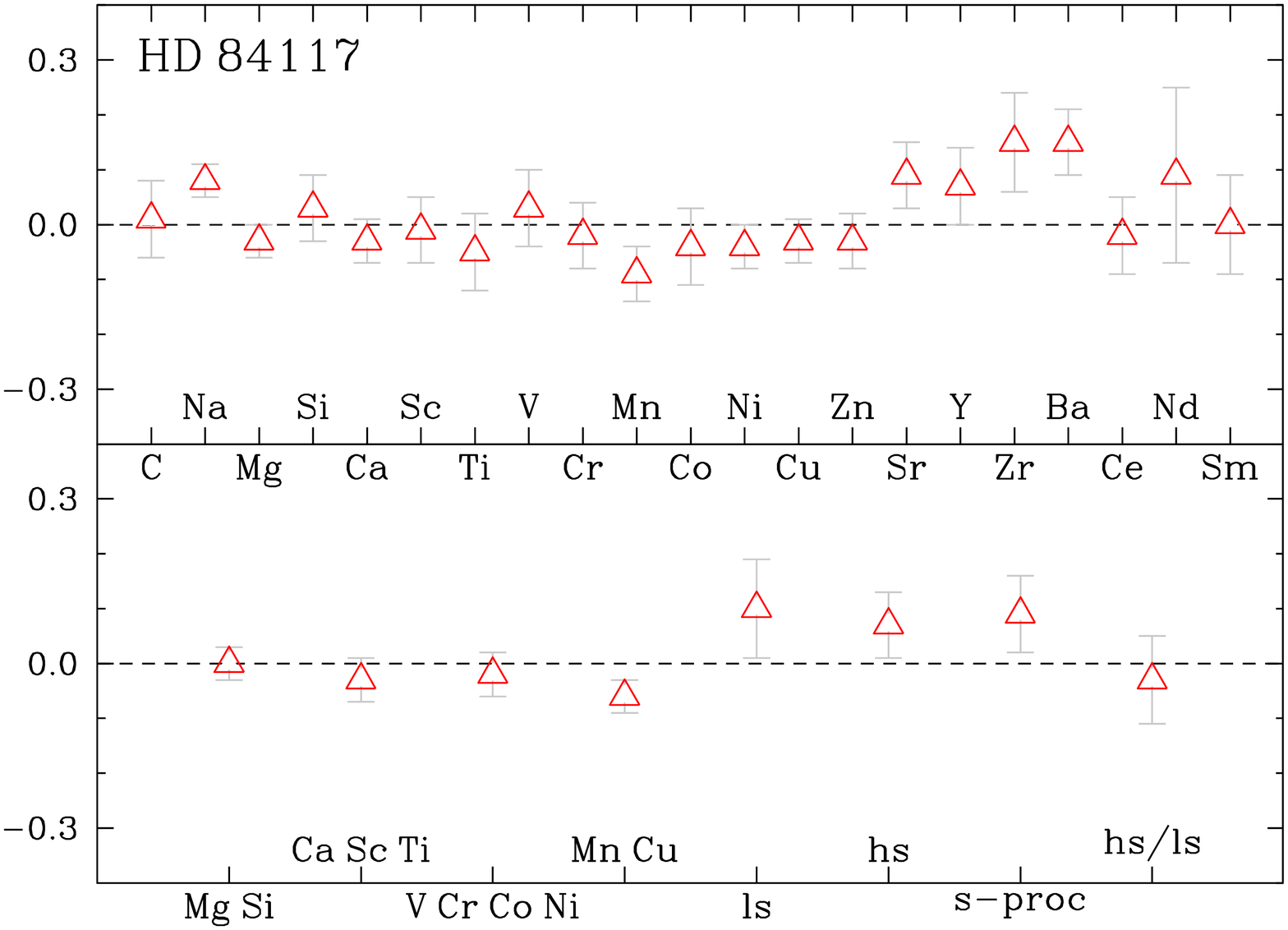}}
\end{minipage} \\
\begin{minipage}[b]{0.33\textwidth}
\centering
\resizebox{\hsize}{!}{\includegraphics[angle=-90]{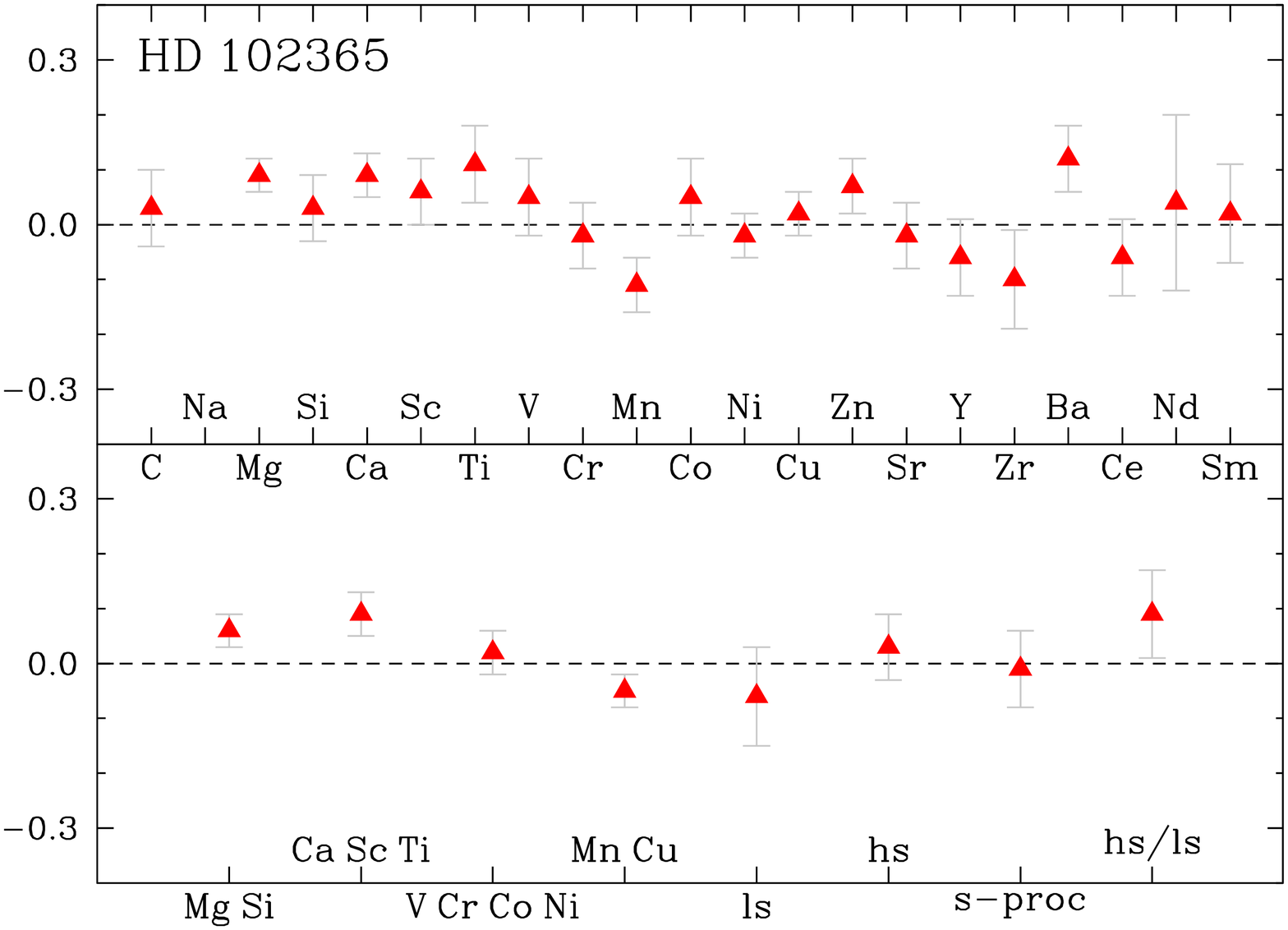}}
\end{minipage}
\begin{minipage}[b]{0.33\textwidth}
\centering
\resizebox{\hsize}{!}{\includegraphics[angle=-90]{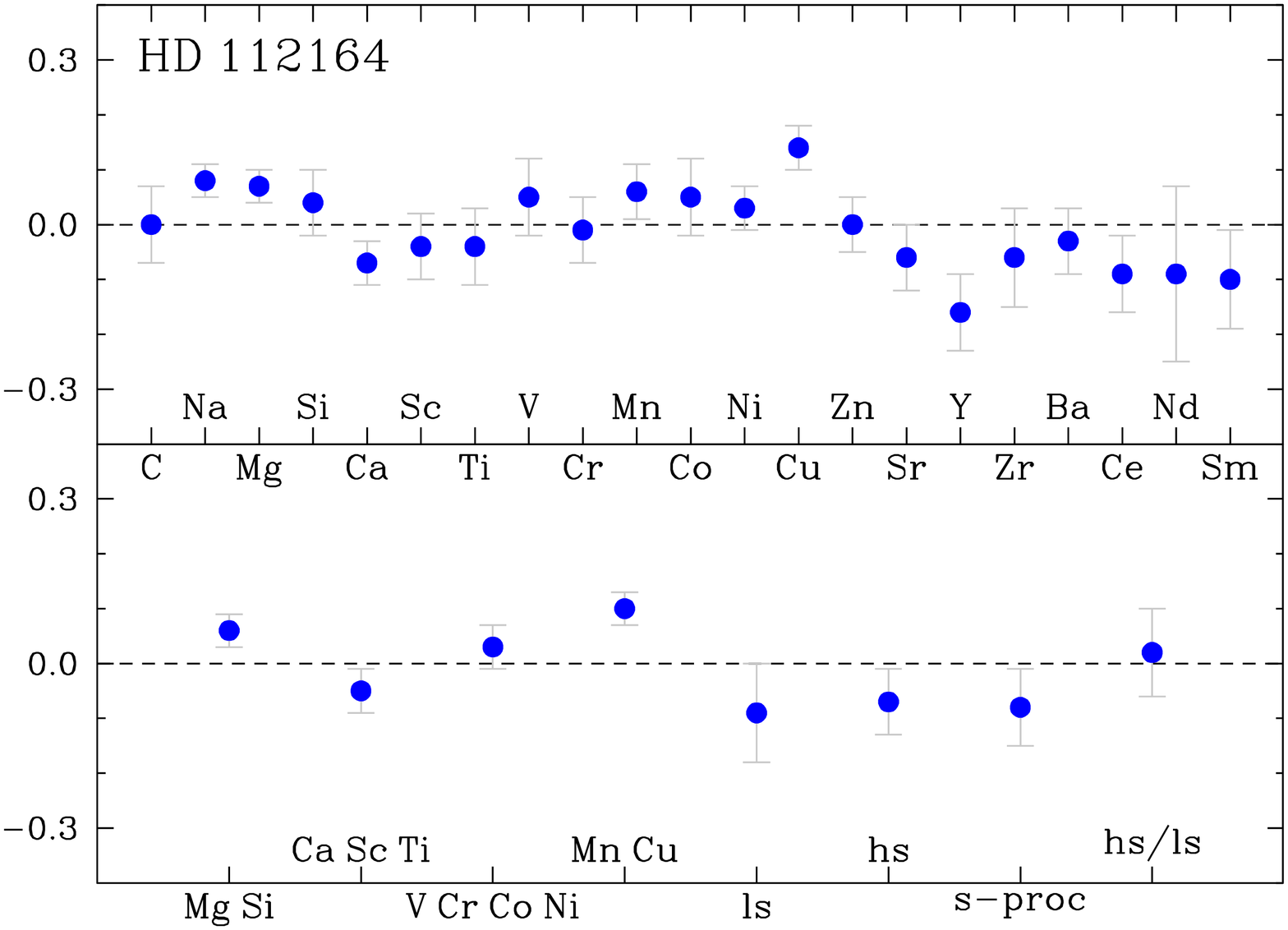}}
\end{minipage}
\begin{minipage}[b]{0.33\textwidth}
\centering
\resizebox{\hsize}{!}{\includegraphics[angle=-90]{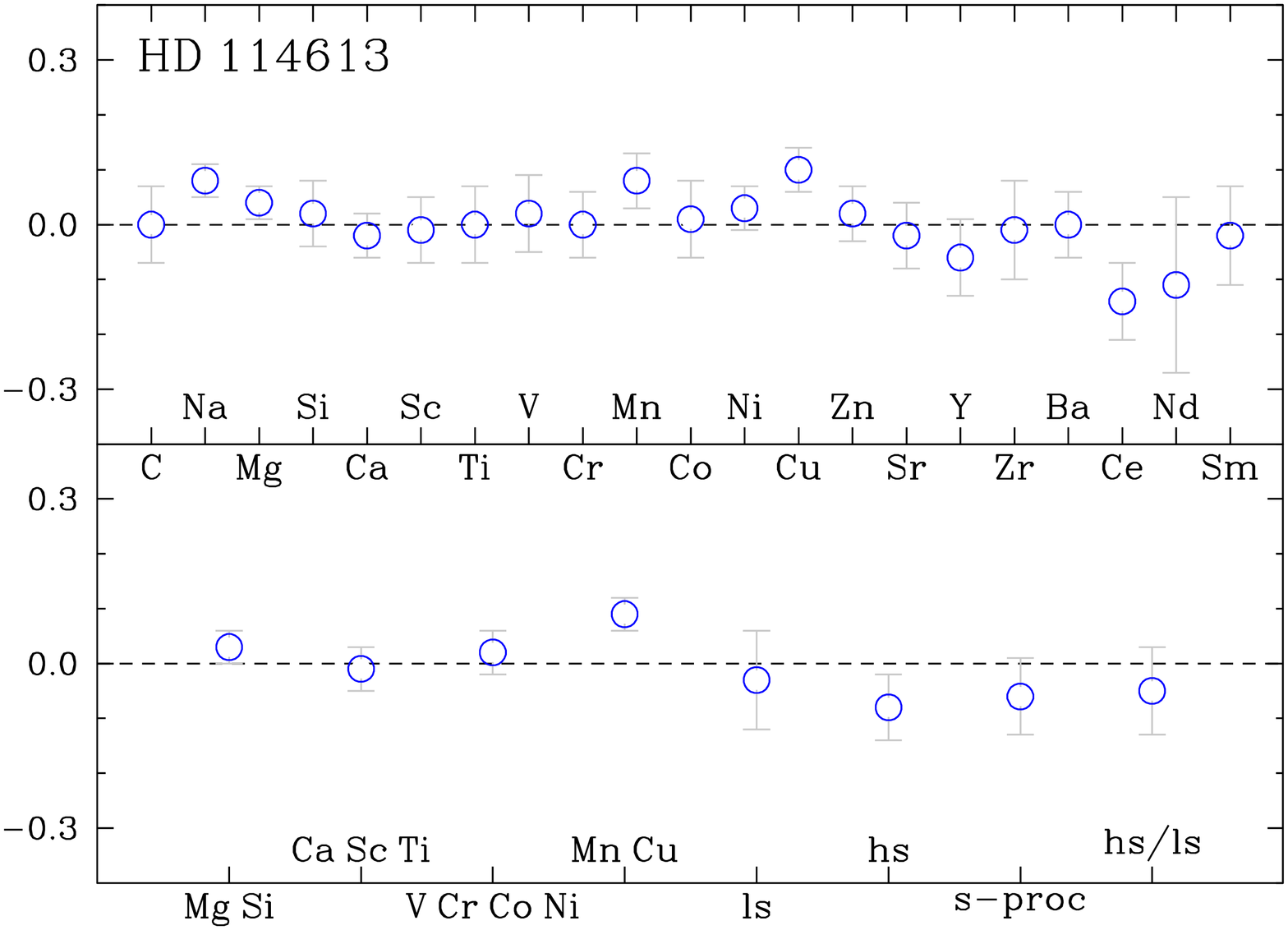}}
\end{minipage} \\
\begin{minipage}[b]{0.33\textwidth}
\centering
\resizebox{\hsize}{!}{\includegraphics[angle=-90]{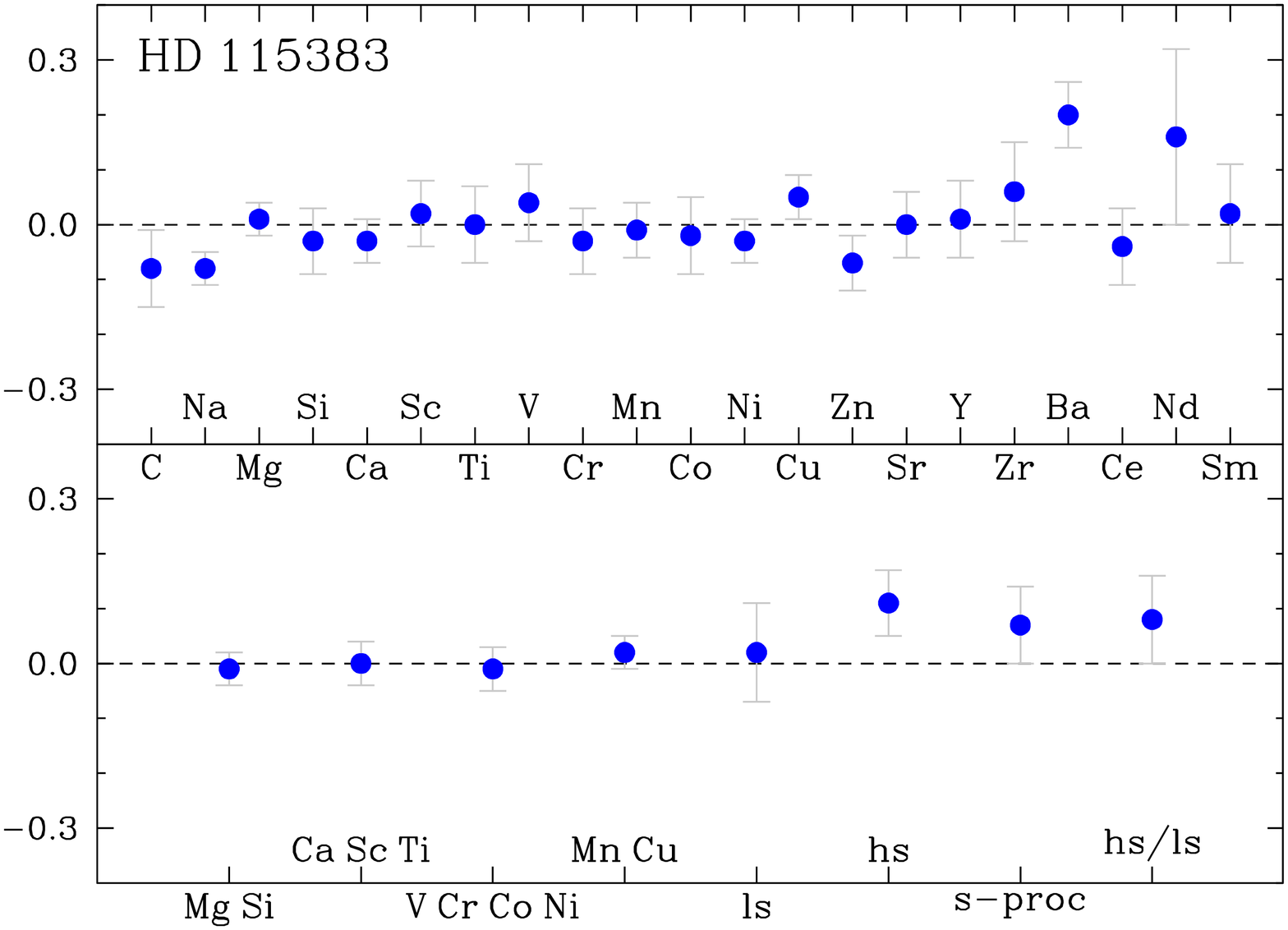}}
\end{minipage}
\begin{minipage}[b]{0.33\textwidth}
\centering
\resizebox{\hsize}{!}{\includegraphics[angle=-90]{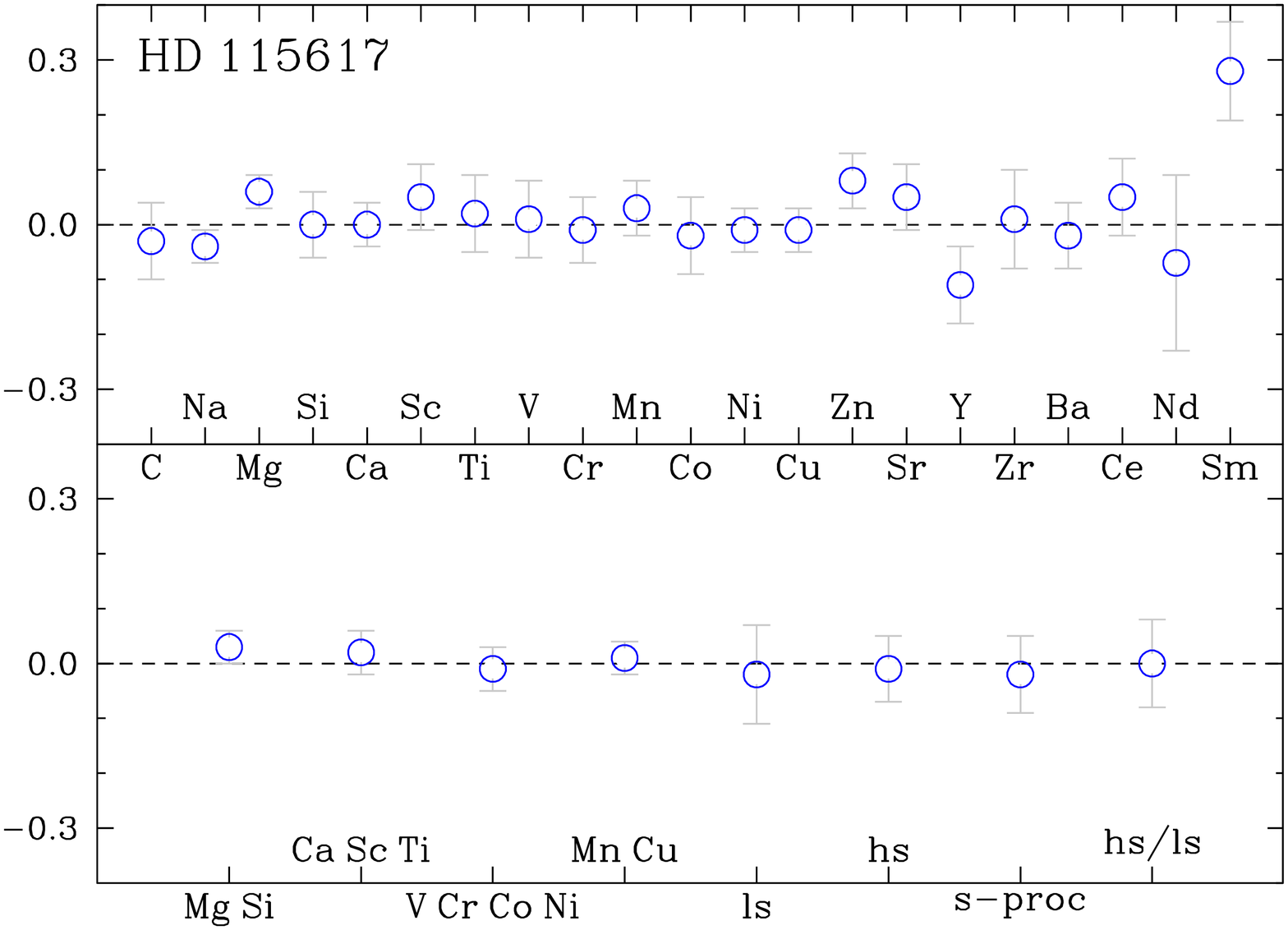}}
\end{minipage}
\begin{minipage}[b]{0.33\textwidth}
\centering
\resizebox{\hsize}{!}{\includegraphics[angle=-90]{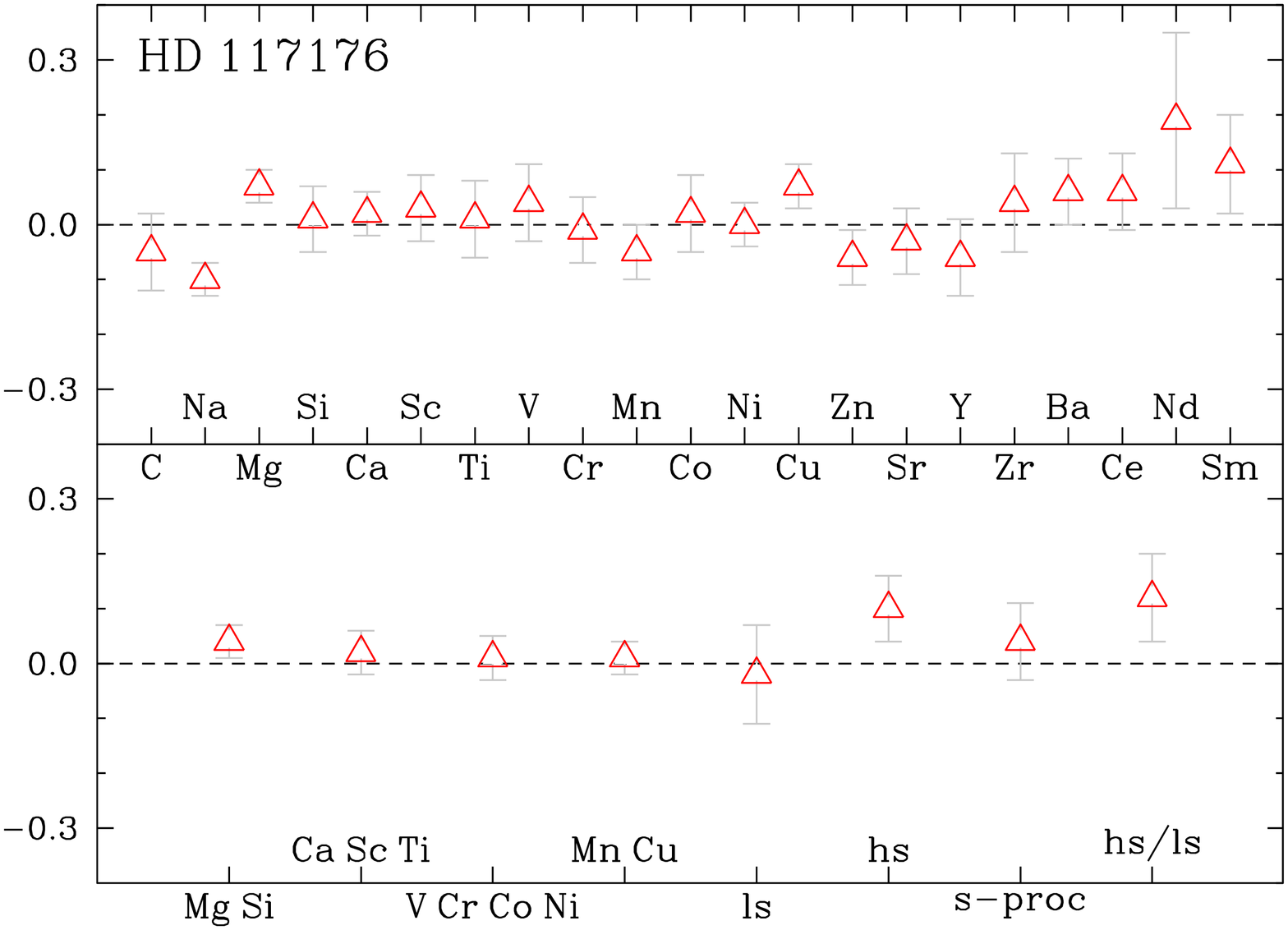}}
\end{minipage}
\caption{Abundance ratios [X/Fe] for individual elements (top of each panel)
         and \mxfe\ for the nucleosynthetic groups (bottom of each panel).
	 The uncertainties adopted are those listed in Tables
	 \ref{ab_err_comp} and \ref{mean_ab_err_comp}. The symbols represent
	 the stellar classification defined in Sect.~\ref{classtree}
	 (see Fig.~\ref{dendrogram}).}
\label{ab_ratios_fig}
\end{figure*}
\addtocounter{figure}{-1}
\begin{figure*}
\begin{minipage}[b]{0.33\textwidth}
\centering
\resizebox{\hsize}{!}{\includegraphics[angle=-90]{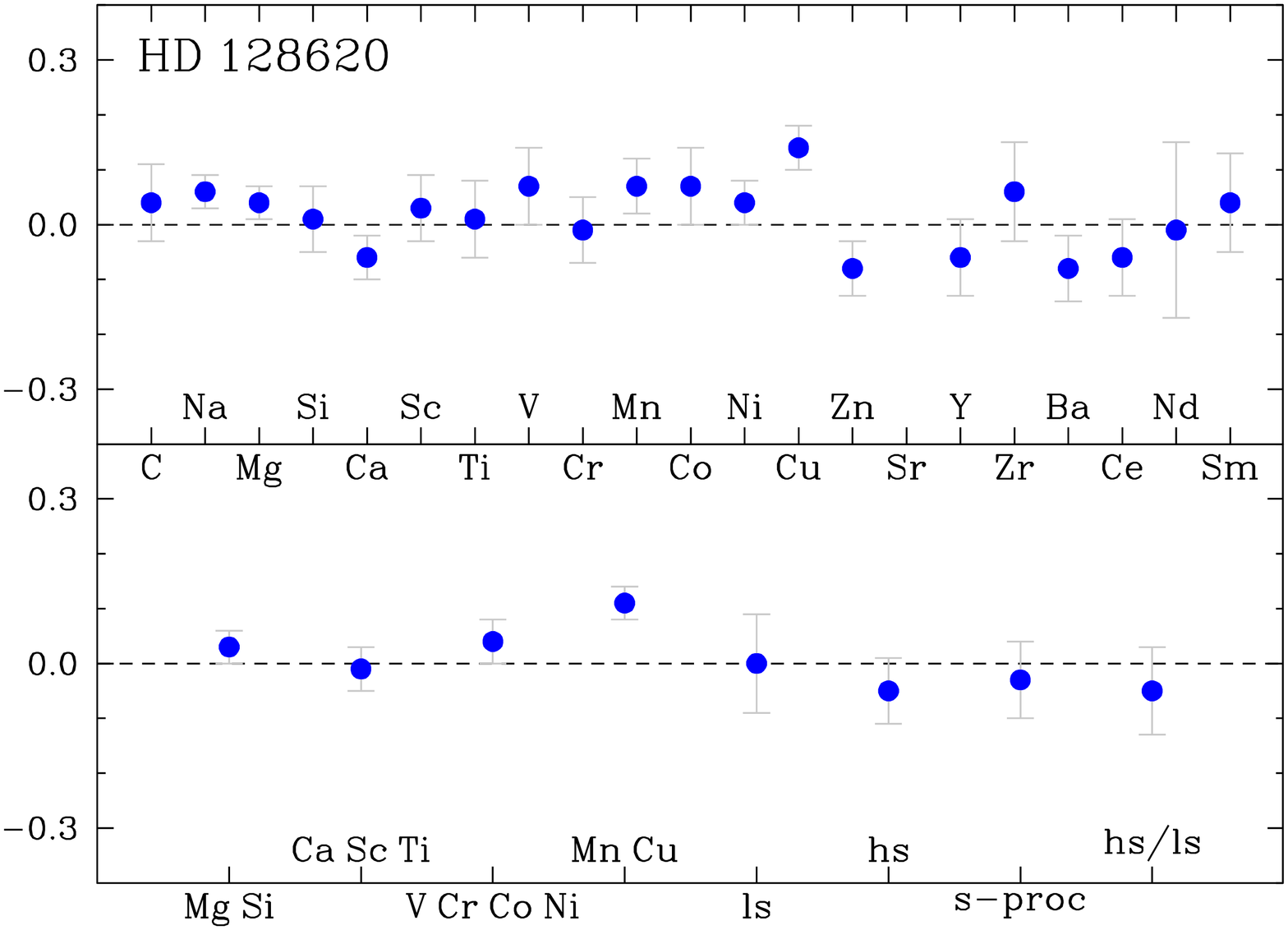}}
\end{minipage}
\begin{minipage}[b]{0.33\textwidth}
\centering
\resizebox{\hsize}{!}{\includegraphics[angle=-90]{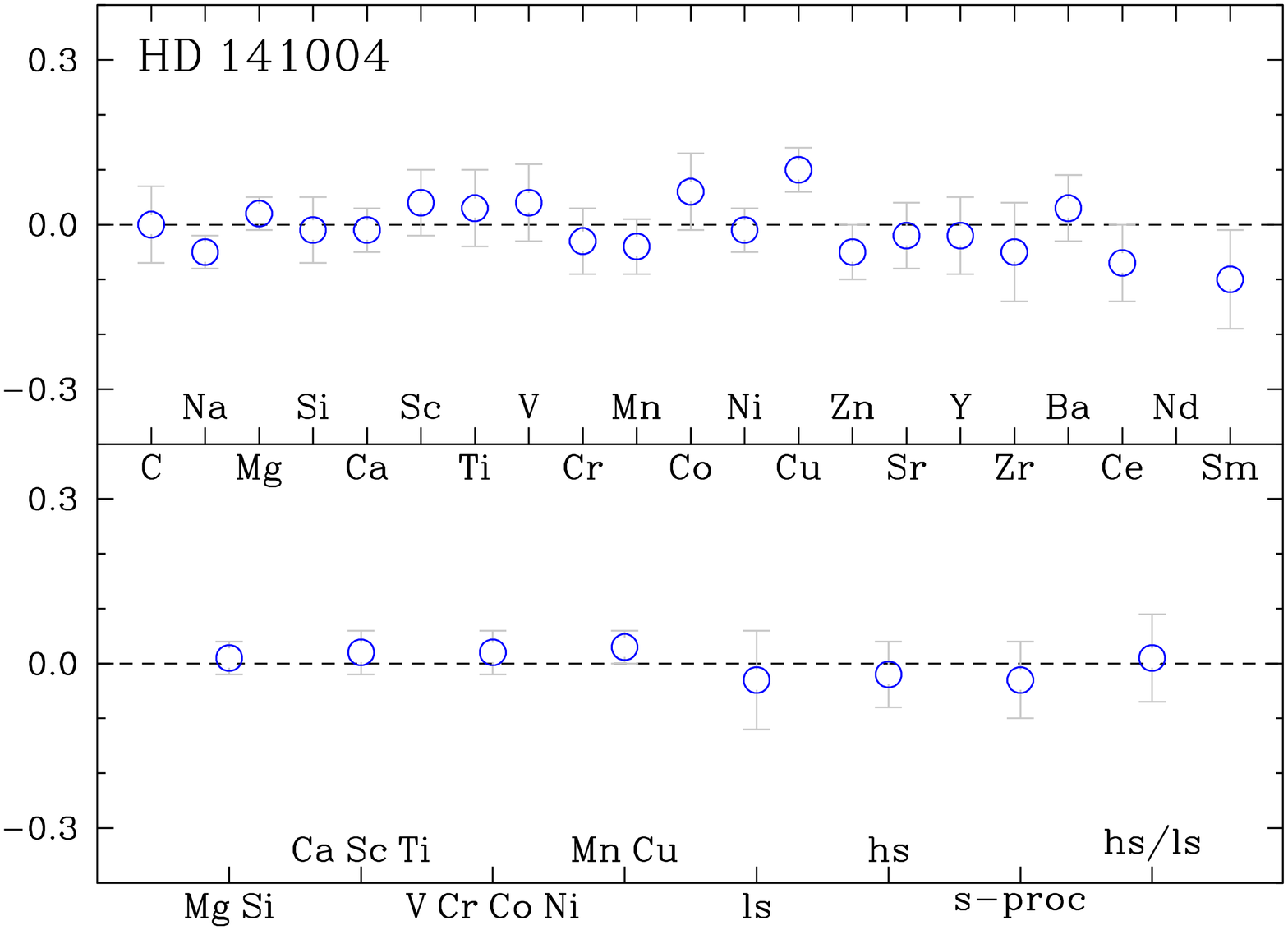}}
\end{minipage}
\begin{minipage}[b]{0.33\textwidth}
\centering
\resizebox{\hsize}{!}{\includegraphics[angle=-90]{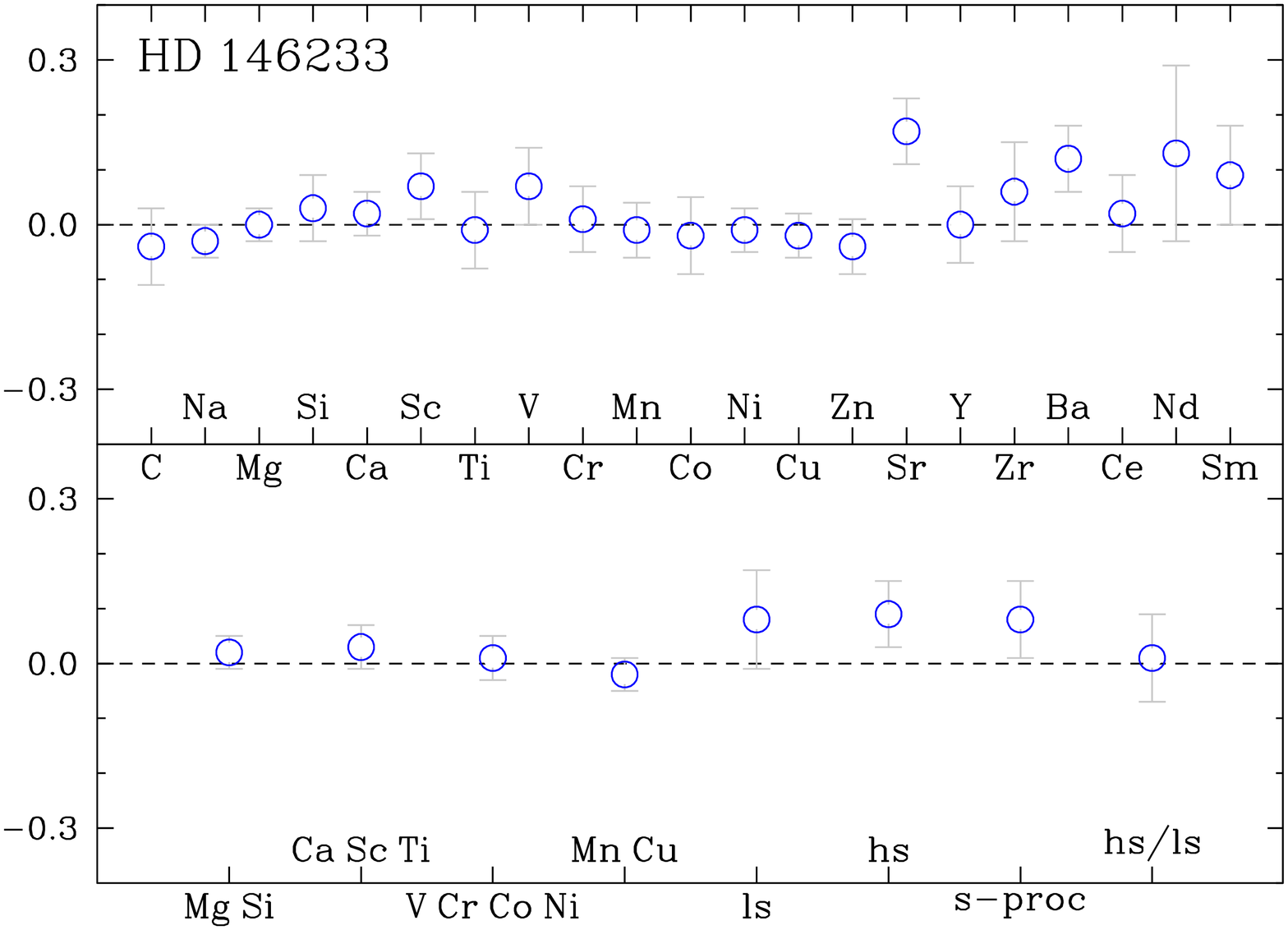}}
\end{minipage} \\
\begin{minipage}[b]{0.33\textwidth}
\centering
\resizebox{\hsize}{!}{\includegraphics[angle=-90]{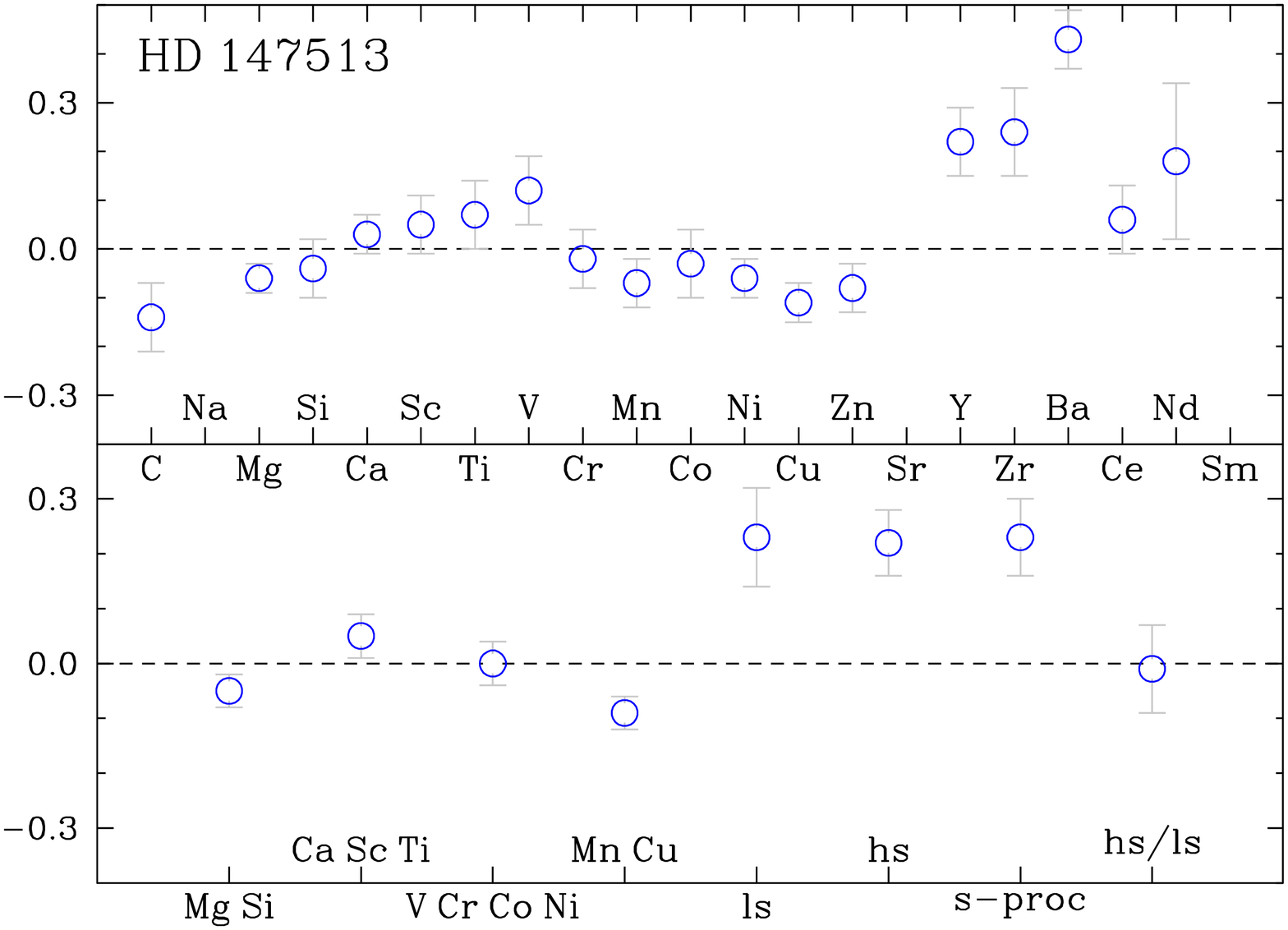}}
\end{minipage}
\begin{minipage}[b]{0.33\textwidth}
\centering
\resizebox{\hsize}{!}{\includegraphics[angle=-90]{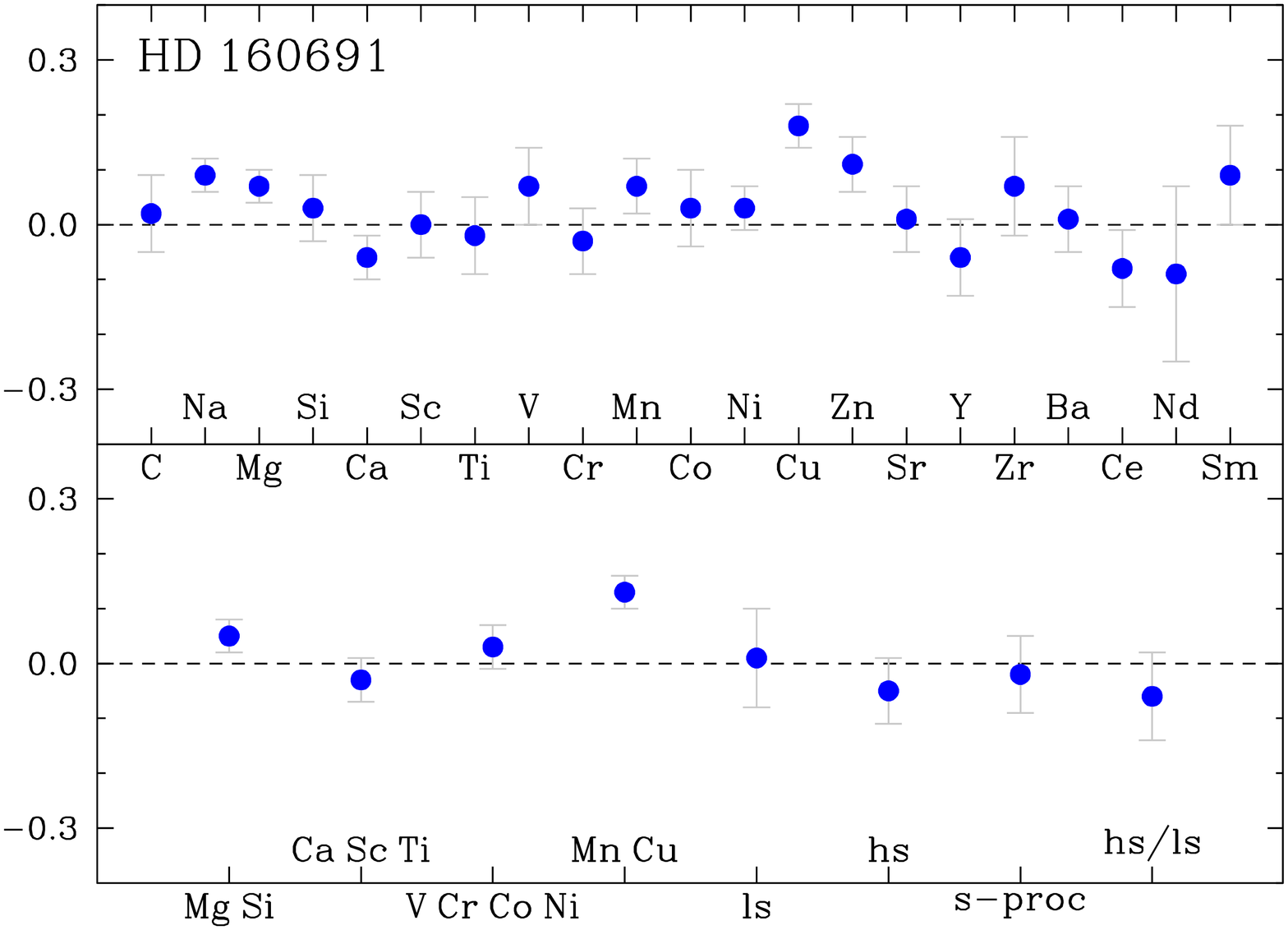}}
\end{minipage}
\begin{minipage}[b]{0.33\textwidth}
\centering
\resizebox{\hsize}{!}{\includegraphics[angle=-90]{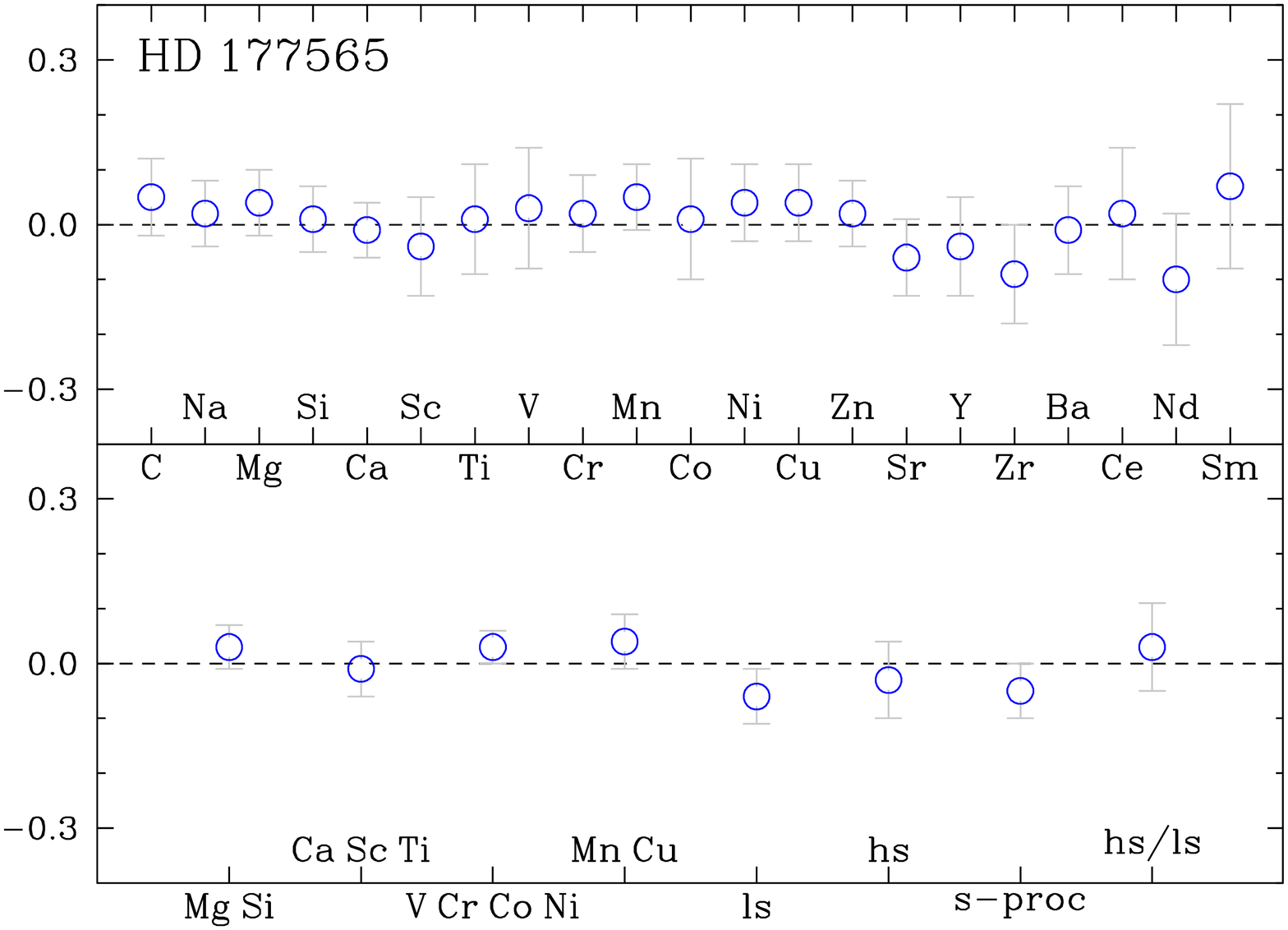}}
\end{minipage} \\
\begin{minipage}[b]{0.33\textwidth}
\centering
\resizebox{\hsize}{!}{\includegraphics[angle=-90]{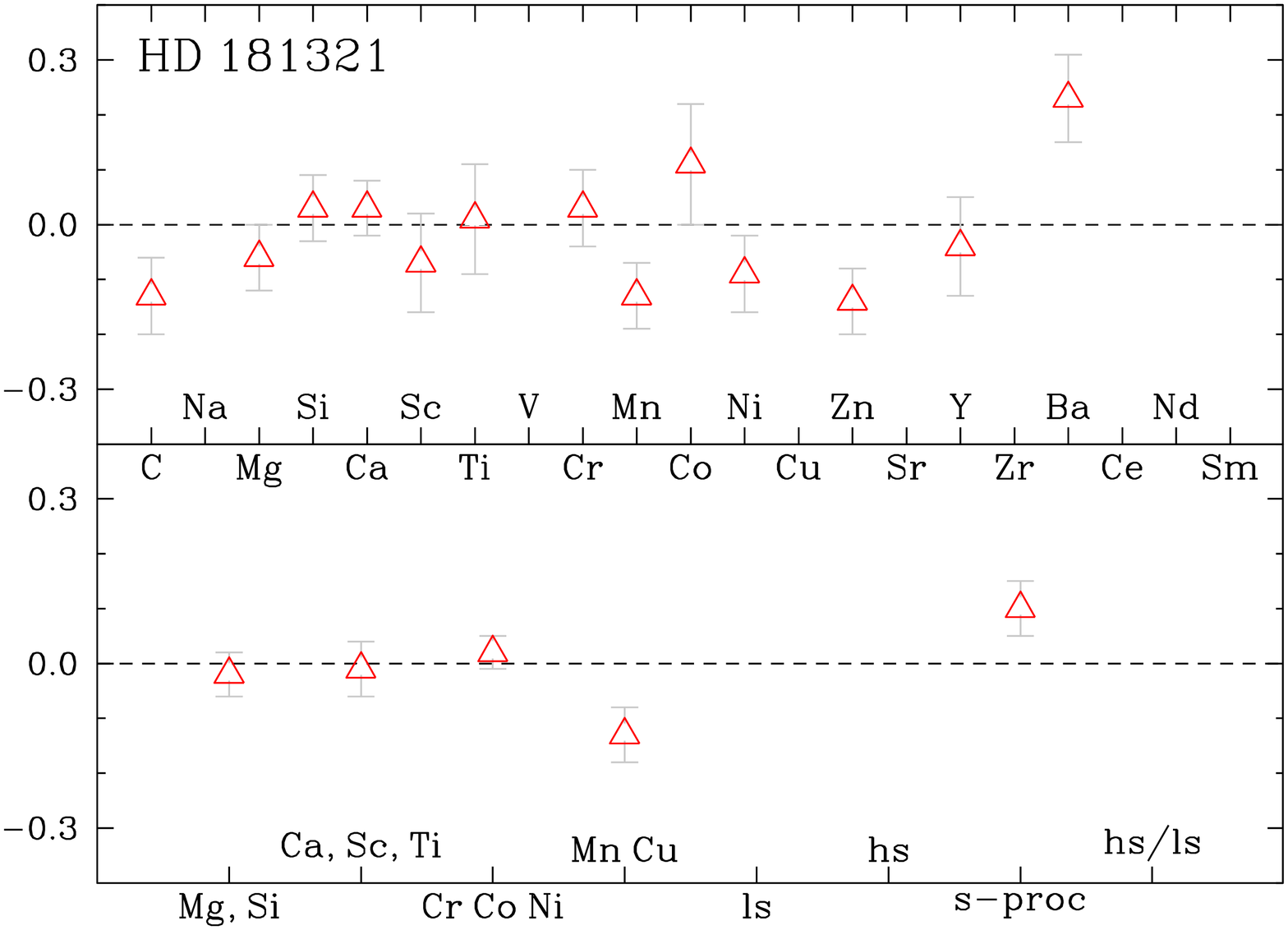}}
\end{minipage}
\begin{minipage}[b]{0.33\textwidth}
\centering
\resizebox{\hsize}{!}{\includegraphics[angle=-90]{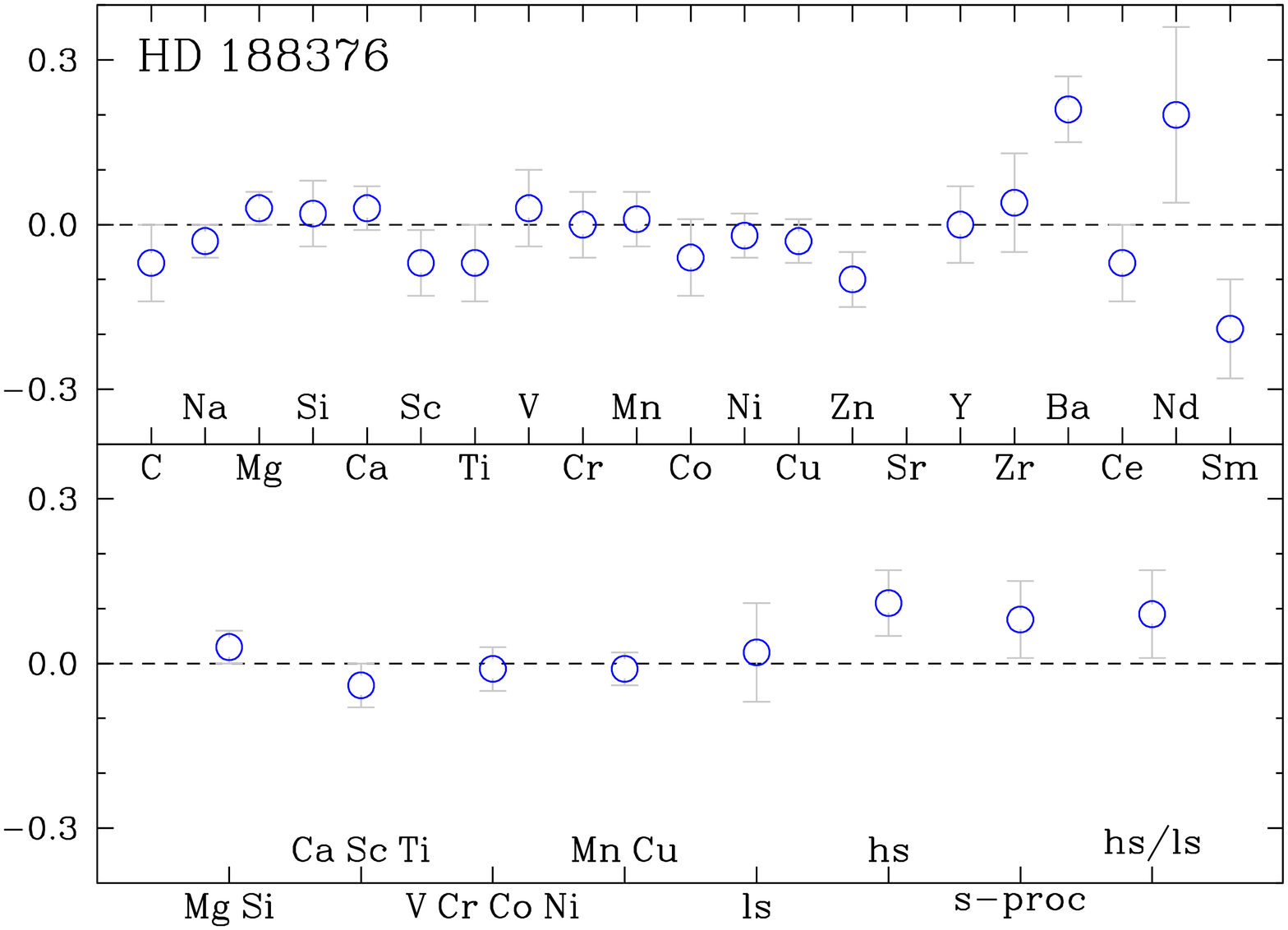}}
\end{minipage}
\begin{minipage}[b]{0.33\textwidth}
\centering
\resizebox{\hsize}{!}{\includegraphics[angle=-90]{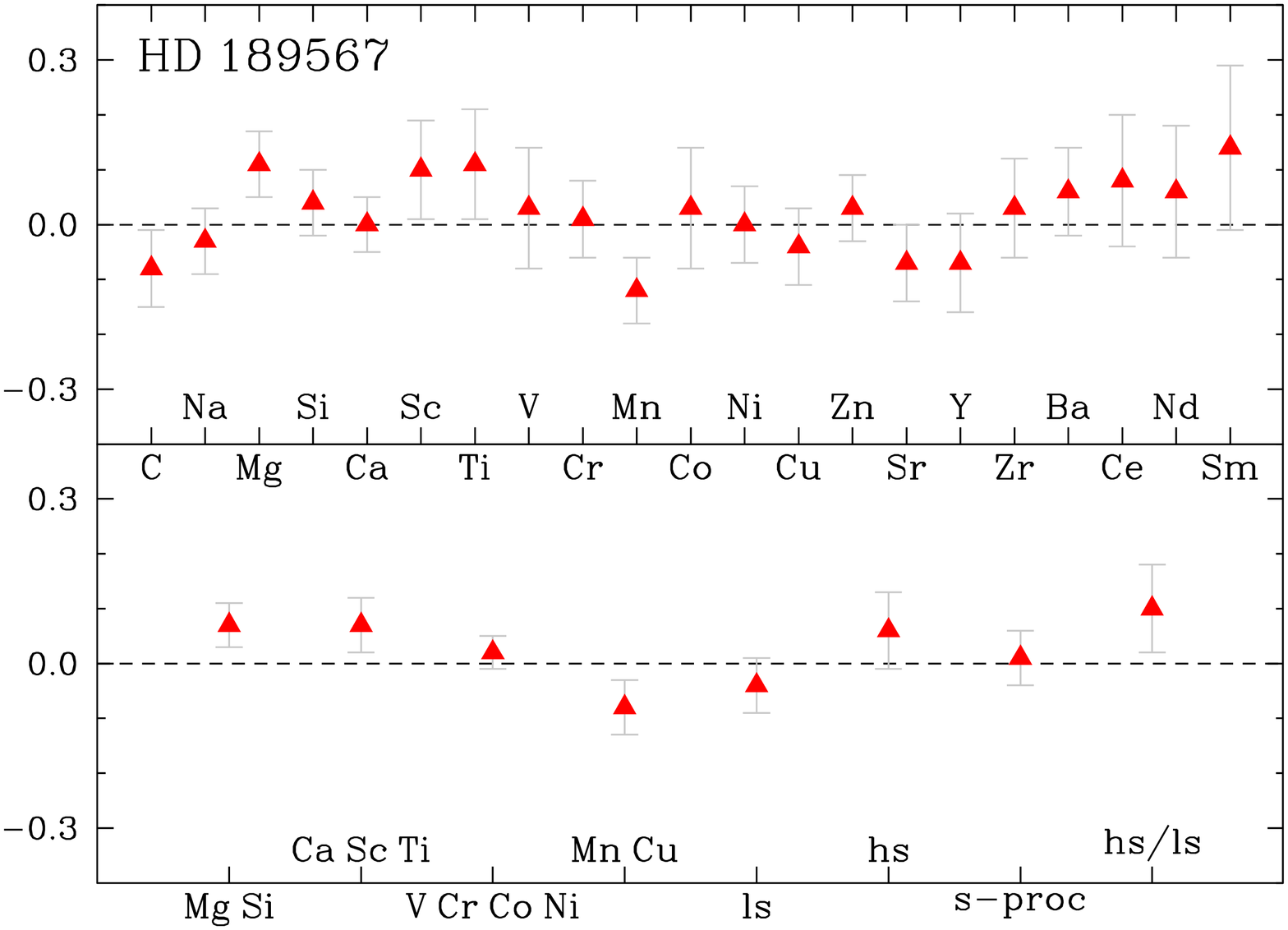}}
\end{minipage} \\
\begin{minipage}[b]{0.33\textwidth}
\centering
\resizebox{\hsize}{!}{\includegraphics[angle=-90]{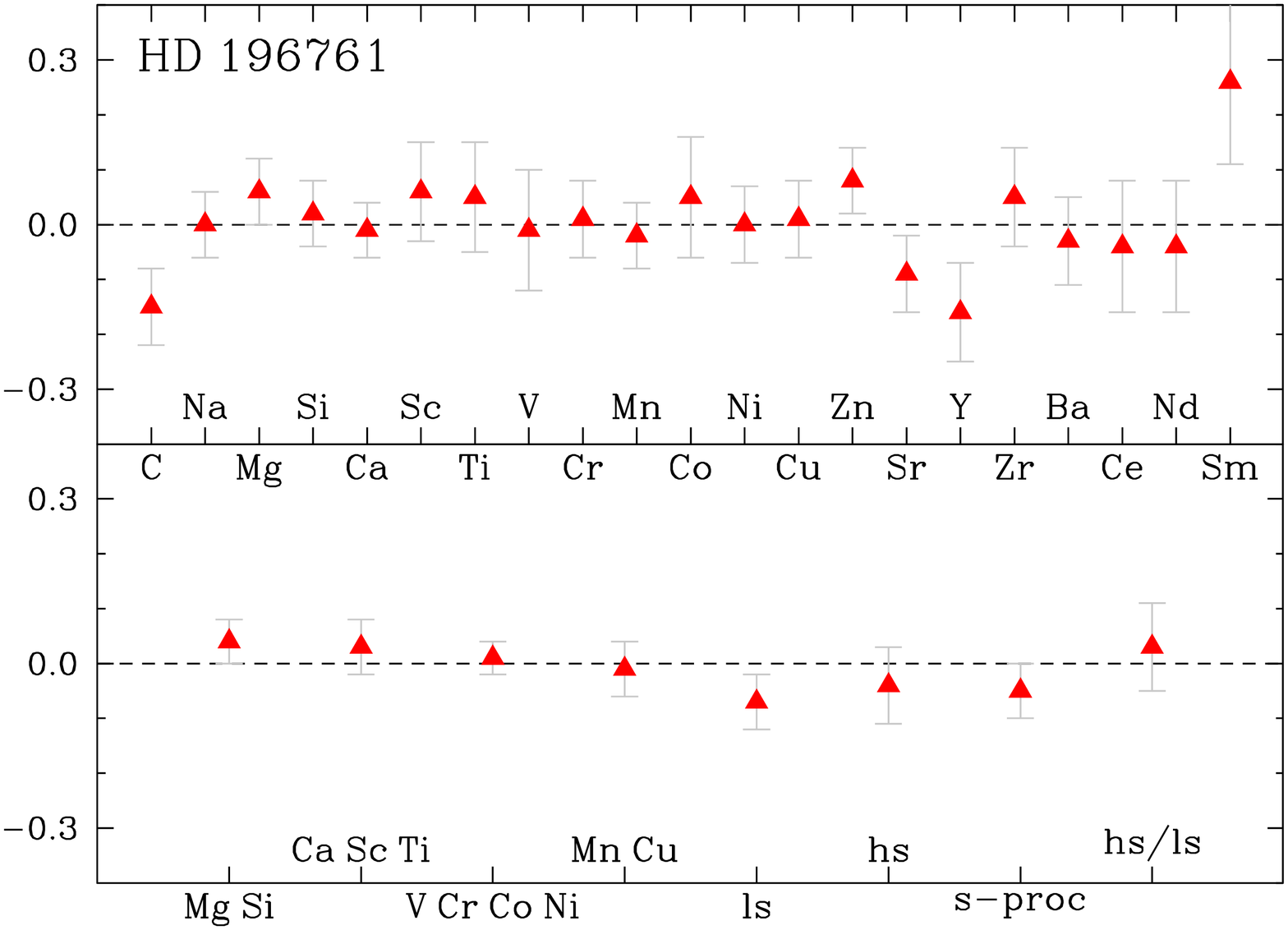}}
\end{minipage}
\caption{continued.}
\end{figure*}

It is also very worthwhile to investigate the chemical abundances in stars
that share similar values of age, metallicity, and Galactic orbit (\Rm\ and
$e$). These subgroup of stars are supposed to share the same physical
conditions of the Galaxy at the time and galactocentric position of their
birth. An example of this includes the stars HD\,43834, HD\,84117,
HD\,141004, and HD\,146233, which also have ages, metallicities, and
Galactic orbits close to the solar values (they were all classified as
intermediate abundance stars in the tree clustering analysis, one in slight
underabundance and the three others in slight overabundance with respect to
the Sun). In spite of this, only HD\,43834 and HD\,141004 show solar
abundances, within the uncertainties, for all (or almost all) elements.
HD\,84117 is deficient in Mn and enriched in Na and in elements of the
s-process (Sr, Y, Zr, and Ba). HD\,146233, proposed by
\citet{PortodeMellodaSilva1997b} as the closest solar twin ever known at
that time, is actually (as also proposed by these authors) enriched in some
elements of the s-process (Sr and Ba) and possibly enriched in Sc, V, and
Sm. Thus, possibly, the investigation of a larger sample of stars at a
similar level of detail as done here could reveal that non-solar abundance
ratios are present even for stars sharing the same place and time of birth.
Whether this reflects intrinsic heterogeneities in their natal interstellar
clouds (in principle a reasonable hypothesis since the elements reflecting
non-solar ratios are related to different nucleosynthetic processes
operating in different timescales) or else is evidence for considerable
radial migration in the Galaxy is a question we plan to address in a
subsequent work involving a larger sample.

Still concerning the relations involving the stellar groups from the
clustering analysis and the kinematic and orbital parameters of our sample,
we can see in Fig.~\ref{cin_diag} that the group of metal-poor and old stars
seems to have larger velocities in the direction of the Galactic centre
($\vert$\Ulsr$\vert >$ 40~\kms) and larger eccentricities ($e \gtrsim 0.2$)
than the other stars. The star HD\,50806 appears to have a singular position
in this figure (in particular, it has the most eccentric orbit among the
sample stars), which is probably related to its membership in the transition
population of thin-thick disc stars. The limitation of our sample does not
allow to verify the results of \citet{RochaPintoetal2006} that metal-poor
and old stars show more orbital radial spread in the Galaxy.


\subsection{Abundance trends as a function of [Fe/H]}

\begin{figure*}
\centering
\begin{minipage}[b]{0.98\textwidth}
\centering
\resizebox{\hsize}{!}{\includegraphics{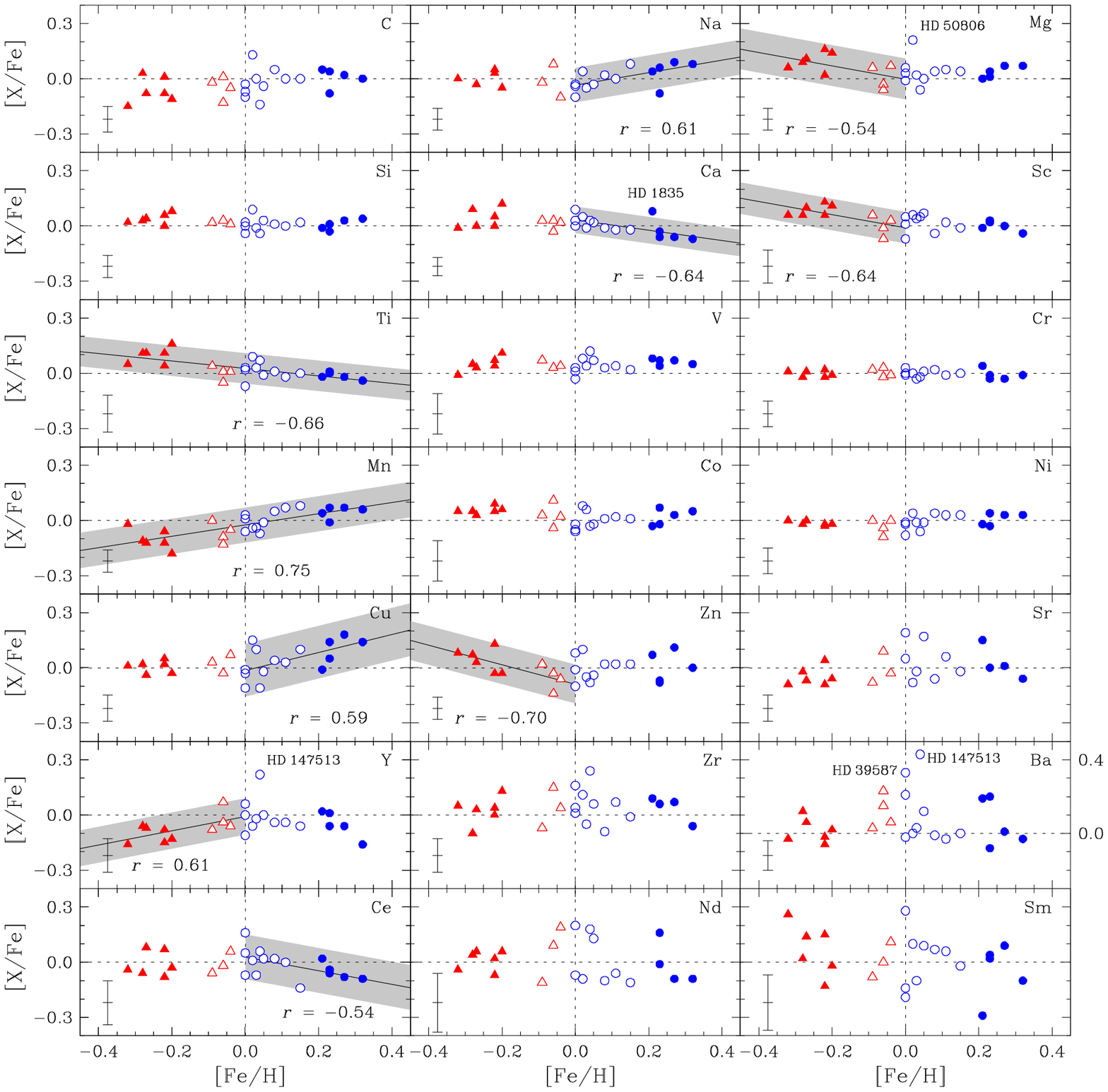}}
\end{minipage} \\[-0.17cm]
\begin{minipage}[b]{0.98\textwidth}
\centering
\resizebox{\hsize}{!}{\includegraphics{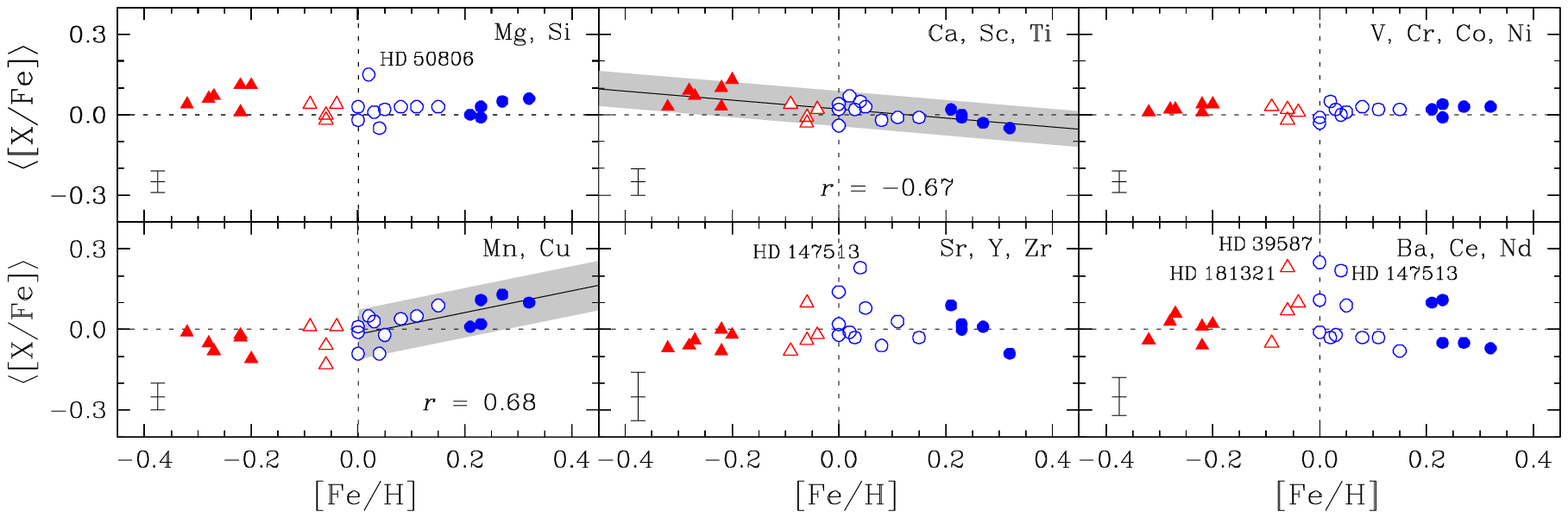}}
\end{minipage}
\caption{Abundance ratios as a function of the stellar metallicity for
         individual elements (top panels) and for nucleosynthetic groups
	 (bottom panels). The ordinate axis has a different scale for
	 [Ba/Fe] due to its larger abundances. The linear regressions (solid
	 line), the 95\% confidence intervals (hashed area), and the
	 cross-correlation coefficients are also shown for $|\,r\,|
	 \geq 0.5$. The symbols follow the classification of
	 Sect.~\ref{classtree} (see Fig.~\ref{dendrogram}).}
\label{ab_feh}
\end{figure*}

Through the analysis of Fig.~\ref{ab_feh} we investigate possible trends in
the abundance ratios as a function of the stellar metallicity. These trends
are more clearly identified if the elements are grouped together, either
based on their nucleosynthetic origin or because they share similar trends
in the diagrams. For this reason, we show in the bottom panels of this
figure the mean abundance \mxfe\ of a few groups of elements as a function
of the metallicity (the same groups plotted in Fig.~\ref{ab_ratios_fig} and
listed in Table~\ref{mean_ab_err_comp}). The stars are also identified
according to the tree clustering analysis. We have fitted linear regressions
on the diagrams in three ranges of metallicity: for stars poorer than the
Sun, for stars of solar metallicity or richer, and for all the sample stars.
We have then computed the cross-correlation coefficients in these three
metallicity ranges and plotted the regressions of the more significant
trends (only if $|\,r\,| \geq 0.5$).

\begin{table}
\centering
\caption[]{Carbon abundance ratios derived from the spectral synthesis of
           the two atomic lines ($\lambda$5052.2 and $\lambda$5380.3) and
	   the two \C2\ band heads ($\lambda$5128 and $\lambda$5165)
	   studied. Column 6 shows the final values adopted.}
\label{carb_ab}
\begin{tabular}{l r@{}l r@{}l r@{}l r@{}l r@{}l}
\hline\hline\noalign{\smallskip}
Star &
\multicolumn{2}{c}{\parbox[c]{0.85cm}{\centering [C/Fe] \\ {\tiny $\lambda 5052$}}} &
\multicolumn{2}{c}{\parbox[c]{0.85cm}{\centering [C/Fe] \\ {\tiny $\lambda 5380$}}} &
\multicolumn{2}{c}{\parbox[c]{0.85cm}{\centering [C/Fe] \\ {\tiny $\lambda 5128$}}} &
\multicolumn{2}{c}{\parbox[c]{0.85cm}{\centering [C/Fe] \\ {\tiny $\lambda 5165$}}} &
\multicolumn{2}{c}{\parbox[c]{0.85cm}{\centering [C/Fe]}} \\
\noalign{\smallskip}\hline\noalign{\smallskip}
HD\,1835   &	0&.07 &    0&.05 &    0&.02 &	 0&.04 &    0&.05 \\
HD\,20807  &    0&.06 &    0&.05 & $-$0&.11 &	 0&.05 &    0&.01 \\
HD\,26491  &    0&.01 &    0&.01 & $-$0&.06 & $-$0&.04 & $-$0&.02 \\
HD\,33021  & $-$0&.05 & $-$0&.04 & $-$0&.14 & $-$0&.19 & $-$0&.11 \\
HD\,39587  & $-$0&.07 & $-$0&.07 & $-$0&.12 & $-$0&.12 & $-$0&.10 \\
HD\,43834  & $-$0&.04 &    0&.01 &    0&.01 &	 0&.02 &    0&.00 \\
HD\,50806  &	0&.21 &    0&.13 &    0&.08 &	 0&.08 &    0&.13 \\
HD\,53705  & $-$0&.07 & $-$0&.04 & $-$0&.14 & $-$0&.07 & $-$0&.08 \\
HD\,84117  & 	0&.09 &    0&.07 & $-$0&.06 & $-$0&.06 &    0&.01 \\
HD\,102365 & 	0&.07 &    0&.12 & $-$0&.08 &	 0&.02 &    0&.03 \\
HD\,112164 & $-$0&.01 & $-$0&.01 &    0&.00 &	 0&.02 &    0&.00 \\
HD\,114613 & $-$0&.01 & $-$0&.02 &    0&.03 &	 0&.00 &    0&.00 \\
HD\,115383 & $-$0&.12 & $-$0&.09 & $-$0&.05 & $-$0&.05 & $-$0&.08 \\
HD\,115617 & $-$0&.11 & $-$0&.04 &    0&.00 &	 0&.04 & $-$0&.03 \\
HD\,117176 & $-$0&.06 & $-$0&.04 & $-$0&.06 & $-$0&.04 & $-$0&.05 \\
HD\,128620 &	0&.01 &    0&.04 &    0&.03 &	 0&.07 &    0&.04 \\
HD\,141004 & $-$0&.03 & $-$0&.03 &    0&.00 &	 0&.07 &    0&.00 \\
HD\,146233 & $-$0&.04 & $-$0&.05 & $-$0&.07 &	 0&.01 & $-$0&.04 \\
HD\,147513 & $-$0&.15 & $-$0&.11 & $-$0&.18 & $-$0&.10 & $-$0&.14 \\
HD\,160691 &	0&.01 & $-$0&.01 &    0&.02 &	 0&.06 &    0&.02 \\
HD\,177565 &	0&.05 &    0&.10 &    0&.02 &	 0&.03 &    0&.05 \\
HD\,181321 & \md{--}  & $-$0&.10 & $-$0&.15 & $-$0&.15 & $-$0&.13 \\
HD\,188376 & $-$0&.16 & $-$0&.06 & $-$0&.02 & $-$0&.05 & $-$0&.07 \\
HD\,189567 & 	0&.00 & $-$0&.05 & $-$0&.15 & $-$0&.10 & $-$0&.08 \\
HD\,196761 & \md{--}  & $-$0&.13 & $-$0&.20 & $-$0&.13 & $-$0&.15 \\
\hline
\end{tabular}
\end{table}

The overall trend of our abundance ratios as a function of the stellar
metallicity normally follows what has been suggested in the literature
concerning the nucleosynthetic origin of the elements and their abundance
evolution in time \citep{Chenetal2000,Reddyetal2003,Bensbyetal2005,
Chenetal2008,Nevesetal2009}. The light metals Ca, Sc, and Ti are
predominantly produced by Type~II Supernovae (SN\,II) at the beginning of
the enrichment history of the Galactic disc. On the other hand, iron and the
iron-peak elements V, Cr, Co, and Ni are predominantly synthesised by
Type~Ia Supernovae (SN\,Ia) in longer time scales. Therefore, it is expected
that the abundance ratio of these light metals with respect to iron
progressively decrease from metal-poor to metal-rich stars, whereas the
abundance ratio of iron-peak elements with respect to iron, all produced at
the same rate, remains constant and close to zero in the whole range of
metallicity. Indeed, this is exactly what is observed in Fig.~\ref{ab_feh}
for [Ca,\,Sc,\,Ti/Fe] and [V,\,Cr,\,Co,\,Ni/Fe], to within our stated
abundance uncertainties.

The light metals Mg and Si may not only be produced by SN\,II considering
that [Mg,\,Si/Fe] flattens out for metallicities higher than $-$0.1~dex. The
same behaviour was found by \citet{Chenetal2000} and \citet{Nevesetal2009},
who suggested that SN\,Ia is possibly contributing. The star HD\,50806 is
clearly enriched in Mg (and in other elements as well) according to
Fig.~\ref{ab_feh}, probably reflecting its membership to the thin-thick disc
transition.

The situation of Mn, Cu, and Zn is somewhat more complex. The hypothesis of
production in SN\,Ia still stands, but this is probably not the unique
source. \citet{AllenPortodeMello2011}, in their study of s-process enriched
stars, suggested that SN\,Ia is the main source of production of manganese,
in opposition to the conclusions of \citet{Feltzingetal2007}, who suggested
that this element is mainly produced by SN\,II. Our results in
Fig.~\ref{ab_feh}, which show an increasing trend of [Mn/Fe] as a function
of [Fe/H] (see also the bottom panel of Fig.~\ref{dendrogram}, in which
there is a sequential crescent ordination of \mxfe\ from the metal-poor
clustering group to the metal-rich one), seem to support the idea of an
extra nucleosynthetic source for the Mn yields. Such an increasing trend is
also usually attributed to a metallicity dependence in the production of Mn
in both SN\,Ia and SN\,II. In this work, Mn and Cu were plotted together
through the mean abundance ratio [Mn,\,Cu/Fe]. Both these elements have
abundances that increase with metallicity, though for Cu this trend is not
as significant as for Mn, and seems to happen only for higher metallicities,
being constant and close to zero for [Fe/H] $<$ 0. Cu and Zn, although being
adjacent elements in the periodic table, stand in the transition between
iron-peak and s-process elements, and their behaviour is in sharp contrast.
A decreasing trend in [Zn/Fe] vs. [Fe/H] is seen for stars poorer than the
Sun, in agreement with Fig.~1 of \citet{AllenPortodeMello2011}.

\begin{table}
\centering
\caption[]{Groups of chemical elements and uncertainties in the mean
           abundance ratios \mxfe. The estimated errors are compared to the
	   dispersions around the mean. For each group, the larger value of
	   each observation run was adopted.}
\label{mean_ab_err_comp}
\begin{tabular}{l l c c c c}
\hline\hline\noalign{\smallskip}
\multicolumn{2}{c}{\multirow{2}{*}{Nucleosynthetic group}} &
\multicolumn{2}{c}{HD\,146233} &
\multicolumn{2}{c}{HD\,26491} \\[0.1cm]
& & $\sigma_{\rm est}$ & $\sigma_{\rm disp}$ &
$\sigma_{\rm est}$ & $\sigma_{\rm disp}$ \\[-0.1cm]
\noalign{\smallskip}\hline\noalign{\smallskip}
light metals    & Mg, Si	& 0.03 & 0.02 & 0.04 & 0.03 \\
light metals    & Ca, Sc, Ti	& 0.03 & 0.04 & 0.05 & 0.01 \\
iron peak       & V, Cr, Co, Ni & 0.03 & 0.04 & 0.02 & 0.03 \\
iron peak       & Mn, Cu	& 0.03 & 0.01 & 0.05 & 0.02 \\
light s-process & Sr, Y, Zr	& 0.04 & 0.09 & 0.05 & 0.01 \\
heavy s-process & Ba, Ce, Nd	& 0.06 & 0.06 & 0.06 & 0.07 \\
\hline
\end{tabular}
\end{table}

\begin{figure*}
\centering
\resizebox{\hsize}{!}{\includegraphics{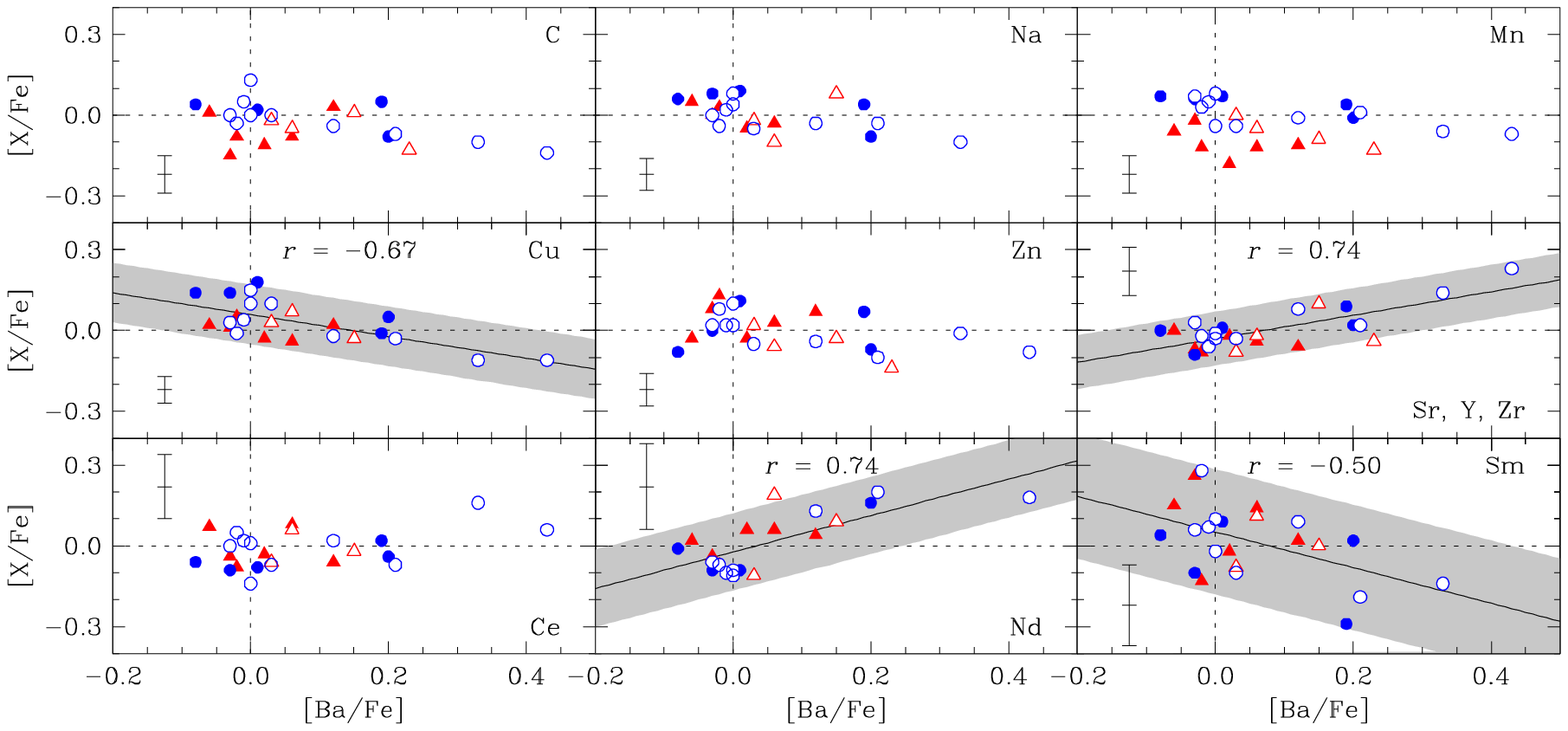}}
\caption{Abundance ratios as a function of [Ba/Fe]. For the elements of the
         light s-process (Sr, Y, and Zr) the mean abundance ratios are
	 plotted. The linear regressions (solid line), the 95\% confidence
	 intervals (hashed area), and the cross-correlation coefficients
	 are also shown for $|\,r\,| \geq 0.5$. The symbols follow the
	 classification of Sect.~\ref{classtree} (see
	 Fig.~\ref{dendrogram}).}
\label{ab_Ba}
\end{figure*}

The elements C and Na were not grouped together with other elements. C is
synthesised in several different sites and behaves similarly to O and N,
with [C/Fe] decreasing with increasing metallicity. This negative trend is
mostly observed in the metal-poor regime ([Fe/H] $< -$0.3), hence not seen
in Fig.~\ref{ab_feh} given the limited metallicity range of our program
stars. Nevertheless, our results agree very well with the recent C abundance
determination performed by \citet{daSilvaetal2011} for solar-like dwarfs. Na
is probably produced, among other processes, in the core of massive stars
and ejected by SN\,II into the interstellar medium. Here we found that
[Na/Fe] is constant and nearly close to zero in the range of metal-poor
stars, with a possible increasing trend for higher metallicities.
\citet{Chenetal2000} suggested that maybe [Na/Fe] is close to zero for the
whole metallicity range of disc stars. \citet{Nevesetal2009}, however, found
that [Na/Fe] is close to zero for thin disc stars for [Fe/H] between $-$0.2
and +0.2, but above solar for other values of metallicity.

The [X/Fe] abundance ratios of elements of the s-process, mainly produced in
TP-AGB of intermediate or low mass stars, and the r-process, produced in
sites with high neutron density such as the final stages of massive stars
(SN\,II, neutron stars), are supposed, respectively, to progressively
increase and decrease from metal-poor stars to higher metallicities. These
facts reflect the production of these elements in different time scales with
respect to iron, the former products of long-lived AGB stars, the latter
arising from short-lived massive stars which explode as supernovae. This is
not clearly observed in the diagrams of Fig.~\ref{ab_feh} for Sm and the
s-process elements because of the short metallicity range. An interesting
result of the analysis of these diagrams are the properties involving some
Ba-enriched stars, which we discuss in the next section.


\subsection{Abundance trends as a function of [Ba/Fe]}
\label{ab_Ba_sect}

A few stars in our sample are remarkably enriched in Ba by more than
3$\sigma$, especially HD\,39587 and HD\,147513. For this reason we also
investigated the behaviour of the abundance ratios [X/Fe] or \mxfe\ as a
function of [Ba/Fe] for some elements or groups of elements showing some
kind of relation with the production of barium (see Fig.~\ref{ab_Ba}).
Once more, we computed the cross-correlations coefficients and we plotted in
the figure the regressions of the most significant trends
($|\,r\,| \geq 0.5$).

\citet{Castroetal1999} proposed the existence of an anticorrelation between
the abundances of Cu and the s-process elements. They found that [Cu/Fe]
decreases with the increasing of [Ba/Fe], suggesting a relation between the
destruction of Cu and the production of Ba (and other s-process elements).
In the recent analysis of Ba-enriched stars of
\citet{AllenPortodeMello2011}, the authors have not supported this scenario,
arguing that Cu seems to be little (or not at all) affected by the
s-process, even though they have acknowledged that some Ba-rich stars do
present anticorrelated abundances of Cu and the s-process elements. Our
results point to a statistically significant decrease in the abundances of
Cu with increasing [Ba/Fe], in accordance with \citet{Castroetal1999}.

Two other iron-peak elements, Mn and Zn, are also shown in Fig.~\ref{ab_Ba}
and no clear correlation is observed. This may indicate that Mn and Cu do
not share the same nucleosynthetic origin. Indeed,
\citet{AllenPortodeMello2011} found that the synthesis of Cu receives a
larger contribution from not so massive stars than Zn, a result roughly in
line with those of \citet{Castroetal1999} and ours. Such results point
towards the necessity of both more extensive observations of the abundances
of Cu and Zn, and more protracted theoretical efforts, in order that a
better understanding of the complex chemical history of these two elements
may be achieved.

\citet{Castroetal1999} also proposed an anticorrelation in the abundances of
C and Na with respect to [Ba/Fe]. Our results do not seem to support this
anticorrelation, though our most Ba-rich stars, the Ursa Major group members
HD39587 and HD147513, are markedly C-deficient.
\citet{PortodeMellodaSilva1997a} attributed the [C/Fe] deficiency of a
{\it barium star} to the $^{13}{\rm C}(\alpha,n)^{16}{\rm O}$ reaction that
occurred in the hot-bottom envelope of its companion during the TP-AGB
phase. However, HD39587 and HD147513 are no longer regarded as true
{\it barium stars}.

Figure~\ref{ab_Ba} shows an evident expected correlation for the light
s-process elements (Sr, Y, and Zr). Correlations involving Nd, another
heavy element of the s-process may also exist, but its abundance
determination has larger uncertainties. An anticorrelation between [Sm/Fe]
and [Ba/Fe] also seems to exist. Sm is a good representative of the
r-process elements, and despite the very large uncertainties in the
determination of such elements, usually showing very few lines in the
spectra of solar-type stars, an interpretation in which this anticorrelation
is due to ever more efficient production of s-process elements in AGB stars,
as compared to the production of the r-process in SN\,II, seems warranted.


\subsection{Abundance trends as a function of age}
\label{ab_age_sect}

\begin{figure*}
\centering
\begin{minipage}[b]{0.98\textwidth}
\centering
\resizebox{\hsize}{!}{\includegraphics{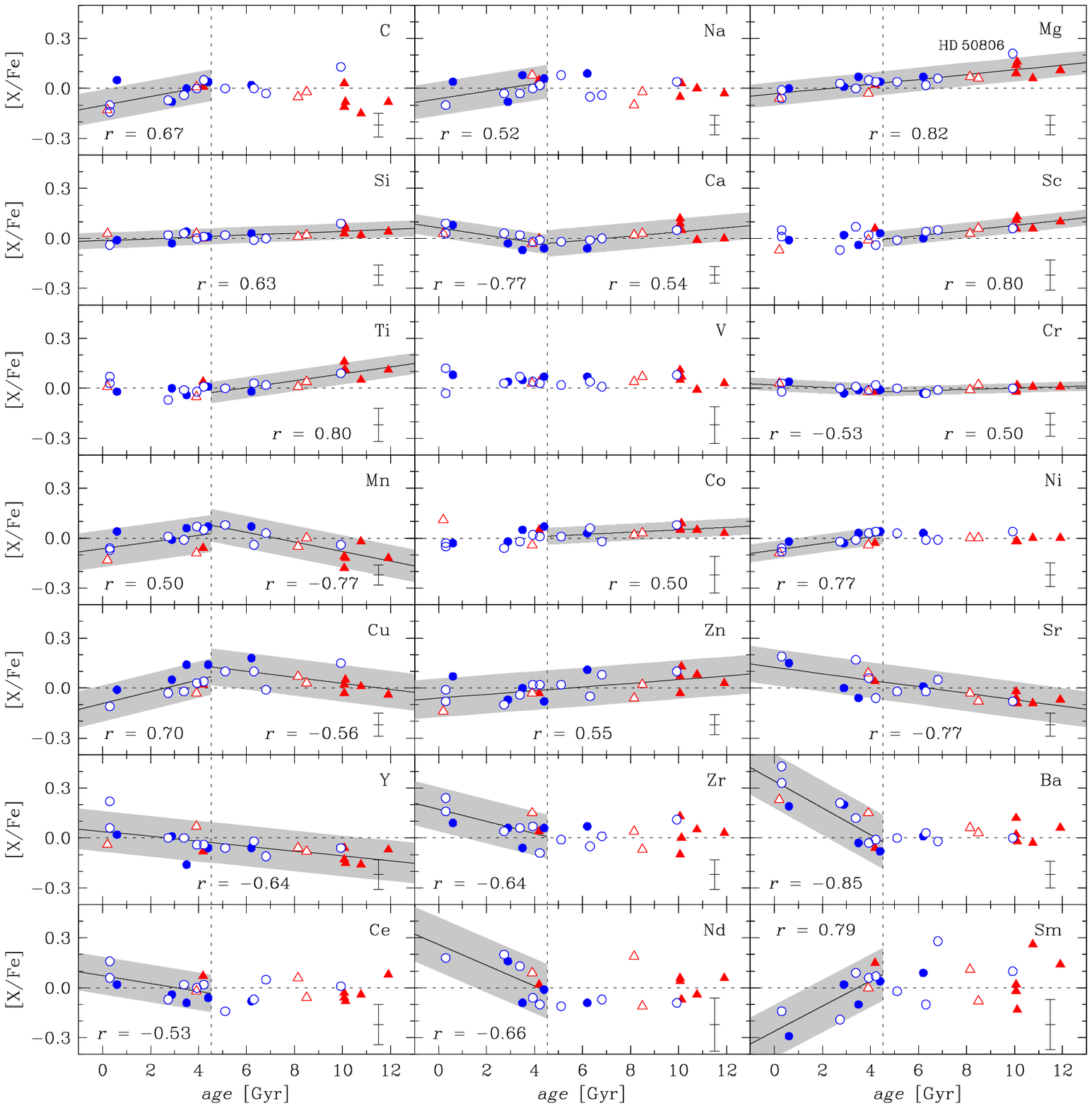}}
\end{minipage} \\[-0.17cm]
\begin{minipage}[b]{0.98\textwidth}
\centering
\resizebox{\hsize}{!}{\includegraphics{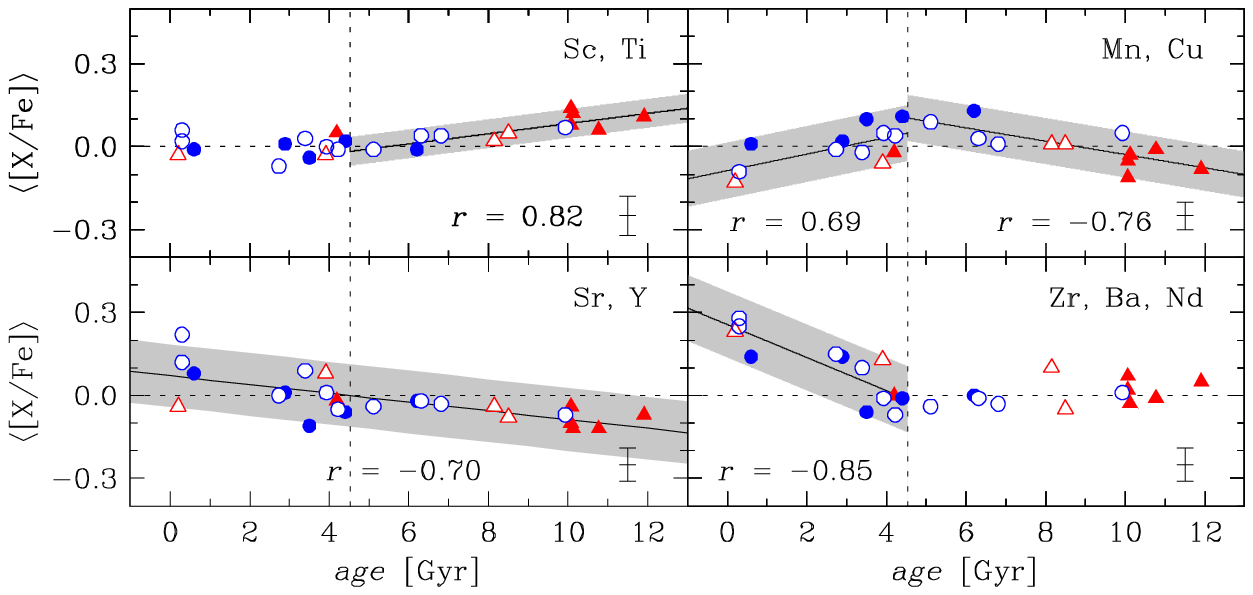}}
\end{minipage}
\caption{Abundance ratios as a function of the stellar age for individual
         elements (top panels) and for nucleosynthetic groups (bottom
	 panels). The vertical dashed line indicates the adopted solar age
	 (4.53~Gyr). The linear regressions (solid line), the 95\%
	 confidence intervals (hashed area), and the cross-correlation
	 coefficients are also shown for $|\,r\,| \geq 0.5$. The symbols
	 follow the classification of Sect.~\ref{classtree} (see
	 Fig.~\ref{dendrogram}).}
\label{ab_age}
\end{figure*}

\begin{figure*}
\centering
\resizebox{\hsize}{!}{\includegraphics{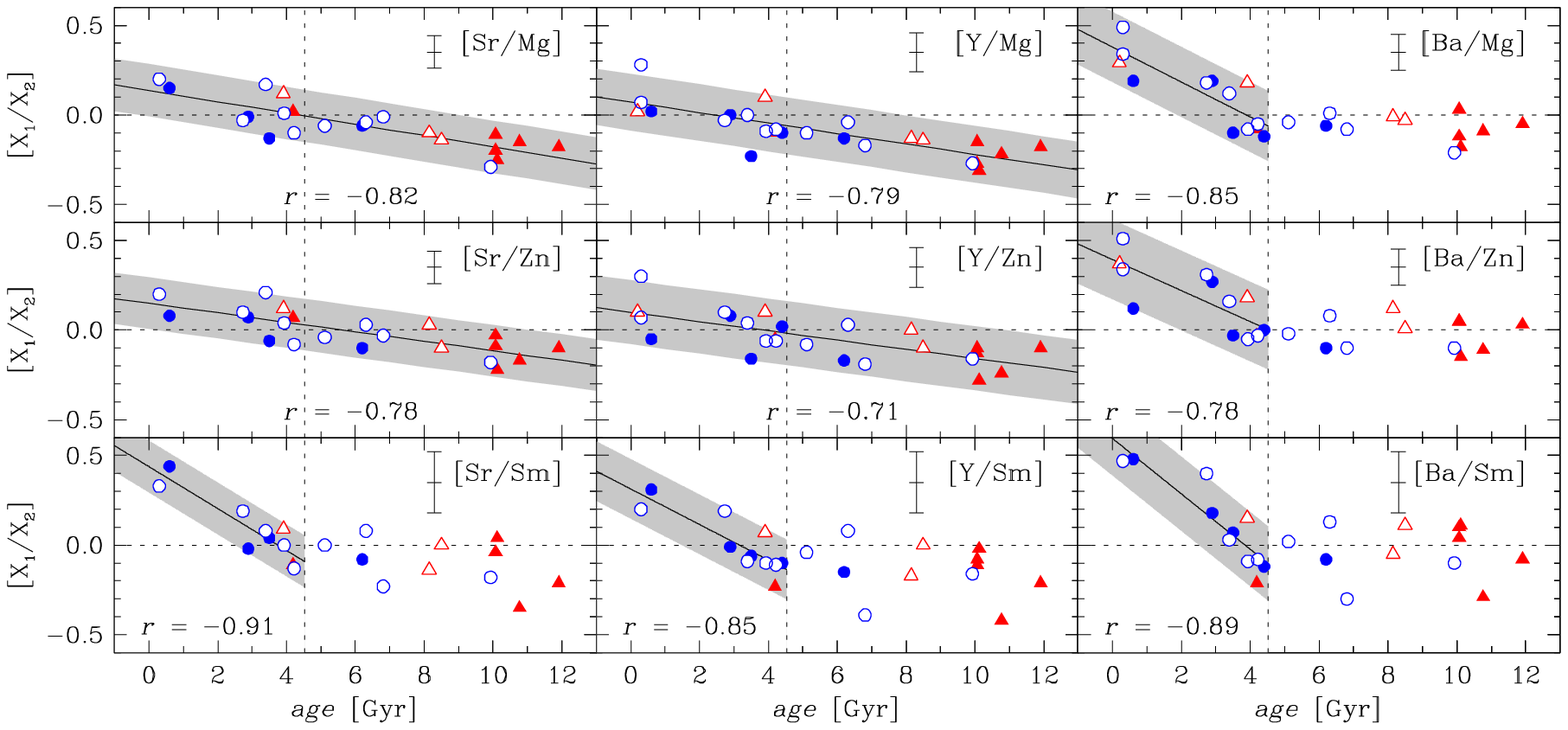}}
\caption{Abundance ratios as a function of the stellar age. The vertical
         dashed line indicates the adopted solar age (4.53~Gyr). The linear
         regressions (solid line), the 95\% confidence intervals (hashed
	 area), and the cross-correlation coefficients are also shown for
	 $|\,r\,| \geq 0.5$. The symbols follow the classification of
	 Sect.~\ref{classtree} (see Fig.~\ref{dendrogram}).}
\label{ab_X_age}
\end{figure*}

One of the motivations of the present paper is to explore the abundance
ratios of elements due to different nucleosynthetic processes and the
stellar ages, taking advantage of the reasonably precise ages that can be
attributed to our program stars. Figure~\ref{ab_age} shows the diagrams
[X/Fe] and \mxfe\ as a function of the stellar age, and Fig.~\ref{ab_X_age}
explores the relation between the abundance ratios [X$_1$/X$_2$] of
different elements and age. Similarly to Fig.~\ref{ab_feh}, we have fitted
linear regressions on the diagrams in three ranges of stellar age: for stars
younger than the Sun ($age <$ 4.53~Gyr), for stars older than the Sun, and
for all the sample stars. We have then computed the cross-correlation
coefficients in these three ranges of stellar age and plotted the
regressions of the more significant trends (only if $|\,r\,| \geq 0.5$). In
these figures there are positive, negative, or flat abundance trends in the
three ranges of age. We notice, however, that the age of the Sun, used as a
reference, was arbitrarily chosen. The exact value of the transition age
when the abundance behaviour changes is not clear from these plots (it is a
value between 4 and 6~Gyr).

\begin{figure}
\centering
\resizebox{0.88\hsize}{!}{\includegraphics[angle=-90]{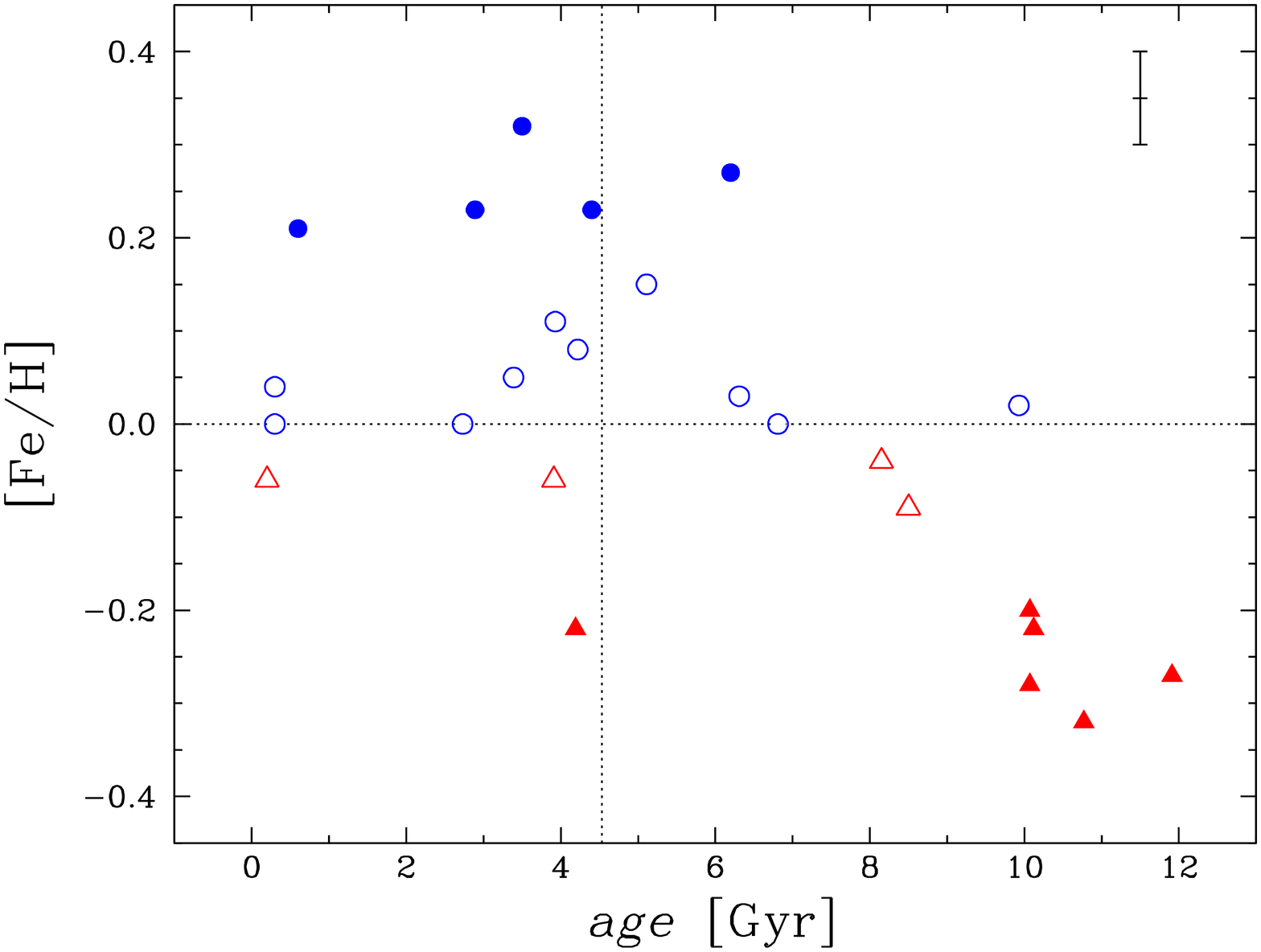}}
\caption{Stellar metallicity as a function of age. The symbols follow the
         classification of Sect.~\ref{classtree} (see
	 Fig.~\ref{dendrogram}).}
\label{ab_age_feh}
\end{figure}

In spite of the long recognition (though not undisputed) of the so-called
age-metallicity relation (see Fig.~\ref{ab_age_feh}), the individual
abundances in Fig.~\ref{ab_age} and those in Fig.~\ref{ab_feh} do not share
exactly the same behaviour, leading us to suggest in the following that the
age-metallicity relation may be a multidimensional concept. For this reason,
we regrouped the elements according to their abundance behaviour with age
(see the bottom panels of Fig.~\ref{ab_age}).

Carbon and sodium do not seem to present any important trends of [X/Fe] with
age. The positive trends observed for Mg, Sc, and Ti (less clearly seen for
Si and Ca) are simply the result of the Galactic chemical evolution. The
production rate of these elements by SN\,II decrease with time since the
formation of the Galactic disc (equivalently to increasing with age)
compared to the increased production of Fe by the longer-lived SN\,Ia as we
approach more recent epochs. Silicon perhaps shows no trend at all; Mg seems
to have a more or less positive linear trend with increasing age; in their
turn Ca, Sc, and Ti sport a more complex behaviour. The statistical
significance of the behaviour of Ca is slight, and not much confidence
should be placed in the apparent [Ca/Fe] decrease with time, followed by an
increase towards more recent times. Taken at face value, this would appear
to lend support to the suggestion that some fraction of the Ca synthesis
might be due to SN\,Ia, in unison with their production of Fe. Sc and Ti
seem to have a significant decrease with time with respect to Fe, but this
decrease stops at a time close to the solar age and flattens thereafter
towards present times.

Among the Fe-peak elements, no important trend is seen in the [X/Fe]
relation with age for V, Cr, Co, and Zn; only a dubious one for Ni in the
interval of young stars. Yet, again, Cu and Mn suggest more underlying
complexity. Even though the statistical significance of the linear
regressions is slight, the abundance ratios to Fe of both these elements
seem first to increase towards the present epoch, and then decrease (a
behaviour that is reinforced when these two elements are plotted together
through the mean abundance ratio [Mn,\,Cu/Fe]). Recalling that
\citet{AllenPortodeMello2011} have found that Mn is mostly due to SN\,Ia,
our result could imply that the relative yield of Mn to Fe in SN\,Ia
decreases with time (and consequently the overall metallicity). The
situation for Cu is less straightforward, as usual.
\citet{AllenPortodeMello2011} suggest that little of the synthesis of Mn,
Cu, and Zn is owed to the main s-process, leaving the action of AGB stars an
unlikely source of such a behaviour. These same authors assert that the
action of the so-called {\it weak} s-process, sited at the He-burning core
of massive stars, has a non negligible contribution to the synthesis of Mn,
Cu, and Zn. One possible explanation for the decrease of the abundance
ratios of [Mn/Fe] and [Cu/Fe] towards more recent times is a decreasing
yield of the weak s-process in their synthesis, as contrasted to the
production of Fe by SN\,Ia. Clearly, an understanding of the detailed
chemical evolution of elements, in both the dimensions of metallicity and
age, of the Fe-peak and its transition with the heavier elements deserves
closer scrutiny, both observationally and theoretically.

We next turn to the [X/Fe]-age relation for the s-process elements. These
present particular interest, since \citet{Castroetal1999} suggested, also
using data from \citet{Edvardssonetal1993}, that [Ba/H] has a steeper
decrease with age than [Fe/H], and therefore that [Ba/Fe] increases towards
modern times \citep[see also][]{Bensbyetal2007}. This fact might be
interpreted, again, as a consequence of the larger yield of the s-process
element synthesis by the long-lived AGB stars in relation to the not
as-long-lived production of Fe by SN\,Ia. Do the other s-process elements
show a behaviour similar to barium? Apparently this is so, but not in a
straightforward way. The lighter s-process elements Sr and Y seem to have a
linear trend of [X/Fe] with age (clearly seen for Sr), increasing towards
present times, as expected. The [Zr/Fe], [Ce/Fe], and [Nd/Fe] ratio,
however, appears flat in the old age regime, possibly increasing only for
stars younger than the Sun. The [Ba/Fe] ratio behaves similarly but the
increase towards younger ages is much sharper and more significant. At face
value, these results point towards the evolution of the relative yields of
such s-process elements with time (and metallicity) in AGB stars, apparently
favouring the heavier species Ba and Nd over lighter ones. The simultaneous
analysis of the [Cu/Fe] and [s-process/Fe] ratios, epitomised, for example,
by the [Ba/Fe] relation with age, suggests an anticorrelation of Cu and Ba
towards younger stars, as found by \citet{Castroetal1999}, adding to the
controversy surrounding the chemical evolution of copper.

Finally, a positive trend with increasing age is observed for [Sm/Fe] in
stars younger than the Sun, again reflecting the smaller number of SN\,II
than SN\,Ia in the present in comparison with the past (in this case the
epoch of the Sun's formation). Significant positive trends are also observed
for Ni and Cu in the regime of younger stars.

Concerning the four groups yielded by the tree clustering analysis and their
relation with age, we can only state that, as expected, the old stars in our
sample are also metal-poor, whereas young stars tend to be metal-rich. Once
more, the case of the star HD\,50806 is evidenced. Although classified in
the intermediate group of stars with slightly over-solar abundances (\mxh\ =
+0.06~dex), it is situated close to the group of metal-poor stars (\mxh\ =
$-$0.24~dex) in Fig.~\ref{ab_age}, which is maybe a consequence of its
population membership.

The bottom panels of Fig.~\ref{ab_age} shows the average of the [X/Fe]
relations with age for selected groups of elements. Thus we see that
grouping Sc and Ti reinforces the relation with age already shown by each
element individually, and the same is seen for the grouping of Sr and Y.
Similarly, averaging the [X/Fe] relations with age for Zr, Ba, and Nd
produces a very steep increase towards present epochs, after a flat relation
from the birth of the Galactic disc up to the solar age.

These results prompted us to investigate the specifics of the [X$_1$/X$_2$]
ratios with age, where X$_1$ and X$_2$ designate elements other than Fe. In
Fig.~\ref{ab_X_age} we explore the time evolution of some elements that
displayed a particular clear [X/Fe] relation with age. Thus, it is apparent
that the [Ba/Mg] and [Ba/Zn] increase steeply towards present epochs, for
stars younger than the Sun, reinforcing their individual and opposite
behaviour in the [X/Fe]-age diagrams. Also, both [Sr/Mg] and [Y/Mg] increase
linearly and significantly from the oldest to the youngest stars; the same
is seen in the [Sr/Zn] and [Y/Zn] ratios. Investigating the [X$_1$/X$_2$]
ratios of s-process to r-process elements, we found steep increases in the
[Ba/Sm], [Sr/Sm], and [Y/Sm] ratios towards younger stars, but only for
objects younger than the Sun. Significant age relations are, therefore,
evidenced in the [X$_1$/X$_2$] ratios of diverse elements, representing a
wide range of nucleosynthetic processes and tentatively allowing the
proposition that the age-metallicity relation is a more complex constraint
to Galactic chemo-dynamical models than hitherto recognised.


\subsection{Abundance trends with condensation temperature}
\label{ab_tc}

Our determination of multi-elemental abundances also provides the study of
possible trends in the abundance ratios [X/Fe] as a function of the
condensation temperature (\tc) of each element. Values of 50\%\,\tc\  (the
temperature when 50\% of an element is in the condensed phase) for a
solar-system composition gas were taken from \citet{Lodders2003} and
\citet{Loddersetal2009}.

For a few stars in our sample we have found some correlations of [X/Fe] with
\tc\  (even after corrections due to Galactic chemical evolution effects
were applied). However, no clear correlation seems to exist when comparing
the slopes for refractory elements (those with \tc\ $\gtrsim$ 900~K) with
several stellar parameters (\tmean, [Fe/H], \logg, $\xi$, mass, and age).
The one involving the metallicity was proposed by \citet{Ramirezetal2010} in
the sense that higher-metallicity stars present more negative slopes. Our
results agree with their paper, but the number of metal-rich stars in our
sample is too small to confirm their conclusions.


\section{Conclusions}
\label{concl}

In this work we have performed a multi-elemental, differentially with
respect to the Sun, spectroscopic analysis of a sample of 25 solar-type
stars in the solar neighbourhood. We have derived their atmospheric
parameters (from various nearly independent criteria and with low internal
errors), masses, ages, kinematical and orbital parameters, and elemental
abundances (derived with very low internal uncertainties) based on
equivalent widths or spectral synthesis.

Despite small in size, our sample was carefully selected to undergo an
homogeneous and detailed analysis based on spectra with high resolution and
high signal-to-noise ratio. We have:
{\it (i)} checked the effective temperatures based on the excitation
equilibrium of neutral iron lines against those from photometric
calibrations and from the H$\alpha$ wings profile,
{\it (ii)} checked the surface gravities computed through the ionisation
equilibrium between \ion{Fe}{I} and \ion{Fe}{II} lines against those
computed based on the evolutionary parameters,
{\it (iii)} derived masses and ages from evolutionary tracks and isochrones
computed considering the metallicity of each star, and
{\it (iv)} applied a differential spectroscopic analysis relative to the
Sun, hence minimising the systematic errors and yielding a mean uncertainty
of 0.06~dex in the abundance ratios.
We thus expect that our determinations have achieved a high level of
precision and accuracy.

We have also applied a statistical study to our abundance results using the
method of tree clustering analysis, through which we looked for groupings of
stars that share similar abundances in the [X/H] space. Although our sample
has a limited range in metallicity, it covers a broad range in age. The
detailed abundance pattern was then investigated through correlations with
kinematics, Galactic orbits, and stellar ages. Our conclusions are thus
summarised:

\begin{itemize}

\item[1-] Four groups were identified, two having over-solar abundances
(with averages +0.26 and +0.06~dex on [X/H]), and two with under-solar
abundance values (on average $-$0.06 and $-$0.24~dex). Possible non-solar
abundance ratios, even for stars which share the same age, Galactic orbit,
and metallicity as the Sun, are suggested. Whether these are due to
heterogeneity in the stellar natal clouds, or by dynamical migration within
the Galactic disc, should be investigated with larger samples. In
particular, the results of \citet{RochaPintoetal2006} that metal-poor and
old stars show more orbital radial spread in the Galaxy could not be
verified given the limitation of our sample;

\item[2-] The presence of Ba-enriched stars in our sample prompted us to
investigate in detail the relation of some elements with Ba. An
anticorrelation between [Cu/Fe] and [Ba/Fe] was found, in line with similar
claims in the literature. The [Sm/Fe] abundance ratios seem to be
anticorrelated with [Ba/Fe], barely at the 95\% confidence level. On the
other hand, previous suggestions of [C/Fe] and [Na/Fe] anticorrelations with
[Ba/Fe] could not be confirmed. Even though the possible connected chemical
evolution of Mn, Cu, and Zn has been recently discussed in the literature,
no significant trend of [Mn/Fe] and [Zn/Fe] with [Ba/Fe] is suggested;

\item[3-] The consideration of the [X/Fe] ratios with age revealed much
differing behaviour of the elements, suggesting that the age-metallicity
relation has more underlying complexity than commonly recognised. The
[Mg/Fe], [Sc/Fe], and [Ti/Fe] decrease towards younger ages. The [Cu/Fe] and
[Mn/Fe] ratios initially increase towards younger stars up to the solar age,
and then decrease towards the youngest objects, a result that may speak of
differing yields in the SN\,Ia production of these elements, related to
metallicity and age, as well as a possible influence of the evolution with
time of the weak s-process yields, operating off of massive stars. The
steepest relation with age was found for the [Ba/Fe] ratio, but only for
ages younger than the solar one, and a similar but less evident behaviour is
seen for Zr, Ce, and Nd. Other heavy s-process elements, however, such as Sr
and Y, show a linearly increasing [X/Fe] towards younger ages, particularly
clearer for Sr. [Sm/Fe] significantly decreases for stars younger than the
Sun. Thus, the [Cu/Ba] ratio clearly decreases towards younger stars, and
the same is seen at a significant level for [Sm/Ba];

\item[4-] Considering the average of elements with similar behaviour with
age considerably reinforces the aforementioned results, particularly for the
[Sc,Ti/Fe], [Mn,Cu/Fe], [Sr,Y/Fe], and [Zr,Ba,Nd/Fe] relations;

\item[5-] The consideration of element ratios not directly involving Fe
shows some marked behaviour. Particularly, the [Ba/Mg], [Ba/Zn], [Ba/Sm],
[Sr/Sm], and [Y/Sm] steeply increase towards younger ages for stars younger
than the Sun. Also, the [Sr/Mg] and [Y/Mg] ratios linearly increase towards
younger ages, and the same is seen for [Sr/Zn] and [Y/Zn];

\item[6-] Possible correlations of the abundances, the condensation
temperatures of the different elements, and the presence of exoplanets in
our program stars was deeply investigated, but no significant correlation
was found.

\end{itemize}

The detailed consideration of precise element abundances derived from
high-quality atmospheric parameters and spectroscopic data, tied to masses,
kinematics, and ages for solar-type stars, generally provides a wealth of
interesting data, contributing towards a broader understanding of the
evolution of the Galaxy in its chemical and dynamical aspects.


\begin{acknowledgements}

R.D.S. thanks the financial support from the Coordena\c c\~ao de
Aperfei\c coamento de Pessoal de N\'ivel Superior (CAPES) in the
form of a fellowship (PROAP/INPE).
L.S.R. thanks the grants (100454/2004-6 and 309326/2009-5) received from the
Brazilian Foundation CNPq, and also Luzia P. Rit\'e and Charles Rit\'e for
their help in data reductions.
G.F.P.M. acknowledges the financial support by CNPq (476909/2006-6 and
474972/2009-7) and FAPERJ (APQ1/26/170.687/2004) grants.
This research has made use of the SIMBAD database, operated at CDS,
Strasbourg, France, and of NASA's Astrophysics Data System.
We acknowledge many fruitful discussions with Verne V. Smith and Katia
Cunha.
We all thank the critical and important report from the referee
Dr. Gustafsson, which has deeply improved this manuscript. 

\end{acknowledgements}

\bibliographystyle{aa}
\bibliography{daSilvaetal2012_aph}

\begin{table*}[p]
\centering
\caption[]{Atomic line parameters of the elements used in the analysis.
           Oscillator strengths (\loggf) and raw $EW$s (before the
	   conversion set out by Eq.~\ref{fit1} and ~\ref{fit2}), given in
	   m\AA, of both the Ganymede spectra observed in the first (Gany 1)
	   and second (Gany 2) runs are listed (except for C, for which the
	   analysis is based on spectral synthesis). Lines with missing $gf$
	   values represent the elements with hyperfine structure (Mg, Sc,
	   V, Mn, Co, and Cu) and the detailed line splitting is shown in
	   Table~\ref{hfs}.}
\label{linelist}
\begin{tabular}{c c c r@{}l r@{}l r@{}l r@{}l | c c c r@{}l r@{}l r@{}l r@{}l}
\hline\hline
 & & & & & & & & & & & & & & & & & & & & & \\[-0.2cm]
\multirow{2}{*}{$\lambda$ {\tiny [\AA]}} &
\multirow{2}{*}{Id.} &
\multirow{2}{*}{\parbox[c]{0.6cm}{\centering $\chi$ {\tiny [eV]}}} &
\mc{4}{c}{Gany 1} & \mc{4}{c|}{Gany 2} &
\multirow{2}{*}{$\lambda$ {\tiny [\AA]}} &
\multirow{2}{*}{Id.} &
\multirow{2}{*}{\parbox[c]{0.6cm}{\centering $\chi$ {\tiny [eV]}}} &
\mc{4}{c}{Gany 1} & \mc{4}{c}{Gany 2} \\[0.1cm]
 & & & \mc{2}{c}{\loggf} &
\mc{2}{c}{$EW$} &
\mc{2}{c}{\loggf} &
\mc{2}{c|}{$EW$} &
 & & & \mc{2}{c}{\loggf} &
\mc{2}{c}{$EW$} &
\mc{2}{c}{\loggf} &
\mc{2}{c}{$EW$} \\[0.1cm]
\hline
 & & & & & & & & & & & & & & & & & & & & & \\[-0.2cm]
5052.167 & \ion{C}{I}   & 7.68 & $-$1&.48 &    &-- & $-$1&.62 &    &-- & 4926.147 & \ion{Ti}{I}  & 0.82 &     &--  &    &-- & $-$2&.17 &   7&.0 \\
5380.322 & \ion{C}{I}   & 7.68 & $-$1&.78 &    &-- & $-$1&.84 &    &-- & 5022.871 & \ion{Ti}{I}  & 0.83 & $-$0&.52 &  77&.3 & $-$0&.35 &  79&.2 \\
6154.230 & \ion{Na}{I}  & 2.10 & $-$1&.52 &  42&.3 & $-$1&.52 &  40&.8 & 5024.842 & \ion{Ti}{I}  & 0.82 & $-$0&.66 &  71&.5 & $-$0&.48 &  73&.8 \\
6160.753 & \ion{Na}{I}  & 2.10 & $-$1&.23 &  61&.8 & $-$1&.29 &  59&.9 & 5071.472 & \ion{Ti}{I}  & 1.46 &     &--  &    &-- & $-$0&.77 &  31&.3 \\
4571.102 & \ion{Mg}{I}  & 0.00 &     &--  & 107&.8 &     &--  & 110&.3 & 5113.448 & \ion{Ti}{I}  & 1.44 & $-$0&.86 &  30&.3 & $-$0&.88 &  27&.6 \\
4730.038 & \ion{Mg}{I}  & 4.34 &     &--  &  78&.2 &     &--  &  70&.9 & 5145.464 & \ion{Ti}{I}  & 1.46 & $-$0&.67 &  38&.8 & $-$0&.64 &  37&.6 \\
5711.095 & \ion{Mg}{I}  & 4.34 &     &--  & 119&.2 &     &--  & 106&.4 & 5147.479 & \ion{Ti}{I}  & 0.00 &     &--  &    &-- & $-$1&.98 &  42&.4 \\
5785.285 & \ion{Mg}{I}  & 5.11 & $-$1&.87 &  50&.5 & $-$1&.82 &  55&.3 & 5152.185 & \ion{Ti}{I}  & 0.02 &     &--  &    &-- & $-$2&.03 &  39&.1 \\
5517.533 & \ion{Si}{I}  & 5.08 & $-$2&.42 &  15&.7 & $-$2&.51 &  13&.0 & 5192.969 & \ion{Ti}{I}  & 0.02 & $-$1&.10 &  88&.3 & $-$1&.02 &  84&.4 \\
5621.607 & \ion{Si}{I}  & 5.08 &     &--  &    &-- & $-$2&.61 &  10&.5 & 5211.206 & \ion{Ti}{I}  & 0.84 & $-$2&.10 &   9&.1 & $-$2&.07 &   9&.2 \\
5665.563 & \ion{Si}{I}  & 4.92 & $-$1&.98 &  42&.2 & $-$1&.96 &  41&.8 & 5219.700 & \ion{Ti}{I}  & 0.02 & $-$2&.32 &  27&.9 & $-$2&.23 &  30&.0 \\
5684.484 & \ion{Si}{I}  & 4.95 &     &--  &    &-- & $-$1&.60 &  62&.3 & 5295.780 & \ion{Ti}{I}  & 1.07 &     &--  &    &-- & $-$1&.60 &  13&.2 \\
5690.433 & \ion{Si}{I}  & 4.93 & $-$1&.80 &  52&.9 & $-$1&.81 &  50&.6 & 5426.236 & \ion{Ti}{I}  & 0.02 &     &--  &    &-- & $-$2&.97 &   7&.8 \\
5701.108 & \ion{Si}{I}  & 4.93 & $-$1&.90 &  46&.3 & $-$1&.97 &  41&.0 & 5471.197 & \ion{Ti}{I}  & 1.44 &     &--  &    &-- & $-$1&.48 &   9&.6 \\
5708.405 & \ion{Si}{I}  & 4.95 &     &--  &    &-- & $-$1&.35 &  79&.1 & 5490.150 & \ion{Ti}{I}  & 1.46 &     &--  &    &-- & $-$1&.00 &  22&.6 \\
5753.622 & \ion{Si}{I}  & 5.61 &     &--  &    &-- & $-$1&.24 &  50&.8 & 5648.567 & \ion{Ti}{I}  & 2.49 & $-$0&.40 &  11&.5 & $-$0&.39 &  11&.1 \\
5772.149 & \ion{Si}{I}  & 5.08 &     &--  &    &-- & $-$1&.56 &  57&.7 & 5679.937 & \ion{Ti}{I}  & 2.47 & $-$0&.65 &   6&.9 & $-$0&.63 &   7&.0 \\
5793.080 & \ion{Si}{I}  & 4.93 & $-$1&.91 &  46&.0 & $-$1&.94 &  42&.6 & 5739.464 & \ion{Ti}{I}  & 2.25 & $-$0&.67 &  10&.5 & $-$0&.75 &   8&.6 \\
6125.021 & \ion{Si}{I}  & 5.61 &     &--  &    &-- & $-$1&.50 &  34&.1 & 5866.452 & \ion{Ti}{I}  & 1.07 &     &--  &    &-- & $-$0&.82 &  49&.6 \\
6131.577 & \ion{Si}{I}  & 5.61 & $-$1&.67 &  25&.8 & $-$1&.65 &  26&.6 & 6064.629 & \ion{Ti}{I}  & 1.05 &     &--  &    &-- & $-$1&.88 &   9&.7 \\
6131.858 & \ion{Si}{I}  & 5.61 & $-$1&.66 &  26&.6 & $-$1&.64 &  27&.0 & 6091.177 & \ion{Ti}{I}  & 2.27 &     &--  &    &-- & $-$0&.44 &  15&.9 \\
6142.494 & \ion{Si}{I}  & 5.62 & $-$1&.44 &  37&.5 & $-$1&.45 &  36&.3 & 6092.798 & \ion{Ti}{I}  & 1.89 & $-$1&.31 &   6&.0 & $-$1&.28 &   6&.1 \\
6145.020 & \ion{Si}{I}  & 5.61 & $-$1&.40 &  40&.8 & $-$1&.36 &  41&.8 & 6098.694 & \ion{Ti}{I}  & 3.06 &     &--  &    &-- & $-$0&.16 &   6&.1 \\
6243.823 & \ion{Si}{I}  & 5.61 & $-$1&.22 &  52&.5 & $-$1&.19 &  52&.7 & 6126.224 & \ion{Ti}{I}  & 1.07 &     &--  &    &-- & $-$1&.40 &  23&.5 \\
6244.476 & \ion{Si}{I}  & 5.61 & $-$1&.26 &  49&.5 & $-$1&.25 &  48&.6 & 6258.104 & \ion{Ti}{I}  & 1.44 & $-$1&.46 &  54&.6 & $-$0&.43 &  52&.3 \\
5261.708 & \ion{Ca}{I}  & 2.52 &     &--  &    &-- & $-$0&.65 &  99&.4 & 4568.345 & \ion{Ti}{II} & 1.22 & $-$2&.85 &  33&.6 & $-$2&.85 &  32&.3 \\
5581.979 & \ion{Ca}{I}  & 2.52 &     &--  &    &-- & $-$0&.68 &  97&.2 & 4583.415 & \ion{Ti}{II} & 1.16 & $-$2&.85 &  36&.2 & $-$2&.84 &  35&.1 \\
5590.126 & \ion{Ca}{I}  & 2.52 & $-$0&.78 &  96&.3 & $-$0&.73 &  93&.9 & 4657.209 & \ion{Ti}{II} & 1.24 &     &--  &    &-- & $-$2&.31 &  55&.8 \\
5867.572 & \ion{Ca}{I}  & 2.93 & $-$1&.62 &  25&.3 & $-$1&.59 &  25&.1 & 4798.539 & \ion{Ti}{II} & 1.08 & $-$2&.75 &  45&.4 & $-$2&.70 &  44&.7 \\
6161.295 & \ion{Ca}{I}  & 2.52 & $-$1&.18 &  71&.6 & $-$1&.08 &  69&.7 & 5211.544 & \ion{Ti}{II} & 2.59 & $-$1&.59 &  32&.1 & $-$1&.54 &  33&.2 \\
6163.754 & \ion{Ca}{I}  & 2.52 &     &--  &    &-- & $-$1&.25 &  83&.4 & 5336.783 & \ion{Ti}{II} & 1.58 & $-$1&.77 &  71&.5 & $-$1&.63 &  73&.4 \\
6166.440 & \ion{Ca}{I}  & 2.52 & $-$1&.18 &  71&.5 & $-$1&.02 &  76&.9 & 5381.020 & \ion{Ti}{II} & 1.57 & $-$1&.91 &  65&.1 & $-$1&.95 &  59&.5 \\
6169.044 & \ion{Ca}{I}  & 2.52 & $-$0&.75 &  99&.4 & $-$0&.70 &  97&.0 & 5418.756 & \ion{Ti}{II} & 1.58 & $-$2&.21 &  49&.7 & $-$2&.17 &  49&.4 \\
6169.564 & \ion{Ca}{I}  & 2.52 & $-$0&.51 & 117&.6 & $-$0&.53 & 119&.8 & 5657.436 & \ion{V}{I}   & 1.06 &     &--  &    &-- &     &--  &   9&.4 \\
6449.820 & \ion{Ca}{I}  & 2.52 &     &--  &    &-- & $-$0&.32 & 127&.8 & 5668.362 & \ion{V}{I}   & 1.08 &     &--  &   8&.7 &     &--  &   6&.7 \\
6455.605 & \ion{Ca}{I}  & 2.52 &     &--  &    &-- & $-$1&.43 &  53&.7 & 5670.851 & \ion{V}{I}   & 1.08 &     &--  &  21&.1 &     &--  &  21&.7 \\
6471.688 & \ion{Ca}{I}  & 2.52 &     &--  &    &-- & $-$0&.64 & 101&.5 & 5727.661 & \ion{V}{I}   & 1.05 &     &--  &  10&.9 &     &--  &  12&.5 \\
6499.654 & \ion{Ca}{I}  & 2.52 &     &--  &    &-- & $-$0&.86 &  87&.2 & 6090.216 & \ion{V}{I}   & 1.08 &     &--  &  35&.6 &     &--  &  34&.4 \\
4743.817 & \ion{Sc}{I}  & 1.45 &     &--  &    &-- &     &--  &   8&.5 & 6135.370 & \ion{V}{I}   & 1.05 &     &--  &  12&.6 &     &--  &  11&.5 \\
5356.091 & \ion{Sc}{I}  & 1.86 &     &--  &    &-- &     &--  &   1&.8 & 6150.154 & \ion{V}{I}   & 0.30 &     &--  &  12&.6 &     &--  &  11&.0 \\
5392.075 & \ion{Sc}{I}  & 1.99 &     &--  &    &-- &     &--  &   7&.0 & 6199.186 & \ion{V}{I}   & 0.29 &     &--  &  15&.1 &     &--  &  14&.0 \\
5484.611 & \ion{Sc}{I}  & 1.85 &     &--  &    &-- &     &--  &   3&.2 & 6216.358 & \ion{V}{I}   & 0.28 &     &--  &    &-- &     &--  &  37&.0 \\
5671.826 & \ion{Sc}{I}  & 1.45 &     &--  &    &-- &     &--  &  19&.3 & 6274.658 & \ion{V}{I}   & 0.27 &     &--  &    &-- &     &--  &   8&.7 \\
6239.408 & \ion{Sc}{I}  & 0.00 &     &--  &    &-- &     &--  &   9&.0 & 6285.165 & \ion{V}{I}   & 0.28 &     &--  &  10&.4 &     &--  &  16&.2 \\
5318.346 & \ion{Sc}{II} & 1.36 &     &--  &    &-- &     &--  &  18&.2 & 4575.092 & \ion{Cr}{I}  & 3.37 &     &--  &    &-- & $-$0&.88 &  13&.6 \\
5357.190 & \ion{Sc}{II} & 1.51 &     &--  &   5&.4 &     &--  &   5&.2 & 4616.120 & \ion{Cr}{I}  & 0.98 &     &--  &    &-- & $-$1&.31 &  91&.9 \\
5526.815 & \ion{Sc}{II} & 1.77 &     &--  &  77&.9 &     &--  &  77&.6 & 4626.174 & \ion{Cr}{I}  & 0.97 &     &--  &    &-- & $-$1&.47 &  84&.8 \\
5657.874 & \ion{Sc}{II} & 1.51 &     &--  &  69&.8 &     &--  &  69&.2 & 4708.019 & \ion{Cr}{I}  & 3.17 &     &--  &    &-- & $-$0&.06 &  58&.1 \\
5684.189 & \ion{Sc}{II} & 1.51 &     &--  &  40&.8 &     &--  &  41&.2 & 4737.355 & \ion{Cr}{I}  & 3.09 &     &--  &    &-- & $-$0&.06 &  62&.0 \\
6245.660 & \ion{Sc}{II} & 1.51 &     &--  &  38&.0 &     &--  &  35&.6 & 4756.137 & \ion{Cr}{I}  & 3.10 &    0&.09 &  74&.6 &    0&.03 &  66&.4 \\
6320.867 & \ion{Sc}{II} & 1.50 &     &--  &   9&.1 &     &--  &   8&.4 & 4801.047 & \ion{Cr}{I}  & 3.12 & $-$0&.28 &  51&.5 & $-$0&.28 &  46&.2 \\
4518.023 & \ion{Ti}{I}  & 0.83 & $-$0&.49 &  76&.2 &     &--  &    &-- & 4936.335 & \ion{Cr}{I}  & 3.11 & $-$0&.35 &  48&.5 & $-$0&.32 &  47&.2 \\
4548.765 & \ion{Ti}{I}  & 0.83 & $-$0&.55 &  73&.7 &     &--  &    &-- & 4964.916 & \ion{Cr}{I}  & 0.94 &     &--  &    &-- & $-$2&.50 &  41&.6 \\
4562.625 & \ion{Ti}{I}  & 0.02 &     &--  &    &-- & $-$2&.73 &  11&.7 & 5200.207 & \ion{Cr}{I}  & 3.38 & $-$0&.58 &  24&.6 & $-$0&.50 &  26&.8 \\
4617.254 & \ion{Ti}{I}  & 1.75 &     &--  &    &-- &    0&.23 &  64&.6 & 5214.144 & \ion{Cr}{I}  & 3.37 & $-$0&.77 &  17&.7 & $-$0&.73 &  18&.4 \\
4758.120 & \ion{Ti}{I}  & 2.25 &    0&.44 &  56&.0 &    0&.26 &  43&.3 & 5238.964 & \ion{Cr}{I}  & 2.71 & $-$1&.43 &  16&.6 & $-$1&.36 &  17&.9 \\
4759.272 & \ion{Ti}{I}  & 2.25 &    0&.47 &  58&.0 &    0&.25 &  45&.9 & 5247.566 & \ion{Cr}{I}  & 0.96 & $-$1&.73 &  83&.1 & $-$1&.61 &  85&.1 \\
4778.259 & \ion{Ti}{I}  & 2.24 & $-$0&.38 &  18&.0 & $-$0&.38 &  17&.2 & 5272.007 & \ion{Cr}{I}  & 3.45 &     &--  &    &-- & $-$0&.36 &  30&.3 \\
\hline
\end{tabular}
\end{table*}
\addtocounter{table}{-1}
\begin{table*}[p]
\centering
  \caption[]{continued.}
\begin{tabular}{c c c r@{}l r@{}l r@{}l r@{}l | c c c r@{}l r@{}l r@{}l r@{}l}
\hline\hline
 & & & & & & & & & & & & & & & & & & & & & \\[-0.2cm]
\multirow{2}{*}{$\lambda$ {\tiny [\AA]}} &
\multirow{2}{*}{Id.} &
\multirow{2}{*}{\parbox[c]{0.6cm}{\centering $\chi$ {\tiny [eV]}}} &
\mc{4}{c}{Gany 1} & \mc{4}{c|}{Gany 2} &
\multirow{2}{*}{$\lambda$ {\tiny [\AA]}} &
\multirow{2}{*}{Id.} &
\multirow{2}{*}{\parbox[c]{0.6cm}{\centering $\chi$ {\tiny [eV]}}} &
\mc{4}{c}{Gany 1} & \mc{4}{c}{Gany 2} \\[0.1cm]
 & & & \mc{2}{c}{\loggf} &
\mc{2}{c}{$EW$} &
\mc{2}{c}{\loggf} &
\mc{2}{c|}{$EW$} &
 & & & \mc{2}{c}{\loggf} &
\mc{2}{c}{$EW$} &
\mc{2}{c}{\loggf} &
\mc{2}{c}{$EW$} \\[0.1cm]
\hline
 & & & & & & & & & & & & & & & & & & & & & \\[-0.2cm]
5287.183 & \ion{Cr}{I}  & 3.44 & $-$0&.90 &  12&.0 & $-$0&.86 &  12&.6 & 5225.525 & \ion{Fe}{I} & 0.11 & $-$4&.81 &  77&.7 & $-$4&.72 &  75&.1 \\
5296.691 & \ion{Cr}{I}  & 0.98 &     &--  &    &-- & $-$1&.40 &  93&.5 & 5242.491 & \ion{Fe}{I} & 3.63 & $-$1&.26 &  86&.0 & $-$1&.16 &  87&.7 \\
5300.751 & \ion{Cr}{I}  & 0.98 & $-$2&.13 &  63&.3 & $-$2&.11 &  56&.7 & 5243.773 & \ion{Fe}{I} & 4.26 & $-$1&.13 &  62&.1 & $-$1&.05 &  63&.8 \\
5304.183 & \ion{Cr}{I}  & 3.46 &     &--  &    &-- & $-$0&.72 &  16&.0 & 5247.049 & \ion{Fe}{I} & 0.09 & $-$5&.00 &  70&.4 & $-$4&.93 &  70&.0 \\
5318.810 & \ion{Cr}{I}  & 3.44 & $-$0&.66 &  19&.3 & $-$0&.65 &  18&.8 & 5250.216 & \ion{Fe}{I} & 0.12 & $-$4&.86 &  75&.4 & $-$4&.95 &  62&.4 \\
5628.621 & \ion{Cr}{I}  & 3.42 &     &--  &    &-- & $-$0&.82 &  14&.6 & 5320.040 & \ion{Fe}{I} & 3.64 & $-$2&.47 &  23&.6 & $-$2&.44 &  23&.5 \\
5648.279 & \ion{Cr}{I}  & 3.82 &    0&.90 &   5&.7 &     &--  &    &-- & 5321.109 & \ion{Fe}{I} & 4.43 & $-$1&.24 &  47&.1 & $-$1&.21 &  45&.8 \\
5784.976 & \ion{Cr}{I}  & 3.32 & $-$0&.45 &  34&.0 & $-$0&.39 &  33&.8 & 5332.908 & \ion{Fe}{I} & 1.56 & $-$3&.07 &  92&.2 & $-$2&.84 &  95&.7 \\
5787.965 & \ion{Cr}{I}  & 3.32 &     &--  &    &-- & $-$0&.12 &  50&.4 & 5379.574 & \ion{Fe}{I} & 3.69 & $-$1&.60 &  64&.9 & $-$1&.56 &  62&.6 \\
6330.097 & \ion{Cr}{I}  & 0.94 & $-$2&.88 &  30&.0 & $-$2&.90 &  26&.9 & 5389.486 & \ion{Fe}{I} & 4.41 & $-$0&.63 &  84&.4 & $-$0&.56 &  83&.3 \\
4588.203 & \ion{Cr}{II} & 4.07 &     &--  &    &-- & $-$0&.73 &  71&.7 & 5395.222 & \ion{Fe}{I} & 4.44 & $-$1&.73 &  22&.9 & $-$1&.73 &  21&.8 \\
4592.049 & \ion{Cr}{II} & 4.07 & $-$1&.30 &  50&.1 & $-$1&.23 &  50&.8 & 5412.791 & \ion{Fe}{I} & 4.43 & $-$1&.76 &  21&.9 & $-$1&.75 &  21&.4 \\
5305.855 & \ion{Cr}{II} & 3.83 &     &--  &    &-- & $-$2&.06 &  27&.7 & 5432.946 & \ion{Fe}{I} & 4.44 & $-$0&.79 &  72&.7 & $-$0&.69 &  74&.4 \\
5308.377 & \ion{Cr}{II} & 4.07 &     &--  &    &-- & $-$1&.81 &  27&.1 & 5436.297 & \ion{Fe}{I} & 4.39 & $-$1&.36 &  42&.5 & $-$1&.31 &  42&.8 \\
5313.526 & \ion{Cr}{II} & 4.07 &     &--  &    &-- & $-$1&.61 &  34&.9 & 5473.168 & \ion{Fe}{I} & 4.19 & $-$1&.99 &  21&.9 & $-$1&.96 &  22&.2 \\
5502.025 & \ion{Cr}{II} & 4.17 &     &--  &    &-- & $-$1&.87 &  21&.8 & 5483.108 & \ion{Fe}{I} & 4.15 & $-$1&.46 &  49&.1 & $-$1&.45 &  47&.1 \\
4626.538 & \ion{Mn}{I}  & 4.71 &     &--  &  30&.5 &     &--  &  26&.8 & 5491.845 & \ion{Fe}{I} & 4.19 & $-$2&.19 &  15&.2 & $-$2&.23 &  13&.8 \\
4739.113 & \ion{Mn}{I}  & 2.94 &     &--  &  61&.6 &     &--  &  63&.3 & 5494.474 & \ion{Fe}{I} & 4.07 & $-$1&.89 &  31&.1 & $-$1&.94 &  27&.6 \\
5004.892 & \ion{Mn}{I}  & 2.92 &     &--  &    &-- &     &--  &  18&.2 & 5522.454 & \ion{Fe}{I} & 4.21 & $-$1&.50 &  44&.2 & $-$1&.44 &  44&.8 \\
5394.670 & \ion{Mn}{I}  & 0.00 &     &--  &    &-- &     &--  &  81&.8 & 5560.207 & \ion{Fe}{I} & 4.43 & $-$1&.12 &  54&.3 & $-$1&.09 &  53&.0 \\
5399.479 & \ion{Mn}{I}  & 3.85 &     &--  &    &-- &     &--  &  40&.0 & 5577.013 & \ion{Fe}{I} & 5.03 & $-$1&.52 &  11&.9 & $-$1&.49 &  12&.5 \\
5413.684 & \ion{Mn}{I}  & 3.86 &     &--  &  24&.9 &     &--  &  25&.7 & 5587.573 & \ion{Fe}{I} & 4.14 & $-$1&.54 &  45&.7 & $-$1&.56 &  41&.8 \\
5420.350 & \ion{Mn}{I}  & 2.14 &     &--  &  86&.6 &     &--  &  90&.3 & 5635.824 & \ion{Fe}{I} & 4.26 & $-$1&.58 &  37&.8 & $-$1&.55 &  37&.1 \\
5432.548 & \ion{Mn}{I}  & 0.00 &     &--  &  52&.3 &     &--  &  53&.8 & 5636.705 & \ion{Fe}{I} & 3.64 & $-$2&.51 &  22&.4 & $-$2&.52 &  21&.1 \\
5537.765 & \ion{Mn}{I}  & 2.19 &     &--  &  41&.1 &     &--  &  37&.3 & 5638.262 & \ion{Fe}{I} & 4.22 & $-$0&.89 &  78&.8 & $-$0&.79 &  79&.4 \\
6013.497 & \ion{Mn}{I}  & 3.07 &     &--  &  87&.0 &     &--  &  86&.9 & 5641.436 & \ion{Fe}{I} & 4.26 & $-$1&.04 &  67&.6 & $-$0&.99 &  66&.1 \\
6021.803 & \ion{Mn}{I}  & 3.07 &     &--  &  97&.0 &     &--  &  99&.0 & 5646.697 & \ion{Fe}{I} & 4.26 & $-$2&.38 &   9&.2 & $-$2&.48 &   5&.5 \\
4523.400 & \ion{Fe}{I}  & 3.65 & $-$1&.85 &  51&.4 &     &--  &    &-- & 5650.019 & \ion{Fe}{I} & 5.10 & $-$0&.78 &  39&.5 & $-$0&.82 &  35&.8 \\
4537.676 & \ion{Fe}{I}  & 3.27 & $-$2&.96 &  17&.6 &     &--  &    &-- & 5652.319 & \ion{Fe}{I} & 4.26 & $-$1&.79 &  27&.6 & $-$1&.79 &  26&.1 \\
4556.925 & \ion{Fe}{I}  & 3.25 & $-$2&.69 &  28&.8 &     &--  &    &-- & 5661.348 & \ion{Fe}{I} & 4.28 & $-$1&.88 &  23&.3 & $-$1&.81 &  24&.6 \\
4585.343 & \ion{Fe}{I}  & 4.61 & $-$1&.57 &  22&.6 & $-$1&.59 &  21&.1 & 5680.240 & \ion{Fe}{I} & 4.19 & $-$2&.34 &  11&.5 & $-$2&.30 &  12&.1 \\
4593.555 & \ion{Fe}{I}  & 3.94 & $-$2&.03 &  29&.2 & $-$2&.00 &  29&.1 & 5701.557 & \ion{Fe}{I} & 2.56 & $-$2&.20 &  90&.2 & $-$2&.13 &  86&.6 \\
4598.125 & \ion{Fe}{I}  & 3.28 & $-$1&.61 &  82&.2 & $-$1&.61 &  76&.3 & 5705.473 & \ion{Fe}{I} & 4.30 & $-$1&.35 &  48&.0 & $-$1&.44 &  40&.8 \\
4602.000 & \ion{Fe}{I}  & 1.61 & $-$3&.32 &  74&.7 & $-$3&.21 &  73&.1 & 5731.761 & \ion{Fe}{I} & 4.26 & $-$1&.13 &  62&.4 & $-$1&.14 &  58&.0 \\
4741.535 & \ion{Fe}{I}  & 2.83 & $-$2&.21 &  73&.2 & $-$2&.08 &  73&.8 & 5738.240 & \ion{Fe}{I} & 4.22 & $-$2&.12 &  16&.6 & $-$2&.19 &  14&.2 \\
4749.961 & \ion{Fe}{I}  & 4.56 & $-$1&.17 &  44&.4 & $-$1&.28 &  36&.2 & 5775.069 & \ion{Fe}{I} & 4.22 & $-$1&.21 &  60&.0 & $-$1&.11 &  62&.1 \\
4793.961 & \ion{Fe}{I}  & 3.05 & $-$3&.40 &  11&.4 & $-$3&.54 &   8&.5 & 5778.463 & \ion{Fe}{I} & 2.59 & $-$3&.53 &  22&.5 & $-$3&.47 &  23&.7 \\
4794.355 & \ion{Fe}{I}  & 2.42 & $-$3&.88 &  14&.8 & $-$3&.86 &  14&.5 & 5784.666 & \ion{Fe}{I} & 3.40 & $-$2&.59 &  29&.1 & $-$2&.53 &  29&.4 \\
4798.273 & \ion{Fe}{I}  & 4.19 & $-$1&.48 &  45&.4 & $-$1&.43 &  45&.5 & 5811.916 & \ion{Fe}{I} & 4.14 & $-$2&.40 &  11&.2 & $-$2&.36 &  11&.8 \\
4798.743 & \ion{Fe}{I}  & 1.61 & $-$4&.19 &  33&.5 & $-$4&.22 &  35&.1 & 5814.805 & \ion{Fe}{I} & 4.28 & $-$1&.85 &  24&.6 & $-$1&.85 &  23&.4 \\
4808.147 & \ion{Fe}{I}  & 3.25 & $-$2&.65 &  31&.2 & $-$2&.59 &  33&.8 & 5835.098 & \ion{Fe}{I} & 4.26 & $-$2&.11 &  16&.0 & $-$2&.10 &  15&.7 \\
4907.733 & \ion{Fe}{I}  & 3.43 & $-$1&.80 &  65&.8 & $-$1&.76 &  63&.5 & 5849.681 & \ion{Fe}{I} & 3.69 & $-$2&.98 &   8&.3 & $-$2&.90 &   9&.5 \\
4908.032 & \ion{Fe}{I}  & 3.93 & $-$1&.56 &  39&.7 & $-$1&.77 &  40&.2 & 5852.222 & \ion{Fe}{I} & 4.55 & $-$1&.26 &  40&.5 & $-$1&.16 &  43&.2 \\
4911.788 & \ion{Fe}{I}  & 3.93 & $-$1&.72 &  45&.6 & $-$1&.61 &  48&.3 & 5855.086 & \ion{Fe}{I} & 4.61 & $-$1&.58 &  22&.9 & $-$1&.49 &  25&.3 \\
4961.915 & \ion{Fe}{I}  & 3.63 & $-$2&.38 &  26&.9 & $-$2&.31 &  28&.6 & 5856.096 & \ion{Fe}{I} & 4.29 & $-$1&.64 &  33&.7 & $-$1&.52 &  37&.3 \\
4962.565 & \ion{Fe}{I}  & 4.18 & $-$1&.33 &  54&.5 & $-$1&.25 &  55&.3 & 5859.596 & \ion{Fe}{I} & 4.55 & $-$0&.70 &  72&.7 & $-$0&.60 &  74&.4 \\
4969.916 & \ion{Fe}{I}  & 4.22 & $-$0&.89 &  77&.6 & $-$0&.78 &  79&.5 & 5916.249 & \ion{Fe}{I} & 2.45 & $-$2&.97 &  57&.0 & $-$2&.89 &  56&.8 \\
5023.189 & \ion{Fe}{I}  & 4.28 & $-$1&.47 &  41&.7 & $-$1&.37 &  44&.3 & 5927.786 & \ion{Fe}{I} & 4.65 & $-$1&.12 &  43&.0 & $-$1&.05 &  44&.5 \\
5025.091 & \ion{Fe}{I}  & 4.26 & $-$1&.87 &  23&.6 & $-$1&.83 &  24&.1 & 5929.666 & \ion{Fe}{I} & 4.55 & $-$1&.17 &  45&.4 & $-$1&.18 &  42&.4 \\
5025.313 & \ion{Fe}{I}  & 4.28 & $-$1&.97 &  19&.2 & $-$1&.78 &  25&.3 & 5930.173 & \ion{Fe}{I} & 4.65 & $-$0&.33 &  91&.9 & $-$0&.26 &  91&.2 \\
5054.647 & \ion{Fe}{I}  & 3.64 & $-$2&.09 &  40&.8 & $-$1&.92 &  44&.4 & 5956.692 & \ion{Fe}{I} & 0.86 & $-$4&.63 &  53&.4 & $-$4&.54 &  53&.5 \\
5067.162 & \ion{Fe}{I}  & 4.22 & $-$0&.98 &  72&.9 & $-$0&.90 &  74&.7 & 6005.551 & \ion{Fe}{I} & 2.59 & $-$3&.50 &  24&.2 & $-$3&.48 &  23&.6 \\
5072.677 & \ion{Fe}{I}  & 4.22 & $-$1&.09 &  66&.0 & $-$0&.98 &  67&.9 & 6007.968 & \ion{Fe}{I} & 4.65 & $-$0&.73 &  66&.0 & $-$0&.71 &  63&.3 \\
5109.649 & \ion{Fe}{I}  & 4.30 & $-$0&.77 &  81&.1 & $-$0&.68 &  83&.3 & 6012.212 & \ion{Fe}{I} & 2.22 & $-$3&.82 &  26&.0 & $-$3&.79 &  25&.5 \\
5127.359 & \ion{Fe}{I}  & 0.93 & $-$3&.57 &  96&.3 & $-$3&.30 & 101&.1 & 6078.499 & \ion{Fe}{I} & 4.79 & $-$0&.36 &  82&.1 & $-$0&.29 &  81&.9 \\
5127.680 & \ion{Fe}{I}  & 0.05 & $-$5&.97 &  24&.3 & $-$5&.84 &  27&.8 & 6079.014 & \ion{Fe}{I} & 4.65 & $-$1&.04 &  47&.6 & $-$0&.98 &  48&.6 \\
5196.065 & \ion{Fe}{I}  & 4.26 & $-$0&.90 &  75&.4 & $-$0&.78 &  77&.3 & 6082.708 & \ion{Fe}{I} & 2.22 & $-$3&.58 &  37&.8 & $-$3&.53 &  37&.3 \\
5197.929 & \ion{Fe}{I}  & 4.30 & $-$1&.52 &  38&.5 & $-$1&.50 &  37&.0 & 6093.666 & \ion{Fe}{I} & 4.61 & $-$1&.37 &  32&.4 & $-$1&.34 &  31&.6 \\
5213.818 & \ion{Fe}{I}  & 3.94 & $-$2&.75 &   7&.7 & $-$2&.67 &   8&.9 & 6098.250 & \ion{Fe}{I} & 4.56 &     &--  &    &-- & $-$1&.75 &  19&.5 \\
5223.188 & \ion{Fe}{I}  & 3.63 & $-$2&.29 &  31&.4 & $-$2&.26 &  31&.0 & 6120.249 & \ion{Fe}{I} & 0.92 & $-$5&.81 &   6&.7 & $-$5&.86 &   5&.8 \\
\hline
\end{tabular}
\end{table*}
\addtocounter{table}{-1}
\begin{table*}[p]
\centering
  \caption[]{continued.}
\begin{tabular}{c c c r@{}l r@{}l r@{}l r@{}l | c c c r@{}l r@{}l r@{}l r@{}l}
\hline\hline
 & & & & & & & & & & & & & & & & & & & & & \\[-0.2cm]
\multirow{2}{*}{$\lambda$ {\tiny [\AA]}} &
\multirow{2}{*}{Id.} &
\multirow{2}{*}{\parbox[c]{0.6cm}{\centering $\chi$ {\tiny [eV]}}} &
\mc{4}{c}{Gany 1} & \mc{4}{c|}{Gany 2} &
\multirow{2}{*}{$\lambda$ {\tiny [\AA]}} &
\multirow{2}{*}{Id.} &
\multirow{2}{*}{\parbox[c]{0.6cm}{\centering $\chi$ {\tiny [eV]}}} &
\mc{4}{c}{Gany 1} & \mc{4}{c}{Gany 2} \\[0.1cm]
 & & & \mc{2}{c}{\loggf} &
\mc{2}{c}{$EW$} &
\mc{2}{c}{\loggf} &
\mc{2}{c|}{$EW$} &
 & & & \mc{2}{c}{\loggf} &
\mc{2}{c}{$EW$} &
\mc{2}{c}{\loggf} &
\mc{2}{c}{$EW$} \\[0.1cm]
\hline
 & & & & & & & & & & & & & & & & & & & & & \\[-0.2cm]
6137.002 & \ion{Fe}{I}  & 2.20 & $-$2&.93 &  72&.4 & $-$2&.83 &  71&.2 & 6455.001 & \ion{Co}{I}  & 3.63 &     &--  &    &-- &     &--  &  14&.4 \\
6151.616 & \ion{Fe}{I}  & 2.18 & $-$3&.35 &  51&.7 & $-$3&.30 &  50&.6 & 4935.831 & \ion{Ni}{I}  & 3.94 & $-$0&.41 &  64&.4 & $-$0&.37 &  72&.0 \\
6159.382 & \ion{Fe}{I}  & 4.61 & $-$1&.86 &  13&.7 & $-$1&.84 &  13&.7 & 4946.029 & \ion{Ni}{I}  & 3.80 & $-$1&.18 &  30&.5 & $-$1&.22 &  27&.5 \\
6173.340 & \ion{Fe}{I}  & 2.22 & $-$2&.95 &  70&.3 & $-$2&.93 &  66&.1 & 4953.200 & \ion{Ni}{I}  & 3.74 & $-$0&.82 &  51&.8 & $-$0&.67 &  56&.1 \\
6187.987 & \ion{Fe}{I}  & 3.94 & $-$1&.71 &  48&.1 & $-$1&.64 &  48&.4 & 5010.934 & \ion{Ni}{I}  & 3.63 & $-$0&.92 &  51&.8 & $-$0&.91 &  49&.1 \\
6199.508 & \ion{Fe}{I}  & 2.56 & $-$4&.25 &   5&.8 & $-$4&.34 &   4&.3 & 5032.723 & \ion{Ni}{I}  & 3.90 & $-$1&.20 &  25&.6 & $-$1&.09 &  28&.8 \\
6200.321 & \ion{Fe}{I}  & 2.61 & $-$2&.45 &  76&.4 & $-$2&.37 &  74&.6 & 5094.406 & \ion{Ni}{I}  & 3.83 & $-$1&.14 &  31&.3 & $-$1&.06 &  33&.1 \\
6213.428 & \ion{Fe}{I}  & 2.22 & $-$2&.68 &  84&.4 & $-$2&.59 &  82&.1 & 5197.157 & \ion{Ni}{I}  & 3.90 & $-$1&.15 &  27&.8 & $-$1&.09 &  28&.8 \\
6219.287 & \ion{Fe}{I}  & 2.20 & $-$2&.52 &  93&.3 & $-$2&.47 &  88&.3 & 5220.300 & \ion{Ni}{I}  & 3.74 & $-$1&.30 &  27&.6 & $-$1&.23 &  29&.4 \\
6226.730 & \ion{Fe}{I}  & 3.88 & $-$2&.10 &  31&.0 & $-$2&.08 &  29&.5 & 5392.330 & \ion{Ni}{I}  & 4.15 & $-$1&.24 &  15&.7 & $-$1&.31 &  13&.3 \\
6240.645 & \ion{Fe}{I}  & 2.22 & $-$3&.37 &  49&.2 & $-$3&.34 &  46&.6 & 5435.866 & \ion{Ni}{I}  & 1.99 & $-$2&.47 &  54&.2 & $-$2&.38 &  54&.1 \\
6265.131 & \ion{Fe}{I}  & 2.18 & $-$2&.65 &  87&.9 & $-$2&.53 &  86&.8 & 5452.860 & \ion{Ni}{I}  & 3.84 & $-$1&.48 &  17&.5 & $-$1&.48 &  18&.5 \\
6271.283 & \ion{Fe}{I}  & 3.33 & $-$2&.70 &  27&.8 & $-$2&.67 &  27&.8 & 5494.876 & \ion{Ni}{I}  & 4.10 & $-$1&.11 &  21&.7 & $-$1&.07 &  22&.7 \\
6297.792 & \ion{Fe}{I}  & 2.22 & $-$2&.77 &  79&.9 & $-$2&.34 &  96&.5 & 5587.853 & \ion{Ni}{I}  & 1.93 & $-$2&.37 &  62&.4 & $-$2&.32 &  60&.4 \\
6315.813 & \ion{Fe}{I}  & 4.07 & $-$1&.66 &  43&.9 & $-$1&.67 &  41&.2 & 5625.312 & \ion{Ni}{I}  & 4.09 & $-$0&.59 &  47&.6 & $-$0&.63 &  42&.9 \\
6322.691 & \ion{Fe}{I}  & 2.59 & $-$2&.47 &  77&.0 & $-$2&.31 &  79&.0 & 5628.354 & \ion{Ni}{I}  & 4.09 &     &--  &    &-- & $-$1&.28 &  15&.8 \\
6358.687 & \ion{Fe}{I}  & 0.86 &     &--  &    &-- & $-$3&.79 &  87&.8 & 5637.128 & \ion{Ni}{I}  & 4.09 & $-$0&.80 &  36&.1 & $-$0&.80 &  34&.8 \\
6380.750 & \ion{Fe}{I}  & 4.19 &     &--  &    &-- & $-$1&.30 &  53&.9 & 5748.346 & \ion{Ni}{I}  & 1.68 & $-$3&.26 &  30&.7 & $-$3&.22 &  30&.5 \\
6385.726 & \ion{Fe}{I}  & 4.73 &     &--  &    &-- & $-$1&.94 &   9&.1 & 5846.986 & \ion{Ni}{I}  & 1.68 & $-$3&.40 &  24&.7 & $-$3&.33 &  26&.3 \\
6392.538 & \ion{Fe}{I}  & 2.28 &     &--  &    &-- & $-$3&.98 &  17&.3 & 6086.276 & \ion{Ni}{I}  & 4.26 & $-$0&.44 &  47&.7 & $-$0&.44 &  45&.4 \\
6393.612 & \ion{Fe}{I}  & 2.43 &     &--  &    &-- & $-$1&.60 & 132&.8 & 6176.807 & \ion{Ni}{I}  & 4.09 & $-$0&.26 &  66&.9 & $-$0&.28 &  61&.6 \\
6430.856 & \ion{Fe}{I}  & 2.18 &     &--  &    &-- & $-$2&.01 & 117&.8 & 6177.236 & \ion{Ni}{I}  & 1.83 & $-$3&.44 &  18&.4 & $-$3&.52 &  14&.8 \\
6498.945 & \ion{Fe}{I}  & 0.96 &     &--  &    &-- & $-$4&.58 &  48&.4 & 6186.709 & \ion{Ni}{I}  & 4.10 & $-$0&.90 &  31&.6 & $-$0&.87 &  31&.6 \\
4576.339 & \ion{Fe}{II} & 2.84 & $-$3&.13 &  66&.0 & $-$3&.03 &  66&.3 & 6191.187 & \ion{Ni}{I}  & 1.68 & $-$2&.30 &  80&.2 & $-$2&.18 &  79&.3 \\
4656.981 & \ion{Fe}{II} & 2.89 &     &--  &    &-- & $-$3&.59 &  41&.2 & 6327.604 & \ion{Ni}{I}  & 1.68 & $-$3&.07 &  41&.2 & $-$3&.04 &  39&.8 \\
4720.149 & \ion{Fe}{II} & 3.20 & $-$4&.49 &   6&.0 & $-$4&.57 &   6&.0 & 6370.357 & \ion{Ni}{I}  & 3.54 &     &--  &    &-- & $-$1&.75 &  18&.5 \\
4993.358 & \ion{Fe}{II} & 2.81 & $-$3&.74 &  40&.0 & $-$3&.69 &  40&.8 & 6378.256 & \ion{Ni}{I}  & 4.15 &     &--  &    &-- & $-$0&.77 &  34&.3 \\
5197.576 & \ion{Fe}{II} & 3.23 & $-$2&.45 &  81&.4 & $-$2&.32 &  82&.3 & 5218.209 & \ion{Cu}{I}  & 3.82 &     &--  &  54&.1 &     &--  &  55&.8 \\
5234.630 & \ion{Fe}{II} & 3.22 & $-$2&.36 &  86&.1 & $-$2&.23 &  86&.8 & 5220.086 & \ion{Cu}{I}  & 3.82 &     &--  &  17&.3 &     &--  &  17&.2 \\
5264.812 & \ion{Fe}{II} & 3.33 & $-$2&.98 &  52&.5 & $-$2&.96 &  50&.7 & 5782.136 & \ion{Cu}{I}  & 1.64 &     &--  &  79&.9 &     &--  &  85&.0 \\
5325.560 & \ion{Fe}{II} & 3.22 & $-$3&.16 &  49&.0 & $-$3&.15 &  47&.3 & 4810.537 & \ion{Zn}{I}  & 4.08 & $-$0&.33 &  76&.1 & $-$0&.27 &  79&.0 \\
5414.075 & \ion{Fe}{II} & 3.22 & $-$3&.60 &  29&.1 & $-$3&.54 &  31&.0 & 4607.338 & \ion{Sr}{I}  & 0.00 &    0&.02 &  48&.8 &    0&.12 &  48&.1 \\
5425.257 & \ion{Fe}{II} & 3.20 & $-$3&.23 &  46&.5 & $-$3&.25 &  43&.6 & 4883.690 & \ion{Y}{II}  & 1.08 &     &--  &    &-- &    0&.06 &  62&.8 \\
5427.826 & \ion{Fe}{II} & 6.72 &     &--  &    &-- & $-$1&.31 &   6&.7 & 4900.124 & \ion{Y}{II}  & 1.03 & $-$0&.29 &  55&.0 & $-$0&.07 &  60&.1 \\
6084.111 & \ion{Fe}{II} & 3.20 & $-$3&.75 &  24&.1 & $-$3&.78 &  22&.7 & 5087.426 & \ion{Y}{II}  & 1.08 & $-$0&.43 &  46&.9 & $-$0&.33 &  48&.5 \\
6149.249 & \ion{Fe}{II} & 3.89 & $-$2&.76 &  38&.5 & $-$2&.73 &  38&.6 & 5200.415 & \ion{Y}{II}  & 0.99 & $-$0&.70 &  39&.0 & $-$0&.71 &  36&.5 \\
6247.562 & \ion{Fe}{II} & 3.89 & $-$2&.37 &  56&.4 & $-$2&.37 &  53&.9 & 5289.820 & \ion{Y}{II}  & 1.03 & $-$1&.77 &   5&.4 &     &--  &    &-- \\
6369.463 & \ion{Fe}{II} & 2.89 & $-$4&.14 &  20&.9 & $-$4&.15 &  20&.3 & 5402.780 & \ion{Y}{II}  & 1.84 & $-$0&.48 &  15&.6 & $-$0&.61 &  15&.0 \\
6383.715 & \ion{Fe}{II} & 5.55 &     &--  &    &-- & $-$2&.07 &  10&.7 & 4739.454 & \ion{Zr}{I}  & 0.65 &    0&.00 &   7&.5 &    0&.04 &   7&.9 \\
6385.458 & \ion{Fe}{II} & 5.55 &     &--  &    &-- & $-$2&.44 &   5&.1 & 4613.921 & \ion{Zr}{II} & 0.97 & $-$0&.62 &  37&.1 & $-$0&.61 &  35&.7 \\
6416.928 & \ion{Fe}{II} & 3.89 & $-$2&.65 &  43&.5 & $-$2&.69 &  40&.5 & 5112.279 & \ion{Zr}{II} & 1.66 & $-$0&.75 &  10&.2 & $-$0&.81 &  12&.3 \\
6456.391 & \ion{Fe}{II} & 3.90 &     &--  &    &-- & $-$2&.24 &  59&.4 & 5853.688 & \ion{Ba}{II} & 0.60 & $-$1&.01 &  64&.0 & $-$0&.84 &  65&.6 \\
4749.662 & \ion{Co}{I}  & 3.05 &     &--  &  50&.0 &     &--  &  40&.0 & 6141.727 & \ion{Ba}{II} & 0.70 &    0&.13 & 120&.0 &    0&.24 & 119&.6 \\
4792.862 & \ion{Co}{I}  & 3.25 &     &--  &  34&.9 &     &--  &  34&.0 & 6496.908 & \ion{Ba}{II} & 0.60 & $-$0&.05 & 101&.0 & $-$0&.07 & 106&.2 \\
4813.479 & \ion{Co}{I}  & 3.21 &     &--  &  48&.0 &     &--  &  48&.6 & 4523.080 & \ion{Ce}{II} & 0.52 &    0&.24 &  21&.4 &     &--  &    &-- \\
5212.691 & \ion{Co}{I}  & 3.51 &     &--  &  18&.4 &     &--  &  20&.4 & 4562.367 & \ion{Ce}{II} & 0.48 &    0&.37 &  27&.9 &     &--  &    &-- \\
5280.629 & \ion{Co}{I}  & 3.63 &     &--  &    &-- &     &--  &  20&.6 & 4628.160 & \ion{Ce}{II} & 0.52 &    0&.21 &  20&.4 &    0&.27 &  21&.8 \\
5342.708 & \ion{Co}{I}  & 4.02 &     &--  &  32&.1 &     &--  &  32&.1 & 4773.959 & \ion{Ce}{II} & 0.92 &    0&.32 &  12&.7 &    0&.31 &  12&.1 \\
5359.192 & \ion{Co}{I}  & 4.15 &     &--  &  11&.0 &     &--  &  10&.4 & 5274.236 & \ion{Ce}{II} & 1.04 &    0&.48 &  14&.3 &    0&.40 &  12&.0 \\
5381.772 & \ion{Co}{I}  & 4.24 &     &--  &   9&.9 &     &--  &   6&.0 & 5089.831 & \ion{Nd}{II} & 0.20 & $-$1&.31 &   2&.7 & $-$1&.23 &   3&.5 \\
5454.572 & \ion{Co}{I}  & 4.07 &     &--  &    &-- &     &--  &  18&.2 & 5319.820 & \ion{Nd}{II} & 0.55 & $-$0&.21 &  14&.6 & $-$0&.17 &  15&.1 \\
5647.234 & \ion{Co}{I}  & 2.28 &     &--  &  15&.1 &     &--  &  14&.8 & 4566.233 & \ion{Sm}{II} & 0.33 & $-$0&.19 &  12&.3 & $-$0&.19 &  12&.0 \\
6000.678 & \ion{Co}{I}  & 3.62 &     &--  &   4&.3 &     &--  &   5&.8 &      &      &  &     &    &    &   &     &    &    &   \\
\hline
\end{tabular}
\end{table*}

\begin{table*}
\centering
\caption[]{Oscillator strengths (\loggf) for lines with hyperfine structure
           computed based on the Ganymede spectra observed in the first
           (Gany 1) and second (Gany 2) runs.}
\label{hfs}
{\scriptsize
\begin{tabular}{c r@{}l r@{}l | c r@{}l r@{}l | c c c | c c c | c c c}
\hline\hline
 & & & & & & & & & & & & & & & & & & \\[-0.2cm]
\multirow{2}{*}{$\lambda$ \tiny [\AA]} & \mc{4}{c|}{\loggf} &
\multirow{2}{*}{$\lambda$ \tiny [\AA]} & \mc{4}{c|}{\loggf} &
\multirow{2}{*}{$\lambda$ \tiny [\AA]} & \mc{2}{c|}{\loggf} &
\multirow{2}{*}{$\lambda$ \tiny [\AA]} & \mc{2}{c|}{\loggf} &
\multirow{2}{*}{$\lambda$ \tiny [\AA]} & \mc{2}{c}{\loggf} \\
 & \mc{2}{c}{Gany 1} & \mc{2}{c|}{Gany 2} &
 & \mc{2}{c}{Gany 1} & \mc{2}{c|}{Gany 2} &
 & Gany 1 & Gany 2 & & Gany 1 & Gany 2 & & Gany 1 & Gany 2 \\[0.1cm]
\hline
 & & & & & & & & & & & & & & & & & & \\[-0.2cm]
\mc{5}{c|}{\bf \ion{Mg}{I} : 4571.102}  & \mc{5}{c|}{\bf \ion{Sc}{II} : 5657.874} & \mc{3}{c|}{\bf \ion{V}{I} : 6274.658}  & \mc{3}{c|}{\bf \ion{Mn}{I} : 5537.765} & \mc{3}{c}{\bf \ion{Co}{I} : 5381.772} \\
4571.078 & $-$6&.67 & $-$6&.49      & 5657.808 & $-$1&.33 & $-$1&.32      & 6274.640 & --      & $-$2.09       & 5537.691 & $-$2.76 & $-$2.79       & 5381.695 & $-$0.47 & $-$0.69      \\
4571.087 & $-$6&.71 & $-$6&.53      & 5657.841 & $-$1&.23 & $-$1&.22      & 6274.658 & --      & $-$2.09       & 5537.710 & $-$2.64 & $-$2.67       & 5381.738 & $-$0.55 & $-$0.77      \\
4571.096 & $-$5&.81 & $-$5&.63      & 5657.874 & $-$1&.44 & $-$1&.43      & 6274.676 & --      & $-$2.09       & 5537.798 & $-$2.61 & $-$2.64       & 5381.772 & $-$0.63 & $-$0.85      \\
\mc{5}{c|}{\bf \ion{Mg}{I} : 4730.038}  & 5657.893 & $-$1&.33 & $-$1&.32      & \mc{3}{c|}{\bf \ion{V}{I} : 6285.165}  & 5537.764 & $-$2.64 & $-$2.67       & 5381.799 & $-$0.74 & $-$0.96      \\
4730.031 & $-$3&.13 & $-$3&.10      & \mc{5}{c|}{\bf \ion{Sc}{II} : 5684.189} & 6285.147 & $-$2.11 & $-$1.86       & 5537.802 & $-$2.28 & $-$2.31       & 5381.824 & $-$0.69 & $-$0.91      \\
4730.038 & $-$3&.17 & $-$3&.14      & 5684.123 & $-$1&.72 & $-$1&.71      & 6285.165 & $-$2.11 & $-$1.86       & \mc{3}{c|}{\bf \ion{Mn}{I} : 6013.497} & \mc{3}{c}{\bf \ion{Co}{I} : 5454.572} \\
4730.046 & $-$2&.27 & $-$2&.24      & 5684.156 & $-$1&.62 & $-$1&.61      & 6285.183 & $-$2.11 & $-$1.86       & 6013.474 & $-$0.75 & $-$0.67       & 5454.495 &  --     & $-$0.38      \\
\mc{5}{c|}{\bf \ion{Mg}{I} : 5711.095}  & 5684.189 & $-$1&.83 & $-$1&.82      & \mc{3}{c|}{\bf \ion{Mn}{I} : 4626.538} & 6013.486 & $-$0.96 & $-$0.88       & 5454.538 &  --     & $-$0.45      \\
5711.074 & $-$2&.69 & $-$2&.69      & 5684.208 & $-$1&.72 & $-$1&.71      & 4626.464 & $-$0.94 & $-$0.99       & 6013.501 & $-$1.10 & $-$1.02       & 5454.572 &  --     & $-$0.54      \\
5711.083 & $-$2&.73 & $-$2&.73      & \mc{5}{c|}{\bf \ion{Sc}{II} : 6245.660} & 4626.504 & $-$0.14 & $-$0.19       & 6013.519 & $-$0.77 & $-$0.69       & 5454.599 &  --     & $-$0.64      \\
5711.091 & $-$1&.83 & $-$1&.83      & 6245.661 & $-$1&.78 & $-$1&.81      & 4626.530 & $-$0.39 & $-$0.44       & 6013.537 & $-$1.35 & $-$1.27       & 5454.624 &  --     & $-$0.59      \\
\mc{5}{c|}{\bf \ion{Sc}{I} : 4743.817}  & 6245.642 & $-$1&.89 & $-$1&.92      & 4626.565 & $-$0.19 & $-$0.24       & \mc{3}{c|}{\bf \ion{Mn}{I} : 6021.803} & \mc{3}{c}{\bf \ion{Co}{I} : 5647.234} \\
4743.751 &     &--  & $-$0&.37      & 6245.609 & $-$1&.68 & $-$1&.71      & 4626.573 & $-$0.49 & $-$0.54       & 6021.764 & $-$1.37 & $-$1.24       & 5647.191 & $-$2.18 & $-$2.17      \\
4743.784 &     &--  & $-$0&.28      & 6245.576 & $-$1&.77 & $-$1&.80      & \mc{3}{c|}{\bf \ion{Mn}{I} : 4739.113} & 6021.780 & $-$1.22 & $-$1.09       & 5647.200 & $-$2.25 & $-$2.24      \\
4743.817 &     &--  & $-$0&.49      & \mc{5}{c|}{\bf \ion{Sc}{II} : 6320.867} & 4739.099 & $-$1.30 & $-$1.23       & 6021.797 & $-$0.39 & $-$0.26       & 5647.234 & $-$2.34 & $-$2.33      \\
4743.836 &     &--  & $-$0&.38      & 6320.884 & $-$2&.52 & $-$2&.57      & 4739.113 & $-$1.44 & $-$1.37       & 6021.806 & $-$0.60 & $-$0.47       & 5647.261 & $-$2.44 & $-$2.43      \\
\mc{5}{c|}{\bf \ion{Sc}{I} : 5356.091}  & 6320.865 & $-$2&.63 & $-$2&.68      & 4739.126 & $-$1.60 & $-$1.53       & 6021.814 & $-$0.47 & $-$0.34       & 5647.291 & $-$2.39 & $-$2.38      \\
5356.025 &     &--  & $-$0&.67      & 6320.832 & $-$2&.42 & $-$2&.47      & 4739.145 & $-$1.15 & $-$1.08       & \mc{3}{c|}{\bf \ion{Co}{I} : 4749.662} & \mc{3}{c}{\bf \ion{Co}{I} : 6000.678} \\
5356.058 &     &--  & $-$0&.58      & 6320.799 & $-$2&.52 & $-$2&.57      & 4739.167 & $-$2.50 & $-$2.43       & 4749.641 & $-$0.63 & $-$0.77       & 6000.607 & $-$2.21 & $-$2.09      \\
5356.091 &     &--  & $-$0&.79      & \mc{5}{c|}{\bf \ion{V}{I} : 5657.436}   & \mc{3}{c|}{\bf \ion{Mn}{I} : 5004.892} & 4749.675 & $-$0.81 & $-$0.85       & 6000.649 & $-$1.50 & $-$1.38      \\
5356.110 &     &--  & $-$0&.68      & 5657.418 &     &--  & $-$1&.34      & 5004.878 & --      & $-$2.08       & 4749.704 & $-$0.92 & $-$1.06       & 6000.678 & $-$1.38 & $-$1.26      \\
\mc{5}{c|}{\bf \ion{Sc}{I} : 5392.075}  & 5657.436 &     &--  & $-$1&.34      & 5004.892 & --      & $-$2.22       & 4749.729 & $-$0.88 & $-$1.02       & 6000.717 & $-$1.55 & $-$1.43      \\
5392.009 &     &--  &    0&.04      & 5657.454 &     &--  & $-$1&.34      & 5004.905 & --      & $-$2.38       & 4749.764 & $-$2.28 & $-$2.42       & 6000.752 & $-$1.54 & $-$1.42      \\
5392.042 &     &--  &    0&.13      & \mc{5}{c|}{\bf \ion{V}{I} : 5668.362}   & 5004.924 & --      & $-$1.93       & \mc{3}{c|}{\bf \ion{Co}{I} : 4792.862} & \mc{3}{c}{\bf \ion{Co}{I} : 6455.001} \\
5392.075 &     &--  & $-$0&.08      & 5668.344 & $-$1&.37 & $-$1&.47      & 5004.946 & --      & $-$3.28       & 4792.811 & $-$2.12 & $-$2.11       & 6454.931 & --  & $-$0.56      \\
5392.084 &     &--  &    0&.03      & 5668.362 & $-$1&.37 & $-$1&.47      & \mc{3}{c|}{\bf \ion{Mn}{I} : 5394.670} & 4792.827 & $-$1.46 & $-$1.45       & 6454.979 & --  & $-$1.35      \\
\mc{5}{c|}{\bf \ion{Sc}{I} : 5484.611}  & 5668.380 & $-$1&.37 & $-$1&.47      & 5394.617 & --      & $-$3.93       & 4792.840 & $-$0.95 & $-$0.94       & 6455.001 & --  & $-$1.22      \\
5484.545 &     &--  & $-$0&.45      & \mc{5}{c|}{\bf \ion{V}{I} : 5670.851}   & 5394.645 & --      & $-$4.01       & 4792.855 & $-$0.53 & $-$0.52       & 6455.022 & --  & $-$1.38      \\
5484.578 &     &--  & $-$0&.36      & 5670.833 & $-$0&.93 & $-$0&.97      & 5394.670 & --      & $-$4.12       & 4792.864 & $-$0.50 & $-$0.49       & 6455.044 & --  & $-$1.31      \\
5484.611 &     &--  & $-$0&.57      & 5670.851 & $-$0&.93 & $-$0&.97      & 5394.689 & --      & $-$4.27       & \mc{3}{c|}{\bf \ion{Co}{I} : 4813.479} & \mc{3}{c}{\bf \ion{Cu}{I} : 5218.209} \\
5484.630 &     &--  & $-$0&.46      & 5670.869 & $-$0&.93 & $-$0&.97      & 5394.703 & --      & $-$4.36       & 4813.428 & $-$1.50 & $-$1.44       & 5218.059 & $-$1.50 & $-$1.40      \\
\mc{5}{c|}{\bf \ion{Sc}{I} : 5671.826}  & \mc{5}{c|}{\bf \ion{V}{I} : 5727.661}   & \mc{3}{c|}{\bf \ion{Mn}{I} : 5399.479} & 4813.451 & $-$1.02 & $-$0.96       & 5218.061 & $-$1.02 & $-$0.92      \\
5671.760 &     &--  & $-$0&.17      & 5727.075 &    0&.71 &    0&.80      & 5399.435 & --      & $-$0.91       & 4813.469 & $-$0.48 & $-$0.42       & 5218.063 & $-$1.15 & $-$1.03      \\
5671.793 &     &--  & $-$0&.07      & 5727.057 &    0&.71 &    0&.80      & 5399.446 & --      & $-$1.10       & 4813.481 & $-$0.41 & $-$0.35       & 5218.065 & $-$0.43 & $-$0.35      \\
5671.826 &     &--  & $-$0&.28      & 5727.038 &    0&.71 &    0&.80      & 5399.479 & --      & $-$1.01       & 4813.492 & $-$0.54 & $-$0.48       & 5218.069 & $-$0.65 & $-$0.55      \\
5671.845 &     &--  & $-$0&.17      & \mc{5}{c|}{\bf \ion{V}{I} : 6090.216}   & 5399.502 & --      & $-$0.63       & \mc{3}{c|}{\bf \ion{Co}{I} : 5212.691} & 5218.071 & $-$0.65 & $-$0.55      \\
\mc{5}{c|}{\bf \ion{Sc}{I} : 6239.408}  & 6090.234 & $-$0&.65 & $-$0&.63      & 5399.536 & --      & $-$1.26       & 5212.614 & $-$1.66 & $-$1.60       & 5218.074 & $-$0.31 & $-$0.21      \\
6239.342 &     &--  & $-$1&.83      & 6090.216 & $-$0&.65 & $-$0&.63      & \mc{3}{c|}{\bf \ion{Mn}{I} : 5413.684} & 5212.656 & $-$0.96 & $-$0.90       & \mc{3}{c}{\bf \ion{Cu}{I} : 5220.086} \\
6239.375 &     &--  & $-$1&.73      & 6090.198 & $-$0&.65 & $-$0&.63      & 5413.613 & $-$1.84 & $-$1.81       & 5212.685 & $-$0.83 & $-$0.77       & 5220.080 & $-$2.26 & $-$2.25      \\
6239.408 &     &--  & $-$1&.94      & \mc{5}{c|}{\bf \ion{V}{I} : 6135.370}   & 5413.653 & $-$1.04 & $-$1.01       & 5212.724 & $-$1.00 & $-$0.94       & 5220.082 & $-$1.78 & $-$1.77      \\
6239.427 &     &--  & $-$1&.83      & 6135.352 & $-$1&.25 & $-$1&.27      & 5413.679 & $-$1.29 & $-$1.26       & 5212.759 & $-$0.99 & $-$0.93       & 5220.084 & $-$1.91 & $-$1.90      \\
\mc{5}{c|}{\bf \ion{Sc}{II} : 5318.346} & 6135.370 & $-$1&.25 & $-$1&.27      & 5413.714 & $-$0.93 & $-$0.90       & \mc{3}{c|}{\bf \ion{Co}{I} : 5280.629} & 5220.086 & $-$1.20 & $-$1.19      \\
5318.280 &     &--  & $-$2&.30      & 6135.388 & $-$1&.25 & $-$1&.27      & 5413.722 & $-$1.39 & $-$1.36       & 5280.559 & --  & $-$0.35       & 5220.090 & $-$1.41 & $-$1.40      \\
5318.313 &     &--  & $-$2&.20      & \mc{5}{c|}{\bf \ion{V}{I} : 6150.154}   & \mc{3}{c|}{\bf \ion{Mn}{I} : 5420.350} & 5280.607 & --  & $-$1.14       & 5220.092 & $-$1.41 & $-$1.40      \\
5318.346 &     &--  & $-$2&.41      & 6150.136 & $-$1&.99 & $-$2&.03      & 5420.277 & $-$2.32 & $-$2.26       & 5280.629 & --  & $-$1.01       & 5220.095 & $-$1.09 & $-$1.08      \\
5318.365 &     &--  & $-$2&.30      & 6150.154 & $-$1&.99 & $-$2&.03      & 5420.301 & $-$2.24 & $-$2.16       & 5280.650 & --  & $-$1.17       & \mc{3}{c}{\bf \ion{Cu}{I} : 5782.136} \\
\mc{5}{c|}{\bf \ion{Sc}{II} : 5357.190} & 6150.172 & $-$1&.99 & $-$2&.03      & 5420.334 & $-$3.10 & $-$3.02       & 5280.672 & --  & $-$1.10       & 5782.032 & $-$3.58 & $-$3.48      \\
5357.124 & $-$2&.71 & $-$2&.76      & \mc{5}{c|}{\bf \ion{V}{I} : 6199.186}   & 5420.376 & $-$2.00 & $-$1.92       & \mc{3}{c}{\bf \ion{Co}{I} : 5342.708}  & 5782.042 & $-$3.89 & $-$3.79      \\
5357.157 & $-$2&.61 & $-$2&.66      & 6199.168 & $-$1&.91 & $-$1&.92      & 5420.429 & $-$1.91 & $-$1.83       & 5342.647 & $-$0.12 & $-$0.10       & 5782.054 & $-$3.19 & $-$3.09      \\
5357.190 & $-$2&.82 & $-$2&.87      & 6199.186 & $-$1&.91 & $-$1&.92      & \mc{3}{c|}{\bf \ion{Mn}{I} : 5432.548} & 5342.690 & $-$0.19 & $-$0.17       & 5782.064 & $-$3.24 & $-$3.14      \\
5357.209 & $-$2&.71 & $-$2&.76      & 6199.204 & $-$1&.91 & $-$1&.92      & 5432.512 & $-$4.37 & $-$4.31       & 5342.724 & $-$0.28 & $-$0.26       & 5782.073 & $-$3.54 & $-$3.44      \\
\mc{5}{c|}{\bf \ion{Sc}{II} : 5526.815} & \mc{5}{c|}{\bf \ion{V}{I} : 6216.358}   & 5432.540 & $-$4.45 & $-$4.39       & 5342.751 & $-$0.38 & $-$0.36       & 5782.084 & $-$2.84 & $-$2.74      \\
5526.749 & $-$0&.97 & $-$0&.95      & 6216.340 &     &--  & $-$1&.38      & 5432.565 & $-$4.56 & $-$4.50       & 5342.776 & $-$0.33 & $-$0.31       & 5782.086 & $-$3.19 & $-$3.09      \\
5526.782 & $-$0&.87 & $-$0&.85      & 6216.358 &     &--  & $-$1&.38      & 5432.584 & $-$4.71 & $-$4.65       & \mc{3}{c|}{\bf \ion{Co}{I} : 5359.192} & 5782.098 & $-$3.19 & $-$3.09      \\
5526.815 & $-$1&.08 & $-$1&.06      & 6216.376 &     &--  & $-$1&.38      & 5432.598 & $-$4.80 & $-$4.74       & 5359.115 & $-$0.51 & $-$0.53       & 5782.113 & $-$2.84 & $-$2.74      \\
5526.834 & $-$0&.97 & $-$0&.95      &      &     &    &     &         &          &         &           & 5359.158 & $-$0.59 & $-$0.61       & 5782.124 & $-$2.84 & $-$2.74      \\
     &     &    &     &         &      &     &    &     &         &          &         &           & 5359.192 & $-$0.67 & $-$0.69       & 5782.153 & $-$2.74 & $-$2.64      \\
     &     &    &     &         &      &     &    &     &         &          &         &           & 5359.219 & $-$0.78 & $-$0.80       & 5782.173 & $-$2.39 & $-$2.29      \\
     &     &    &     &         &      &     &    &     &         &          &         &           & 5359.244 & $-$0.73 & $-$0.75       &          &     &          \\
\hline
\end{tabular}
}
\end{table*}

\begin{table*}
\centering
\caption[]{Elemental abundance relative to iron. Our results for carbon
           abundances are presented in Table~\ref{carb_ab}.}
\label{ab_ratios}
\begin{tabular}{l r@{}l r@{}l r@{}l r@{}l r@{}l r@{}l r@{}l r@{}l r@{}l r@{}l}
\hline\hline\noalign{\smallskip}
Star &
\md{[Na/Fe]} & \md{[Mg/Fe]} & \md{[Si/Fe]} & \md{[Ca/Fe]} & \md{[Sc/Fe]} &
\md{[Ti/Fe]} & \md{[V/Fe] } & \md{[Cr/Fe]} & \md{[Mn/Fe]} & \md{[Co/Fe]} \\
\noalign{\smallskip}\hline\noalign{\smallskip}
HD\,1835   &    0&.04 &    0&.00 & $-$0&.01 &    0&.08 & $-$0&.01 & $-$0&.02 &    0&.08 &    0&.04 &    0&.04 & $-$0&.03 \\
HD\,20807  &    0&.05 &    0&.02 &    0&.00 &    0&.00 &    0&.06 &    0&.04 &    0&.04 & $-$0&.02 & $-$0&.06 &    0&.05 \\
HD\,26491  & $-$0&.02 &    0&.06 &    0&.02 &    0&.03 &    0&.06 &    0&.04 &    0&.07 &    0&.02 &    0&.00 &    0&.03 \\
HD\,33021  & $-$0&.05 &    0&.14 &    0&.08 &    0&.12 &    0&.11 &    0&.16 &    0&.11 & $-$0&.01 & $-$0&.18 &    0&.06 \\
HD\,39587  & $-$0&.10 & $-$0&.01 & $-$0&.04 &    0&.09 &    0&.01 &    0&.03 & $-$0&.03 &    0&.03 & $-$0&.06 & $-$0&.05 \\
HD\,43834  &    0&.00 &    0&.05 &    0&.00 & $-$0&.02 &    0&.02 & $-$0&.02 &    0&.04 & $-$0&.01 &    0&.07 &    0&.02 \\
HD\,50806  &    0&.04 &    0&.21 &    0&.09 &    0&.05 &    0&.06 &    0&.09 &    0&.08 &    0&.00 & $-$0&.04 &    0&.08 \\
HD\,53705  &    0&.03 &    0&.16 &    0&.06 &    0&.05 &    0&.13 &    0&.11 &    0&.07 &    0&.02 & $-$0&.12 &    0&.09 \\
HD\,84117  &    0&.08 & $-$0&.03 &    0&.03 & $-$0&.03 & $-$0&.01 & $-$0&.05 &    0&.03 & $-$0&.02 & $-$0&.09 & $-$0&.04 \\
HD\,102365 &     &--  &    0&.09 &    0&.03 &    0&.09 &    0&.06 &    0&.11 &    0&.05 & $-$0&.02 & $-$0&.11 &    0&.05 \\
HD\,112164 &    0&.08 &    0&.07 &    0&.04 & $-$0&.07 & $-$0&.04 & $-$0&.04 &    0&.05 & $-$0&.01 &    0&.06 &    0&.05 \\
HD\,114613 &    0&.08 &    0&.04 &    0&.02 & $-$0&.02 & $-$0&.01 &    0&.00 &    0&.02 &    0&.00 &    0&.08 &    0&.01 \\
HD\,115383 & $-$0&.08 &    0&.01 & $-$0&.03 & $-$0&.03 &    0&.02 &    0&.00 &    0&.04 & $-$0&.03 & $-$0&.01 & $-$0&.02 \\
HD\,115617 & $-$0&.04 &    0&.06 &    0&.00 &    0&.00 &    0&.05 &    0&.02 &    0&.01 & $-$0&.01 &    0&.03 & $-$0&.02 \\
HD\,117176 & $-$0&.10 &    0&.07 &    0&.01 &    0&.02 &    0&.03 &    0&.01 &    0&.04 & $-$0&.01 & $-$0&.05 &    0&.02 \\
HD\,128620 &    0&.06 &    0&.04 &    0&.01 & $-$0&.06 &    0&.03 &    0&.01 &    0&.07 & $-$0&.01 &    0&.07 &    0&.07 \\
HD\,141004 & $-$0&.05 &    0&.02 & $-$0&.01 & $-$0&.01 &    0&.04 &    0&.03 &    0&.04 & $-$0&.03 & $-$0&.04 &    0&.06 \\
HD\,146233 & $-$0&.03 &    0&.00 &    0&.03 &    0&.02 &    0&.07 & $-$0&.01 &    0&.07 &    0&.01 & $-$0&.01 & $-$0&.02 \\
HD\,147513 &     &--  & $-$0&.06 & $-$0&.04 &    0&.03 &    0&.05 &    0&.07 &    0&.12 & $-$0&.02 & $-$0&.07 & $-$0&.03 \\
HD\,160691 &    0&.09 &    0&.07 &    0&.03 & $-$0&.06 &    0&.00 & $-$0&.02 &    0&.07 & $-$0&.03 &    0&.07 &    0&.03 \\
HD\,177565 &    0&.02 &    0&.04 &    0&.01 & $-$0&.01 & $-$0&.04 &    0&.01 &    0&.03 &    0&.02 &    0&.05 &    0&.01 \\
HD\,181321 &     &--  & $-$0&.06 &    0&.03 &    0&.03 & $-$0&.07 &    0&.01 &     &--  &    0&.03 & $-$0&.13 &    0&.11 \\
HD\,188376 & $-$0&.03 &    0&.03 &    0&.02 &    0&.03 & $-$0&.07 & $-$0&.07 &    0&.03 &    0&.00 &    0&.01 & $-$0&.06 \\
HD\,189567 & $-$0&.03 &    0&.11 &    0&.04 &    0&.00 &    0&.10 &    0&.11 &    0&.03 &    0&.01 & $-$0&.12 &    0&.03 \\
HD\,196761 &    0&.00 &    0&.06 &    0&.02 & $-$0&.01 &    0&.06 &    0&.05 & $-$0&.01 &    0&.01 & $-$0&.02 &    0&.05 \\
\noalign{\smallskip}\hline\noalign{\smallskip}
 &
\md{[Ni/Fe]} & \md{[Cu/Fe]} & \md{[Zn/Fe]} & \md{[Sr/Fe]} & \md{[Y/Fe] } &
\md{[Zr/Fe]} & \md{[Ba/Fe]} & \md{[Ce/Fe]} & \md{[Nd/Fe]} & \md{[Sm/Fe]} \\
\noalign{\smallskip}\hline\noalign{\smallskip}
HD\,1835   & $-$0&.02 & $-$0&.01 &    0&.07 &    0&.15 &    0&.02 &    0&.09 &    0&.19 &    0&.02 &     &--  & $-$0&.29 \\
HD\,20807  & $-$0&.03 &    0&.02 & $-$0&.03 &    0&.04 & $-$0&.08 &    0&.04 & $-$0&.06 &    0&.07 &    0&.02 &    0&.15 \\
HD\,26491  &    0&.00 &    0&.03 &    0&.02 & $-$0&.08 & $-$0&.08 & $-$0&.07 &    0&.03 & $-$0&.06 & $-$0&.11 & $-$0&.08 \\
HD\,33021  & $-$0&.02 & $-$0&.03 & $-$0&.03 & $-$0&.06 & $-$0&.13 &    0&.13 &    0&.02 & $-$0&.03 &    0&.06 & $-$0&.02 \\
HD\,39587  & $-$0&.08 & $-$0&.11 & $-$0&.01 &    0&.19 &    0&.06 &    0&.16 &    0&.33 &    0&.16 &     &--  & $-$0&.14 \\
HD\,43834  &    0&.03 &    0&.03 &    0&.02 &    0&.06 & $-$0&.04 &    0&.07 & $-$0&.03 &    0&.00 & $-$0&.06 &    0&.06 \\
HD\,50806  &    0&.04 &    0&.15 &    0&.10 & $-$0&.08 & $-$0&.06 &    0&.11 &    0&.00 &    0&.01 & $-$0&.09 &    0&.10 \\
HD\,53705  & $-$0&.02 &    0&.05 &    0&.13 & $-$0&.09 & $-$0&.15 &    0&.00 & $-$0&.02 & $-$0&.08 & $-$0&.07 & $-$0&.13 \\
HD\,84117  & $-$0&.04 & $-$0&.03 & $-$0&.03 &    0&.09 &    0&.07 &    0&.15 &    0&.15 & $-$0&.02 &    0&.09 &    0&.00 \\
HD\,102365 & $-$0&.02 &    0&.02 &    0&.07 & $-$0&.02 & $-$0&.06 & $-$0&.10 &    0&.12 & $-$0&.06 &    0&.04 &    0&.02 \\
HD\,112164 &    0&.03 &    0&.14 &    0&.00 & $-$0&.06 & $-$0&.16 & $-$0&.06 & $-$0&.03 & $-$0&.09 & $-$0&.09 & $-$0&.10 \\
HD\,114613 &    0&.03 &    0&.10 &    0&.02 & $-$0&.02 & $-$0&.06 & $-$0&.01 &    0&.00 & $-$0&.14 & $-$0&.11 & $-$0&.02 \\
HD\,115383 & $-$0&.03 &    0&.05 & $-$0&.07 &    0&.00 &    0&.01 &    0&.06 &    0&.20 & $-$0&.04 &    0&.16 &    0&.02 \\
HD\,115617 & $-$0&.01 & $-$0&.01 &    0&.08 &    0&.05 & $-$0&.11 &    0&.01 & $-$0&.02 &    0&.05 & $-$0&.07 &    0&.28 \\
HD\,117176 &    0&.00 &    0&.07 & $-$0&.06 & $-$0&.03 & $-$0&.06 &    0&.04 &    0&.06 &    0&.06 &    0&.19 &    0&.11 \\
HD\,128620 &    0&.04 &    0&.14 & $-$0&.08 &     &--  & $-$0&.06 &    0&.06 & $-$0&.08 & $-$0&.06 & $-$0&.01 &    0&.04 \\
HD\,141004 & $-$0&.01 &    0&.10 & $-$0&.05 & $-$0&.02 & $-$0&.02 & $-$0&.05 &    0&.03 & $-$0&.07 &     &--  & $-$0&.10 \\
HD\,146233 & $-$0&.01 & $-$0&.02 & $-$0&.04 &    0&.17 &    0&.00 &    0&.06 &    0&.12 &    0&.02 &    0&.13 &    0&.09 \\
HD\,147513 & $-$0&.06 & $-$0&.11 & $-$0&.08 &     &--  &    0&.22 &    0&.24 &    0&.43 &    0&.06 &    0&.18 &     &--  \\
HD\,160691 &    0&.03 &    0&.18 &    0&.11 &    0&.01 & $-$0&.06 &    0&.07 &    0&.01 & $-$0&.08 & $-$0&.09 &    0&.09 \\
HD\,177565 &    0&.04 &    0&.04 &    0&.02 & $-$0&.06 & $-$0&.04 & $-$0&.09 & $-$0&.01 &    0&.02 & $-$0&.10 &    0&.07 \\
HD\,181321 & $-$0&.09 &     &--  & $-$0&.14 &     &--  & $-$0&.04 &     &--  &    0&.23 &     &--  &     &--  &     &--  \\
HD\,188376 & $-$0&.02 & $-$0&.03 & $-$0&.10 &     &--  &    0&.00 &    0&.04 &    0&.21 & $-$0&.07 &    0&.20 & $-$0&.19 \\
HD\,189567 &    0&.00 & $-$0&.04 &    0&.03 & $-$0&.07 & $-$0&.07 &    0&.03 &    0&.06 &    0&.08 &    0&.06 &    0&.14 \\
HD\,196761 &    0&.00 &    0&.01 &    0&.08 & $-$0&.09 & $-$0&.16 &    0&.05 & $-$0&.03 & $-$0&.04 & $-$0&.04 &    0&.26 \\
\noalign{\smallskip}\hline
\end{tabular}
\end{table*}

\end{document}